\documentclass[aip,reprint,twocolumn, superscriptaddress,preprintnumbers, showpacs, floatfix]{revtex4-1}

\usepackage{ifthen} 
\usepackage{graphicx}
\usepackage{amsmath}
\usepackage{amssymb}
\usepackage{braket}
\usepackage{cancel}
\usepackage{hyperref}
\hypersetup{backref,pdfpagemode=FullScreen,colorlinks=true,allcolors=blue}
\usepackage{color}
\usepackage{textcomp}
\usepackage[caption=false]{subfig}

\def\ketbra#1#2{\lvert#1\rangle\langle#2\rvert}

\begin{document}

\title{A self-consistent ground-state 
formulation of the first-principles Hubbard U 
parameter validated on one-electron self-interaction error}

\author{Glenn Moynihan}
\email{omuinneg@tcd.ie}
\affiliation{School of Physics, CRANN and AMBER, Trinity College Dublin, Dublin 2, Ireland}

\author{Gilberto Teobaldi}
\affiliation{Stephenson Institute for Renewable Energy and  Department of Chemistry, The University of Liverpool, L69 3BX Liverpool, United Kingdom}
\affiliation{Beijing Computational Science Research Center, Beijing 100094, China}

\author{David D. O'Regan}
\affiliation{School of Physics, CRANN and AMBER, Trinity College Dublin, Dublin 2, Ireland}

\date{\today{}}

\begin{abstract}
In electronic structure methods based on
the correction of approximate density-functional theory
(DFT) for systematic inaccuracies, Hubbard $U$ parameters may
be used 
to quantify and amend the self-interaction 
errors ascribed to selected subspaces.
Here, in order to enable the  
accurate, computationally convenient 
calculation of $U$  by means of DFT algorithms that
locate the ground-state by direct total-energy
 minimization, we introduce a 
reformulation of the successful 
linear-response method for $U$ in terms of the  
fully-relaxed constrained ground-state density.
Defining $U$ as an implicit functional of the ground-state density
implies the comparability of 
DFT + Hubbard $U$ (DFT+$U$) 
total-energies, and related properties, as external parameters such as
ionic positions are varied together with their 
corresponding first-principles  $U$ values.
Our approach  provides a framework in which 
to address the partially unresolved  question of self-consistency over  
$U$, for which  plausible 
schemes have been proposed, and to precisely define the 
energy associated with subspace many-body self-interaction error.
We demonstrate that DFT+$U$ 
precisely corrects the total energy
for self-interaction error under ideal conditions, 
but only if a simple  
self-consistency condition is applied.
Such parameters also promote to first-principles  a 
recently proposed DFT+$U$ based method
for enforcing Koopmans' theorem.

\end{abstract}


\maketitle

\section{Introduction}\label{introduction}
Approximate density-functional theory 
(DFT)~\cite{PhysRev.136.B864,PhysRev.140.A1133}  
is a central element in the simulation of 
many-body atomistic systems and an indispensable partner to 
experiment~\cite{RevModPhys.87.897,Esch752,
jain2016computational}.
DFT is prone, however, within its commonplace 
local-density (LDA)~\cite{PhysRevB.23.5048},
generalized-gradient (GGA)~\cite{PhysRevLett.77.3865},
and hybrid~\cite{doi:10.1063/1.464913,Yanai200451,doi:10.1063/1.1412605} 
exchange-correlation (xc) approximations,
to significant systematic errors~\cite{PhysRevB.57.1505,Cohen792}.
The most widely encountered of these
is the many-electron self-interaction error 
(SIE)~\cite{PhysRevB.23.5048}, or delocalization~\cite{Cohen792} 
error, which is manifested as a spurious curvature
in the total-energy profile
of a system with respect to its total electron 
number~\cite{PhysRevLett.49.1691}. 
The SIE contributes to inaccuracies in 
insulating band gaps~\cite{PhysRevB.93.195208}, 
charge-transfer energies~\cite{PhysRevLett.95.146402,:/content/aip/journal/jcp/126/20/10.1063/1.2743004},  
activation barriers~\cite{doi:10.1021/jp049908s} ,
 binding and formation energies,  
as well as in spin-densities and their moments.
While the nature of SIE is well understood,
it remains persistently challenging to reliably avoid its introduction using 
approximate xc functionals of a computationally tractable, explicit
analytical form, even if exact exchange is incorporated 
 (see, e.g.,  the B3LYP curve in 
Fig. 2 of Ref.~\onlinecite{doi:10.1021/cr200107z}).

DFT+$U$  (DFT + Hubbard $U$)~\cite{PhysRevB.43.7570, 
PhysRevB.44.943,PhysRevB.48.16929,PhysRevB.57.1505,PhysRevB.58.1201,PhysRevB.71.035105}
is a computationally efficient~\cite{PhysRevB.85.085107} and
formally straightforward method 
that has matured as a corrective approach for SIE
in systems where it may be reasonably attributed
to particular selected subspaces~\cite{PhysRevLett.97.103001,:/content/aip/journal/jcp/133/11/10.1063/1.3489110,:/content/aip/journal/jcp/145/5/10.1063/1.4959882}.
Originally designed to capture Mott-Hubbard 
physics in transition-metal oxides~\cite{PhysRevB.44.943,PhysRevB.57.1505,PhysRevB.58.1201,0953-8984-9-4-002}, 
it now sees very diverse 
applications~\cite{QUA:QUA24521, doi:10.1021/jp3107809,PhysRevB.93.085135,doi:10.1021/jz3004188,PhysRevLett.113.086402}.
It has gained a transparent interpretation as an efficient corrective
method for  SIE  
particularly since the work of Kulik, Cococcioni, and co-workers in 
Ref.~\onlinecite{PhysRevLett.97.103001}.
The DFT+$U$ corrective energy term
is often invoked in its  rotationally-invariant, 
simplified form~\cite{PhysRevB.71.035105,PhysRevB.84.115108,PhysRevB.93.085135,PhysRevLett.97.103001}, given by
\begin{equation}
E_U \left[ \hat n^{I\sigma} \right]=\sum_{I,\sigma} \frac{U^I}{2}
\textrm{Tr}
\left[ \hat{n}^{I \sigma}
- \hat{n}^{I \sigma}  \hat{n}^{I \sigma}
\right],
\label{Eq:dft+u}
\end{equation}
where the density-matrices
$\hat{n}^{I \sigma} = \hat{P}^I \hat{\rho}^\sigma \hat{P}^I$
are those for the subspaces $I$
over which the SIE is to be  corrected.
Here, the Kohn-Sham density-matrix 
 $\hat{\rho}^\sigma$ corresponds to the 
spin indexed by $\sigma$, which we hereafter suppress for simplicity, 
and the idempotent subspace projection operators 
$\hat{P}^I=\sum_m\ketbra{\varphi^I_m}{\varphi^{I m}}$ 
are usually built from fixed, spin-independent, 
orthonormal, localized orbitals,
which may also be nonorthogonal~\cite{PhysRevB.83.245124} 
and  self-consistent~\cite{PhysRevB.82.081102}.

The quadratic term of Eq.~\ref{Eq:dft+u} 
alters the intra-subspace
self-interaction,
on a one-electron basis in the frame of the individual
orthonormal eigenstates of $\hat{n}^{I}$, 
which may generally 
be expected to change as a result.
The linear term then imposes the condition 
that the correction to the total-energy should vanish 
for each subspace eigenstate as its corresponding 
eigenvalue $n^I_i$ approaches zero or one, 
implying that the xc functional is 
assumed to be correct for such eigenstates.
This mirrors the well-known result 
that the total-energy of open systems at integer filling 
is reasonably well described by conventional 
approximate xc functionals~\cite{PhysRevB.49.6736}.
While the linear term does not 
directly affect the SIE explicitly,  
it represents an important boundary
condition on the SIE correction.
Simultaneously , the corresponding correction to the potential 
$\hat{v}_U^I = U^I(\hat{1}-2\hat{n}^I)/2$
vanishes at eigenvalues of one-half and, 
when a Kohn-Sham gap is symmetry-allowed,  
the occupancy-dependence of the potential 
acts to energetically split states 
lesser and greater in occupancy than one-half  
by an energy interval 
on the order of $U$~\cite{QUA:QUA24521}.

\subsection{A one-electron litmus test:  how DFT+$U$ affects  
$H_2^+$}

As it exhibits no  multi-reference or static correlation
error effects, by definition, 
but a straightforwardly variable bonding regime,  
the dihydrogen cation
$H_2^+$ is perhaps the ideal system for the study of pure SIE, also
known as delocalization error~\cite{Cohen792}.
It serves as a convenient test bed for the exploration of 
system-specific  additive corrections, such as DFT+$U$,
and more generally for density-functionals which are, at
least in part, implicitly defined via parameters  to be
calculated, such as the self-consistent Hubbard $U$. 
Particularly subject to  the ideal  population analysis 
and non-overlapping subspace conditions available 
in the dissociated limit, $H_2^+$ will  allow us
to draw firm conclusions regarding 
the numerous plausible but
 different strategies 
currently in use for 
defining self-consistency over the Hubbard $U$.

\begin{figure}[t]
  \captionsetup[subfigure]{labelformat=empty}
  \subfloat[]{
  	\centering
  	\includegraphics[width=\columnwidth]{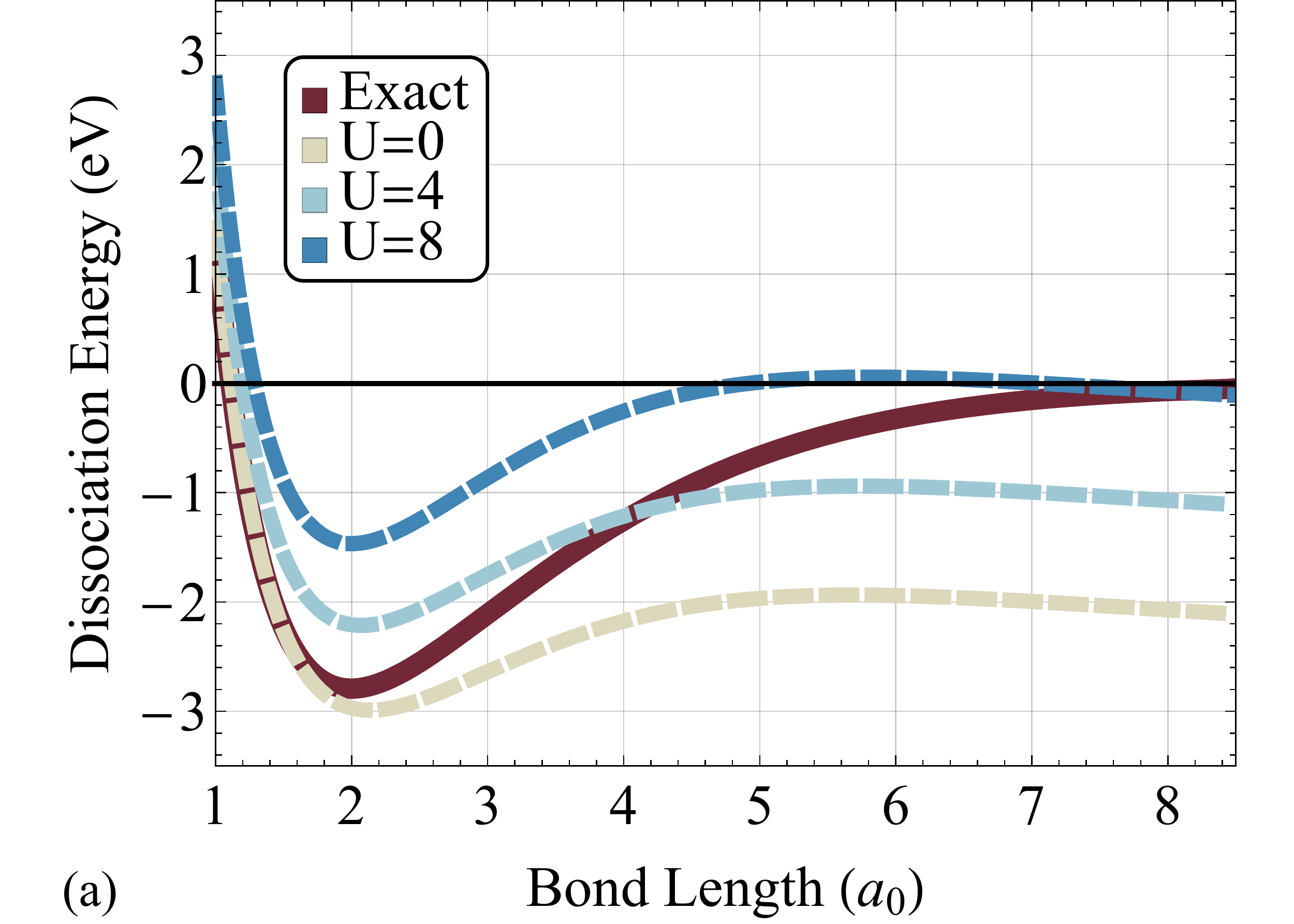}
	\label{figure1a}}
    \newline
  \subfloat[]{
  	\centering
	\includegraphics[width=\columnwidth]{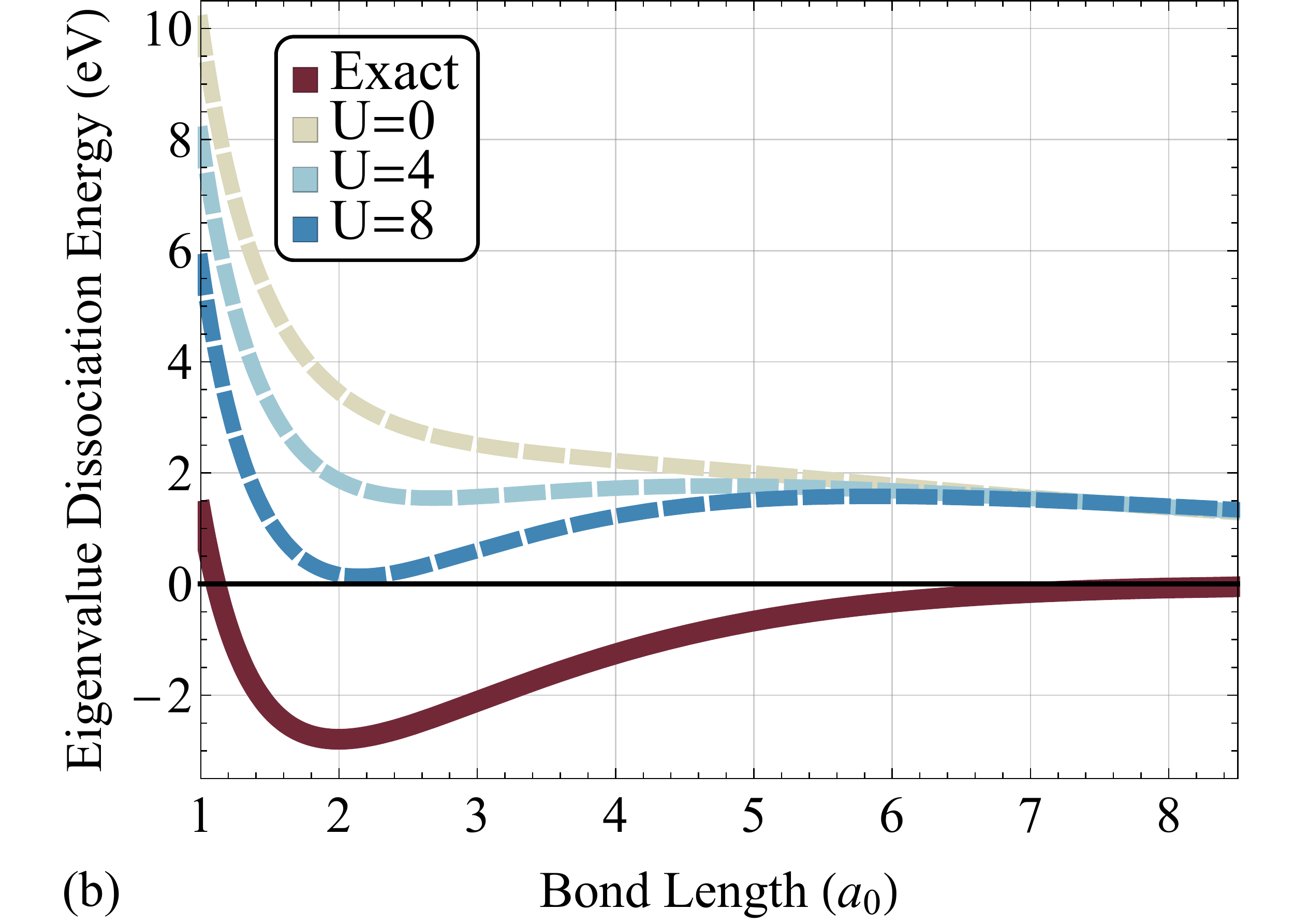}
	\label{figure1b}}
\caption{(Color online) 
The dissociation energy curve 
of  $H_2^+$ calculated from 
(a) the total-energy and 
(b) the occupied Kohn-Sham eigenvalue 
using the 
exact xc-functional (solid), 
and PBE+$U$ with 
$U= 0, 4, 8$~eV (dashed).
PBE+$U$ successfully resolves the SIE in the total-energy,
for a strongly bond-length dependent Hubbard $U$,
but it does not address the inaccuracy in the Kohn-Sham
eigenvalue, or non-compliance with Koopmans' condition, 
which grows with bond-length.}
\end{figure}

The action of the DFT+$U$ functional under varying
bonding conditions may be observed 
in the dissociation curves of $H_2^+$ depicted in 
Fig.~\ref{figure1a}.
Here, the total-energy error in approximate DFT,
specifically the PBE xc-functional~\cite{PhysRevLett.77.3865},
is seen to grow significantly with bond-length as 
the electron count on each atom 
approaches one-half.
The result of the exact xc-functional, 
in which the Hartree and xc energies and potentials cancel, 
is indicated by the solid line, 
and the results of 
DFT(PBE)+$U = 0, 4, 8$~eV, 
are indicated by the  dashed 
lines~\footnote{Calculations were performed 
using the DFT+$U$ 
functionality~\cite{PhysRevB.85.085107}
available in the \textsc{ONETEP} linear-scaling
DFT package~\cite{:/content/aip/journal/jcp/122/8/10.1063/1.1839852} 
with a hard ($0.65$~a$_0$ cutoff) 
norm-conserving pseudopotential~\cite{PhysRevB.41.1227}, 
$10$~a$_0$ Wannier function cutoff radii, and open 
boundary conditions~\cite{:/content/aip/journal/jcp/110/6/10.1063/1.477923}.
DFT+$U$ was applied simultaneously to each an atom,
using a separate $1s$  orbital subspace
 centred on each, defined using 
 the occupied Kohn-Sham state of the pseudopotential
 for neutral hydrogen. The correct symmetry of $H_2^+$
 was maintained for all values of $U$
 given a symmetric initial guess, i.e., we observed
 no tendency for the charge to localize on a single ion.}.
The Hubbard $U$ parameter required to correct the
PBE total-energy to the exact value 
varies over approximately $8$~eV from the fully bonded to 
dissociated limits, highlighting the 
importance of chemical environment
dependent, and not just species-dependent,  $U$
parameters, as previously  shown, 
e.g., in Refs.~\onlinecite{doi:10.1063/1.3660353,
doi:10.1063/1.4865831,PhysRevB.90.115105}.

A critical and perhaps defining characteristic 
of an SIE-free system is 
its compliance with Koopmans' 
condition~\cite{PhysRevB.82.115121,
PhysRevB.90.075135}, 
and in a one-electron system such as  $H_2^+$
this implies that the total-energy and the 
occupied Kohn-Sham eigenvalue $\varepsilon$
should differ only by the ion-ion energy. 
Thus, the dissociation curve of $H_2^+$ 
should be equivalently accessible by calculating the
total energies for the dimer and its constituent atoms directly, 
or by using total energies  derived from the occupied eigenvalue
and the expression  
$E = \varepsilon + E_\textrm{ion-ion}$.
Fig.~\ref{figure1b} illustrates the
 strikingly poor results of DFT+$U$ when combined
 with this latter 
procedure~\footnote{In all dissociation curves presented, 
the fully dissociated reference energy was fixed to
the total-energy 
of a single exact-functional hydrogen atom placed at the
midpoint of the dimer, which is equal to 
the occupied Kohn-Sham eigenvalue of the same system, or
two half-charged exact hydrogen atoms.
In this way, only the SIE specific to the PBE dimer is analysed, 
without the SIE present in  isolated, 
half-charged PBE hydrogen atoms.}.
A $U \gtrsim 4$~eV is required for the eigenvalue-derived
dissociation curve to exhibit a local minimum.
We observe that the non-compliance with  
Koopmans' condition (disagreement between
Figs.~\ref{figure1a} and~\ref{figure1b}) broadly decreases both with 
bond-length and with the Hubbard $U$, 
but that the effect of DFT+$U$ on the eigenvalue is lost entirely
in the dissociated limit  since both factors drive the 
subspace occupancy to $1/2$.
Across the dissociation curve, the Hubbard $U$
 required to enforce compliance
with Koopmans' condition, and that needed to attain the 
exact result typically differ substantially.

The facile correction of the total-energy of this system 
at each bond length with a varying but reasonable  $U$ value
sharply contrasts  with the inefficacy of DFT+$U$ for
fixing the occupied eigenvalue. 
At dissociation, the latter is not
significantly affected by DFT+$U$,  
suggesting an intrinsic limitation in the linear term  
of that correction.
We have previously introduced a generalized
DFT+$U$ functional in Ref.~\onlinecite{PhysRevB.94.220104},
in which  the linear term was amended to 
 enforce Koopmans condition.
We will return to put this approach on a 
 first-principles footing using
self-consistent variational Hubbard $U$ 
parameters in Section~\ref{sec:u1+u2}.

\subsection{The Hubbard $U$ as a first-principles
response property:  
motivations for seeking a variational formalism}

The  Hubbard $U^I$ 
are external parameters that
define the SIE correction strength applied to each subspace
in DFT+$U$.
They may be thought of as subspace-averaged
SIEs quantified \emph{in situ}~\cite{PhysRevB.71.035105,
PhysRevLett.97.103001}. 
Historically and to this day, 
the Hubbard $U$ has frequently been determined 
via the empirical fitting of calculated 
DFT+$U$ observables to experimental data,  
typically spectral~\cite{:/content/aip/journal/jcp/127/24/10.1063/1.2800015,Morgan20075034}; 
structural~\cite{PhysRevB.70.125426,PhysRevB.75.035115,:/content/aip/journal/jcp/127/24/10.1063/1.2800015} or energetic~\cite{PhysRevB.73.195107,C1EE01782A}.
This approach is pragmatic and in many cases very successful, 
but it is clearly inapplicable when the necessary
experimental data is unavailable or difficult to measure.
If the Hubbard $U$ parameters are instead themselves
calculated as properties of the electronic structure, 
however, becoming no longer free parameters but auxiliary
variables, 
in effect, then DFT+$U$ is restored to a first-principles status.
If the Hubbard $U$ are calculated strictly as
variational ground-state density-functional properties, even implicitly, 
i.e., one of the central developments of the present work, 
then DFT+$U$ as whole becomes a 
fully self-contained, variational first-principles method. 
Only under the latter condition would we expect the 
fully rigorous direct comparability of the total energies, 
and their derived thermodynamic observables,  calculated from
different DFT+$U$ calculations with different system-specific
 $U$ parameters. 

In this work, we build upon
the very widely-used~\cite{PhysRevB.70.235121,Esch752,Zhou20041144,Zhou2004181,PhysRevLett.97.103001,PhysRevLett.102.026101,PhysRevB.75.045121,PhysRevB.91.195138,PhysRevB.92.085111,PhysRevB.94.035120,nmat3568,doi:10.1021/jp070549l,doi:10.1021/acscatal.6b01907,PhysRevB.94.035136} 
and  successful linear-response method proposed in 
Ref.~\onlinecite{PhysRevB.71.035105},
in which Cococcioni and de Gironcoli 
demonstrated that a small number of DFT calculations is sufficient
to calculate first-principles Hubbard $U$ parameters 
by finite-differences, as well as upon
 the earlier linear-response scheme proposed by 
Pickett and co-workers~\cite{PhysRevB.58.1201}, 
and aspects of the modified constrained LDA scheme
of Aryasetiawan and co-workers~\cite{PhysRevB.74.125106}.
In this linear-response DFT+$U$
method~\cite{PhysRevB.71.035105}, 
a small, external uniform perturbation of strength $\alpha$ 
is applied  to the subspace of interest and    
 the interacting response function $\chi$,
and its non-interacting Kohn-Sham equivalent $\chi_0$, 
are computed, respectively, from the first derivatives 
of the self-consistent and non-self-consistent 
subspace total occupancies $N^I=\textrm{Tr}[\hat n^{I}]$ 
with respect to $\alpha$.
The scalar  Dyson equation 
$U^I = (\chi_0^{-1}- \chi^{-1})^I$ 
yields  $U$  for a single subspace model,
which may be further improved 
under self-consistency~\cite{PhysRevLett.97.103001,PhysRevB.81.245113,cococcioni4}.
The Hubbard $U$ parameters appropriate to a generalized
model in which inter-subspace parameters 
$V$ are included~\cite{campo2010extended}, 
as well as the inter-subspace $V$ themselves, 
may also be calculated~\cite{PhysRevB.71.035105} 
by treating the Dyson equation as a site-indexed 
matrix equation~\footnote{In this work, since we find it necessary to use only
single-site DFT+$U$ with no +$V$ term, 
we treat the two $1s$ atomic subspaces as
decoupled, each comprising the majority of the
screening bath for the other. Hence we use 
a pair of scalar Dyson equations 
(identical by symmetry, i.e., only one is treated
numerically) to calculate the Hubbard $U$, 
rather than selecting the diagonal  of 
the $2 \times 2$ site-indexed  Hubbard $U$.}.

To date, based on our extensive literature search, 
the linear-response method has only been used
in conjunction with the self-consistent field (SCF) algorithms
very typically used to solve the Kohn-Sham equations for
smaller isolated and periodic systems. 
In the SCF case, it is convenient to calculate 
the non-interacting response function $\chi_0$ 
following the first iteration of the SCF cycle
as prescribed in Ref.~\onlinecite{PhysRevB.71.035105}, i.e., 
following the initial charge re-organization 
induced by the external potential $\hat{v}_\textrm{ext}
= \alpha^I \hat{P}^I $, but before any 
update of the remaining terms in the Kohn-Sham potential 
is carried out.
This technique  is impractical to implement, however, 
in codes that utilize a direct minimization of 
the total-energy with respect to the
density, Kohn-Sham orbitals, or density-matrix 
to locate the ground-state, since there
it is not efficient or customary to nest the density and 
potential update processes.
These codes comprise a growing number of linear-scaling 
or large-system adapted packages, where explicit Hamiltonian
diagonalization is typically avoided 
altogether where possible,  such as
{\sc ONETEP}~\cite{:/content/aip/journal/jcp/122/8/10.1063/1.1839852,hine2009linear,hine2010linear}, 
{\sc CONQUEST}~\cite{Gillan200714,0953-8984-14-11-303},
{\sc Siesta}~\cite{0953-8984-14-11-302,0953-8984-20-6-064208}, 
{\sc BigDFT}~\cite{Genovese2011149}, 
{\sc OpenMX}~\cite{PhysRevB.72.045121}, 
and {\sc CP2K}~\cite{doi:10.1063/1.2841077}, among others, 
albeit that the SCF technique 
may also be available in some of these.
We are therefore motivated to seek a linear-response formalism
for the Hubbard $U$ that is readily compatible with
direct-minimization DFT and large systems, 
particularly for linear-scaling DFT+$U$~\cite{PhysRevB.85.085107}.

In this work, we develop and present 
a minimal revision of the established
`SCF linear-response' approach 
(terminology specific to this article, 
introduced for the avoidance of ambiguity)
 for the Hubbard $U$ 
parameters, one based on the response of the fully relaxed
ground-state density subject to a varying perturbation.
We have implemented our `variational linear-response' 
method for the $U$ in the 
linear-scaling direct-minimization DFT package 
{\sc ONETEP}~\cite{:/content/aip/journal/jcp/122/8/10.1063/1.1839852,hine2009linear,hine2010linear}, 
where the cost of the method itself scales with the number of 
targeted subspaces to be assessed, multiplied by the total 
number of atoms present. 
It is thus readily applicable to systems 
that are both spatially disordered 
and electronically challenging.
More generally, the variational linear-response
method is equally applicable to 
direct-minimization and SCF DFT codes irrespective
of the basis set used, and it
 may prove helpful in cases where the
SCF non-interacting response $\chi_0$ is
numerically problematic~\cite{QUA:QUA24521}. 
We find that it provides a convenient  framework 
in which to analyze 
a number of different criteria that have been 
proposed for defining the self-consistent  Hubbard
$U$ and with it, for the particular case of the variational linear-response
Hubbard $U$ at least, we identify a 
well-defined best choice of  self-consistency criterion  
supported by numerical results.

\subsection{Article outline}

In Section~\ref{kernel_selection}, we investigate 
the conditions that must hold 
for a first-principles Hubbard $U$ parameter 
to correct SIE subspace-locally by means of
Eq.~\ref{Eq:dft+u}.
Arriving at a simple, variational linear-response 
formulation in terms of 
fully-relaxed constrained density and its
resulting properties, 
we make the calculation of Hubbard $U$ parameters
accessible to direct-minimization DFT codes.
In Section~\ref{self_consistency_scheme}, 
we address, for specific case of variational 
linear-response,
the question: which of the 
previously-proposed  and available 
Hubbard $U$ self-consistency criteria, if any is necessary, 
is suitable for correcting the SIE-affected 
total energy by means of DFT+$U$?
In Section~\ref{results}, we further analyse  
our results by means of numerically stringent DFT+$U$
calculations along the dissociation curve of
$H_2^+$, 
an ideal system for studying one-electron SIE~\cite{vuckovic2015hydrogen,:/content/aip/journal/jcp/127/16/10.1063/1.2800022}.
Finally, in our concluding Section~\ref{conclusions}, we discuss
the theoretical relationship between 
the SCF and variational linear-response formalisms, 
the relevance  to the comparability
of total-energies and other thermodynamics quantities
from DFT+$U$ calculations with system-specific first-principles
Hubbard $U$ parameters, and our outlook on the 
practicability of such parameters.

\section{A variational ground-state approach to the 
linear-response Hubbard U parameter}
\label{kernel_selection}

In order to calculate the Hubbard $U$ 
parameter required to subtract 
the many-electron SIE 
attributed to a particular subspace, 
by means of Eq.~\ref{Eq:dft+u},
we may define the parameter for each subspace
as the net average electronic interaction acting within it.
More specifically, we seek only the interactions
at leading order in the subspace density-matrices, 
that is those coupling to $( n_i^{I} )^2$, 
in order to comply with Eq.~\ref{Eq:dft+u}, 
Thus, for a particular site, 
we  define the $U$ on the basis of the
interaction kernel 
$\hat{f}_\textrm{int} = 
\delta^2 E_\textrm{int} / \delta \hat{\rho}^2$ only, 
i.e., not $\hat{g}_\textrm{int} = 
\delta^3 E_\textrm{int} / \delta \hat{\rho}^3$ etc.,
where $E_\textrm{int}$ is the interacting 
contribution to the  total-energy.
Furthermore, we require only the components of 
the interaction for each subspace that arise due to density variations
within it, so that $\hat{f}_\textrm{int}$ must be appropriately projected.
In order to illustrate the requirements of such a
projection, let us consider
some candidate formulae for Hubbard 
$U$ parameters which do not meet them.

 The many-electron SIE of an approximate xc functional, 
 applied to an an open quantum system
 that does not interact with its bath for particle exchange, 
 is characterized by the spurious non-zero second total-energy
 derivative with respect to the total occupancy.
 We may apply this definition to the individual DFT+$U$ subspaces,
 with occupancies given by $N = \textrm{Tr} [ \hat{n}]$
 (suppressing subspace indices), 
 under the reasonable assumption that 
 the subspace-bath interactions are negligible compared to 
 the interactions within the subspace.
 By defining the Hubbard $U$ for each subspace
 as the net value of the latter interaction, 
in a precise sense yet to be determined, 
 the DFT+$U$ functional should 
 act to correct the many-body SIE by 
 subtracting the individual one-electron SIE  
 of each eigenstate of the subspace density-matrix.

Immediately, we may rule out 
as a Hubbard $U$ parameter the 
straightforward fully interacting curvature $d^2 E / d N^{2}$,
discussed in Ref.~\onlinecite{PhysRevB.58.1201}, 
which may be calculated as $ - d \alpha / d N $. 
Here, $\hat{v}_\textrm{ext} = \alpha \hat{P}$ 
is the external potential inducing the occupancy change.
As discussed  in Ref.~\onlinecite{PhysRevB.71.035105}, 
this term comprises a substantial non-interacting contribution, 
which, in accordance with Dyson equations quite generally, 
 is superfluous to the definition of an interaction and must be
 subtracted.
On the other hand, one may suggest 
 the direct subspace projection of the interaction kernel
 (Hartree, xc, any other electronic interaction terms),
 denoted here for a single site by 
 $ \hat{P}( \delta^2 E_\textrm{int} / 
\delta \hat{n}^{ 2} ) \hat{P}  = \hat{P} \hat{f}_\textrm{int} \hat{P} $. 
Any bare interaction of this kind 
neglects the potentially substantial screening effects 
of density-matrix variations outside the subspace. 
Thus, it is also an unsuitable starting point for measuring 
many-body SIE,  ruling it out. 
The  $U$ must be bath-screened, yet remain bare of 
intra-subspace screening.

More interesting 
 is the curvature of the interaction term in the total-energy,  
$d^2 E_\textrm{int} / d  N^{2}$, 
and the reasons for its 
non-suitability are perhaps more subtle.
Since the Hellman-Feynman theorem 
cannot be applied to $E_\textrm{int}$ alone,
its first total derivative with respect to $N$, i.e., 
\begin{equation}
\frac{dE_\textrm{int}}{dN} = \frac{\partial E_\textrm{int}}{\partial N} + \textrm{Tr}\left[ \frac{\delta E_\textrm{int}}{\delta \hat{\rho}} \frac{d \hat{\rho}}{ d N} \right]
\end{equation}
yields not only the partial derivative 
(vanishing due to no explicit $N$-dependence in
$E_\textrm{int}$, and only an implicit 
dependence via the changes to the total
density-matrix $\hat{\rho}$), 
but it also  comprises a term proportional to 
the  interaction potential $\delta E_\textrm{int} / \delta \hat{\rho} $, 
bath-screened
since $d \hat{\rho} / d N$ couples to the external potential
$\hat{v}_\textrm{ext} = \alpha \hat{P}$.
The second total derivative
$ d^2 E_\textrm{int} / d N^2  $
 incorporates screening again, 
and the resulting twice-screened objects are  unphysical. 
This problem here is the opposite, in a sense, to that of
$ \hat{P}\hat{f}_\textrm{int} \hat{P}$, 
from which one may surmise the 
correct definition is an intermediate case, 
where screening effects due to the complement 
of the subspace at hand should be incorporated, but only once.
This motivates us to work not from the energy,  
but  from the potential, i.e., from the unscreened functional derivative
of the energy with respect to the density-matrix, and to 
differentiate by $N$.

As a functional derivative, i.e., a generalized partial derivative,
the interaction term  in the Kohn-Sham potential 
$\hat{v}_\textrm{int} = \delta E_\textrm{int} 
/ \delta \hat{\rho}$
 is bare of screening, 
 as is its subspace projection 
 $\hat{P}(  \delta E_\textrm{int} 
/ \delta \hat{n} ) \hat{P} = 
\hat{P} \hat{v}_\textrm{int} \hat{P} $. 
The quantity then given by
$  \textrm{Tr} [ \hat{P}( d \hat{v}_\textrm{int} / d \hat{n}) \hat{P}] /
\textrm{Tr} [  \hat{P} ]^2$ 
seems to fulfil many of the requirements for
a valid Hubbard parameter, namely, that it is 
 a subspace-averaged, once-screened
interaction that is non-extensive, 
i.e. it does not scale extensively with the subspace 
eigenvalue count $\mathrm{Tr} [ \hat{P} ]$.
In practice, however, 
the screened kernel $ d \hat{v}_\textrm{int} / d \hat{n} $ 
is cumbersome to calculate,
even in orbital-free density-functional theory,
and, more importantly,
it includes screening effects 
due to density-matrix rearrangements 
within the subspace, which make it unsuitable
as a quantifier for the subspace-bare interaction 
to be explicitly corrected by DFT+$U$.
Instead, the object that we required is 
the average, net, subspace-bare 
but bath (i.e., environment) screened 
self-interaction of the subspace.
We may meet these specifications
by taking the total derivative 
with respect to the total subspace occupancy $N$,
and,  finally, by defining 
\begin{align}
U \equiv \frac{d v_\textrm{int} }{  d  N},
 \quad 
\mbox{where}
\quad
v_\textrm{int} \equiv \frac{ \textrm{Tr} [\hat{v}_\textrm{int} \hat{P}]
}{ \textrm{Tr} [  \hat{P} ] }
\end{align} 
is the conveniently calculated, 
non-extensive,
subspace-averaged interaction potential
(comprising Hartree, xc, etc.).
Here, since the uniform potential $\alpha$ 
used in the linear-response method 
induces to first-order no microscopic 
density-variations within the subspace
except for the uniform shift,
 the screening processes within 
the subspaces are effectively suppressed, 
much as in the  
constrained random phase approximation~\cite{PhysRevB.74.125106,PhysRevB.77.085122,PhysRevB.80.155134}.

In practice, 
as we return to discuss around Eq.~\ref{Eq:linear_response_2}.
the proposed variational linear-response $U$ for a 
single-site model may still be computed 
using the Dyson equation, but with
the response functions 
$\chi = d N / d \alpha $ and $\chi_0 = d N / d v_\textrm{KS}$, 
where $v_\textrm{KS} \equiv \textrm{Tr}
 [\hat{v}_\textrm{KS} \hat{P}]
/ \textrm{Tr} [  \hat{P} ] $.  
Here, both $\chi$ and $\chi_0$ are to calculated at the end 
of the minimization procedure 
from the same set of 
constrained ground-state densities defined by $\alpha$.
Thus, while $\chi$ is identical to that used in the
SCF linear-response  introduced in 
Ref.~\onlinecite{PhysRevB.71.035105}, 
 our $\chi_0$ formula is somewhat different
(at least formally, the numerical differences remain unclear).
The SCF and variational linear-response 
formalisms are equally compatible with
SCF and direct-minimization DFT, 
as well as with the matrix Dyson equation required, e.g., for
calculating longer-ranged 
inter-subspace parameters $V$~\cite{campo2010extended} 
and their corresponding Hubbard $U$ values.

\subsection{The subspace contribution to total-energy curvature}
The subspace contribution 
to the interacting part of the total-energy SIE, 
specifically that corresponding to the
variational linear-response Hubbard $U$,  
is the  integral of the 
the interacting part of the Kohn-Sham 
potential over the subspace occupancy 
up to its ground-state value.
To the same effect, we may use the negative of the integral over 
the external potential $\alpha$ needed to fully deplete
that occupancy back to zero, as in 
\begin{align}
E^\textrm{SIE}_{\textrm{int}}
\left( N \right) ={}&\int_0^{N} v_{\textrm{int}}\left(N'\right)\ dN' \\
		={}& - \int^{\infty}_0 v_{\textrm{int}}\left(N'\left(\alpha\right)\right)
		\frac{dN'}{d\alpha}\ d\alpha  \nonumber \\  \nonumber
={}&  \int^{\infty}_0 v_{\textrm{int}}\left(N'\left(\alpha\right)\right)
		\left( \left. \frac{d^2 E_\textrm{total}}{d N''^2} \right|_{N''  \left( \alpha \right)} \right)^{-1} d\alpha
		\end{align}

In the final line,  we make the connection to the occupancy
curvature of the total-energy $E_\textrm{total}$ using
the result for the constrained DFT system, 
$d E_\textrm{total} / d N = - \alpha$.
Although $E^\textrm{SIE}_{\textrm{int}}$ does not 
appear anywhere in our DFT+$U$ implementation in practice, 
 we emphasise that for a single-site model, 
 in the variational linear-response
 formalism  at least, it is
 $d^2 E_\textrm{SIE} / d N^2 = U $, and not the total-energy
 curvature
$d^2  E_\textrm{total} / d N^2 = - \chi^{-1}$, which yields
the  parameter required for DFT+$U$. 
Considering the difference of the 
energy curvatures ascribed to 
the bath-screened subspace and the overall global system, 
both as a function of subspace occupancy, we find
that $d^2 ( E_\textrm{total} - E_\textrm{SIE}  ) / d N^2
= - \chi^{-1} - U = - \chi_0^{-1} \ge 0 $, where the latter inequality
was proven in Ref.~\onlinecite{PhysRevB.94.035159}.
This result is reminiscent of the findings 
of Kulik \emph{et al.} in Ref~\onlinecite{:/content/aip/journal/jcp/145/5/10.1063/1.4959882}, 
to wit, that while the application of 
DFT+$U$ can only be expected  to mitigate subspace SIE, 
and at the very least it cannot disimprove the global SIE, albeit 
for a different sense of global pertaining to total occupancy.

The quantity $E^\textrm{SIE}_{\textrm{int}}$ differs from the
full subspace contribution to the total-energy SIE by a
non-interacting contribution required to restore Koopmans' 
condition. 
It is interesting to assume, for a moment, that the SIE kernel
$U = d v_\textrm{int} / d N$ is Hartree-dominated
and hence approximately constant, 
so that $v_\textrm{int} ( N ) \approx U N$
and $E^\textrm{SIE}_{\textrm{int}} \approx U N^2 / 2$.
If we further assume that the subspace is SIE-free
at the nearest integer occupancy,  $N_0$, as well as at
$N_0 \pm 1$, with a linear (i.e., non-interacting) interpolation term being
required  between these points,
then we may make the curvature-preserving modification 
$E^\textrm{SIE} \approx ( U / 2 ) [ ( N - N_0)^2 - 
| N - N_0 | ] $. Taking the negative of this energy to estimate
a total-energy correction, and considering 
single-orbital, single-spin sites, we effectively re-derive
the $E_U$ of Eq.~\ref{Eq:dft+u}.
Even non-self-consistently, 
this turns out to be an acceptable energy correction  
for $H_2^+$ in the dissociated limit, with two subspaces of 
$U \approx 8$~eV and $N \approx 1/2$, yielding 
$ - E_\textrm{SIE} \approx 2$~eV~$\approx E_\textrm{exact}
- E_\textrm{PBE}$.
%

\section{Self-Consistency over the Hubbard $U$}\label{self_consistency_scheme}

Beginning with the work of 
Kulik and co-workers in Ref~\onlinecite{PhysRevLett.97.103001},
and in later works~\cite{PhysRevB.81.245113,campo2010extended,cococcioni4}, 
it has been demonstrated that 
a self-consistently calculated $U$ can be required 
for certain systems where 
the nature of the electronic states  
(and corresponding response properties) 
in the DFT+$U$ ground-state differ 
qualitatively from those of the DFT ground-state~\cite{PhysRevB.57.1505,PhysRevB.83.075112,PhysRevB.84.115108}. 
In self-consistency schemes  generally, 
incremental values of $U_\textrm{in}$ are applied to the subspace
at hand,  
with varying ground-state orbitals and densities as a result,  
and a new first-principles $U_\textrm{out}$ is  computed for each
$U_\textrm{in}$.
The  numerical relationship 
$U_\textrm{out} ( U_\textrm{in} )$  
is then used to select the  
self-consistent $U$, using a pre-defined  criterion.
Its clear conceptual elegance aside, a self-consistent $U$ 
has been shown to provide  
improvements in transition-metal 
chemistry~\cite{kulik2008self,:/content/aip/journal/jcp/133/11/10.1063/1.3489110,kulik2011transition,PhysRevB.84.115108,PhysRevLett.106.118501,Youmbi20141,2053-1591-3-8-086104,doi:10.1021/acs.jctc.6b00937,Hsu201019}, 
biological systems~\cite{doi:10.1021/jp070549l},
photovoltaics~\cite{PhysRevB.78.241201,PhysRevLett.105.146405,doi:10.1021/jp1041316}, 
and high-density energy storage~\cite{PhysRevB.93.085135}.
While many researchers have used an original, 
linear-extrapolation type $U_\textrm{scf}$
in their studies~\cite{kulik2008self,:/content/aip/journal/jcp/133/11/10.1063/1.3489110,doi:10.1021/jp070549l,2053-1591-3-8-086104,Youmbi20141,doi:10.1021/acs.jctc.6b00937,PhysRevB.78.241201,PhysRevLett.105.146405,doi:10.1021/jp1041316,Hsu201019,PhysRevB.80.075102}, 
others have used the equality between  
$U_\textrm{in}$ and $U_\textrm{out}$ 
as an alternative self-consistency condition~\cite{doi:10.1021/jp070549l,campo2010extended,PhysRevB.84.115108,PhysRevB.93.085135,PhysRevLett.106.118501,doi:10.1063/1.4947240}.
The majority of published first-principles $U$ calculations 
involve no self-consistency over the  parameter at all, and there may 
even be a case to be made that none is ordinarily warranted. 
The resolution of this ambiguity is, in itself, 
an intriguing open challenge in abstract DFT, but 
it particularly demands investigation in the present context of
the variational linear-response $U$ since, ideally, the optimal scheme 
to match that method should be established from the outset.
On the basis of this study, however, we  emphasise that
we cannot draw  conclusions
concerning the  self-consistency schemes for 
$U$ parameters calculated by 
means of any other methods.

In order to compute variational linear-response 
$U_\textrm{out}$ 
for a single subspace already subject to a DFT+$U$
term of strength $U_\textrm{in}$, 
the subspace-averaged interaction $v_\textrm{int}$
must incorporate the DFT+$U$ potential $\hat{v}_U$,
as well as the usual Hartree + xc term $\hat{v}_\textrm{Hxc}$.
Each component in the subspace-averaged
interaction potential 
$ v_\textrm{KS} - v_\textrm{ext} \equiv 
v_\textrm{int}=v_\textrm{Hxc}+v_{U_\textrm{in}}$
must be defined in such a manner that 
does not scale extensively with the orbital count of 
the subspace, $\textrm{Tr} [ \hat{P} ] $.
For Hartree + xc, the appropriate average is 
$v_\textrm{Hxc}=\textrm{Tr}[\hat 
v_\textrm{Hxc}\hat P]/\textrm{Tr}[\hat P]$
(the operator $\hat v_\textrm{Hxc}$ may approximately 
scale with $N$ but the averaging scheme does not), 
while the average differential to the external potential is, similarly,
$d v_\textrm{ext}=\textrm{Tr}[\hat 
d v_\textrm{ext} \hat P]/\textrm{Tr}[\hat P] 
=\textrm{Tr}[ 
d \alpha \hat P \hat P]/\textrm{Tr}[\hat P] = d \alpha$ by the 
idempotency of $\hat{P}$.
Unlike $\hat{v}_\textrm{Hxc}$, which acts on one
state but is generated by all occupied states, 
 the  DFT+$U$ potential 
$\hat{v}_{U_\textrm{in}} = U_\textrm{in} (\hat{P}-
2\hat{P}\hat{\rho}\hat{P})/2$ is intrinsically both specific to 
and due to 
each subspace occupancy matrix eigenvector individually.
Thus, we find that the simple trace 
$v_{U_\textrm{in}} = \textrm{Tr} [ \hat{v}_{U_\textrm{in}} ] 
= U_\textrm{in} ( \mathrm{Tr} [ \hat{P} ] - 2 N ) / 2 $
is that which scales appropriately  with $N$ or,
put another way, $v_{U_\textrm{in}} $ would be the average
DFT+$U$ potential acting on a  subspace eigenvector were there
$\textrm{Tr} [\hat{P}] $ copies of that eigenvector, and thus is  
comparable with $v_\textrm{Hxc}$.
The factor $ \mathrm{Tr} [ \hat{P} ] $ separating   
the definitions of $v_\textrm{Hxc}$ and $\hat{v}_{U_\textrm{in}}$
 is consistent with DFT+$U$ correcting the Hartree + xc generated 
many-body subspace SIE, which is assumed to be
proportional to $N^{2} \approx 
 ( \mathrm{Tr} [ \hat{P} ] \langle n_i \rangle ) ^{2} $, by 
 only $ \mathrm{Tr} [ \hat{P} ]$ 
 one-electron SIE corrector
terms on the order of $\langle n_i \rangle^{2}$.
Finally we may write, for the single-site variational
linear-response Hubbard $U_\textrm{out}$
in the presence of a non-zero  $U_\textrm{in}$, 
that 
\begin{align}
\label{Eq:linear_response_2}
U_\textrm{out}=&\chi_0^{-1}- \chi^{-1}=\frac{d v_\textrm{KS}  -  d v_{\textrm{ext}}}{d N} 
=\frac{d v_\textrm{int}}{d N}\nonumber\\
=&\frac{d v_\textrm{Hxc} }{d N}-U_\textrm{in}
=f^{\hat{P}}_\textrm{Hxc}(U_\textrm{in})-U_\textrm{in},
\end{align}
where $f^{\hat{P}}_\textrm{Hxc} (U_\textrm{in}) \equiv d 
v_\textrm{Hxc}/d N$
is the subspace-averaged, subspace-bare but
bath-screened Hxc interaction
calculated at the fully-relaxed DFT+$U_\textrm{in}$ ground-state.

From Eq.~\ref{Eq:linear_response_2} 
we may readily identify 
three unique self-consistency criteria.
The first is a very plausible self-consistency criterion, 
first proposed in Ref.~\onlinecite{campo2010extended} 
and later utilized in Refs.~\onlinecite{PhysRevB.84.115108,PhysRevB.93.085135}, 
which requires that $U_\textrm{out}=U_\textrm{in}$
and thus gives  
$U_\textrm{in}= f^{\hat{P}}_\textrm{Hxc}(U_\textrm{in})/2$.
This $U_\textrm{in}$, denoted here as $U^{(1)}$, 
appears to account for, i.e., cancel away one-half 
of the subspace SIE that remains at that DFT+$U_\textrm{in}$.
The second criterion is given by $U_\textrm{out}=0$, 
denoted here  by $U^{(2)}$, which
dictates that    
$U_\textrm{in}= f^{\hat{P}}_\textrm{Hxc}(U_\textrm{in})$, 
implying that
$U_\textrm{in}$  fully cancels 
the subspace-related SIE 
computed at the same DFT+$_\textrm{in}$ ground-state.
The third condition, denoted by $U^{(3)}$, 
matches (albeit with a different underling linear-response
procedure) the original 
self-consistency scheme~\cite{PhysRevLett.97.103001}
where it is denoted $U_\textrm{scf}$. 
Here, the $U_\textrm{out} ( 0 )$ of the DFT+$U$ electronic structure
is calculated by a linear-extrapolation of  
$U_\textrm{out} ( U_\textrm{in} )$ for sufficiently large $U_\textrm{in}$
to obtain a good fit,
back to $U_\textrm{in} = 0$~eV.

For our present purposes, it is reasonable to assume that 
a DFT+$U$ corrected electronic structure has been well-obtained 
at $U^{(2)}$, and thus performing the 
linear extrapolation for $U^{(3)}$ around  $U^{(2)}$, 
we find that
\begin{equation}
U^{(3)}=U^{(2)} \left(1-  \left. \frac{d f^{\hat{P}}_\textrm{Hxc} }{ dU_\textrm{in}  } \right|_{U^{(2)}} \right).
\end{equation} 
From this, a clear interpretation of
$U^{(3)}$ as screened version of $U^{(2)}$ emerges,  
in a generalized sense of screening in which,
instead of an externally applied potential 
being attenuated by relaxation of the electronic structure, 
it is instead the externally applied \emph{interaction correction}
which is attenuated.
A normal dielectric screening operator 
measures the rate of change of 
the potential with respect to an external perturbation, taking 
the form 
$\hat{\epsilon}^{-1}= d\hat{v}_\textrm{KS} / d
\hat{v}_\textrm{ext} = \hat{1}+
\hat{f}_\textrm{Hxc} \hat{\chi}$.
A generalized screening function here instead  
measures the rate of reduction in  subspace-averaged
SIE with respect to $U_\textrm{in}$, and is given by
$\epsilon_U^{-1}= - dU_\textrm{out} / dU_\textrm{in} = 1-df^{\hat{P}}_\textrm{Hxc}/dU_\textrm{in}$.
Therefore, while we  require a DFT+$U$
correction with parameter  $U^{(2)}$
to  cancel the subspace-averaged SIE 
including all  self-consistent
response effects in the electronic structure, 
when we have done so we have 
in fact removed an SIE (with respect to DFT) 
of magnitude $U^{(3)}= \epsilon_U^{-1} U^{(2)}$, 
which is typically smaller in magnitude than $U^{(2)}$.
There is a numerically relevant 
 distinction between the external `bare'  $U_\textrm{in}$ 
that we  apply using DFT+$U$,  and the `screened' SIE quantifier 
$U_\textrm{out}$ that we then measure.

The SIE measure  $U^{(3)}$, calculated around
the $U^{(2)}$ ground-state, is  of particular interest, e.g., for
quantifying the change in SIE in a subspace in response to an
external parameter such as atomic position, or if comparing
the SIE of an atom in two different charge states.
We also expect $U^{(3)}$ to be suitable as
an input Hubbard $U$ parameter
for non-self-consistent protocols such as a post-processing
DFT+$U$ band-structure correction based on the DFT density, 
or a DFT + dynamical mean-field theory (DMFT) calculation
with no density
self-consistency. $U^{(3)}$ linearly accounts
for the resistance to SIE reduction that \emph{would} be met
were density self-consistency in response to $U$ allowed.
On the basis of the above analysis, however, we conclude that
the criterion $U^{(2)}$ represents the appropriate 
self-consistency scheme for the variational linear-response
method, wherever the standard self-consistent response of the
density occurs upon application of DFT+$U$.
The value of $U^{(2)}$ may be efficiently obtained, e.g., 
by the bisection method.
The three self-consistency conditions are 
summarized  in Table~\ref{table1}.

\begin{table}[h]
\begin{tabular*}{\columnwidth}{@{\extracolsep{\fill}}cll}
\hline\hline
Notation & 	Criterion 		& Formula derived from Eq.~\ref{Eq:linear_response_2}\\
\hline
$U^{(1)}$	&\;$U_\textrm{out}=U_\textrm{in}$ & \;$U_\textrm{in}= f^{\hat{P}}_\textrm{Hxc}(U_\textrm{in})/2$\\
$U^{(2)}$	&\;$U_\textrm{out}=0$	& \;$U_\textrm{in}= f^{\hat{P}}_\textrm{Hxc}(U_\textrm{in})$\\
$U^{(3)}$	&\;$U_\textrm{out}(0)$	& \;$U_\textrm{out}(0)=U^{(2)}(1-d f^{\hat{P}}_\textrm{Hxc}/dU_\textrm{in}|_{U^{(2)}})$\\
\hline\hline
\end{tabular*}
\caption{Summary of  three first-principles Hubbard $U$ self-consistency criteria derived from Eq.~\ref{Eq:linear_response_2}.}
\label{table1}
\end{table}

\section{Numerical results}\label{results}
\subsection{Self-consistent $U$ schemes 
applied to dissociating $H_2^+$}

In order to  assess the potency of 
DFT+$U$ for correcting SIE under varying bonding
conditions without the complicating
effects of static correlation error, 
we calculated  self-consistent Hubbard $U$
values and the resulting DFT+$U$ electronic structure
along the binding curve of the dissociating 
one-electron dimer $H_2^+$.
A further advantage of the one-electron system is that 
the PBE and exact (i.e., for one electron, simply no Hartree or xc)
 functionals are available 
using the same first-principles code and pseudopotential, 
which ensures the accurate comparability 
of energies across the parameter space.
Stringent numerical conditions were applied, with an extremely 
accurate small-core norm-conserving PBE pseudopotential  
and a plane-wave equivalent kinetic energy 
cutoff of approximately $2650~$eV, yielding deviations from
$0.5$~Ha within
w (x) and y (z) on the isolated-atom
total-energy and occupied Kohn-Sham 
eigenvalue, respectively, for the exact (PBE in parentheses) functional.
The dissociated limit is of particular interest 
 for confirming the relative appositeness of
Hubbard $U$ self-consistency schemes that may yield 
numerically similar results since, in that limit,  
the neutral-atom PBE $1s$ orbitals used to define
each of the two  DFT+$U$ subspaces 
spatially overlap (i.e., double-count)
and spill the total charge minimally, and the DFT+$U$ population
analysis for the PBE dimer becomes  ideal.
Furthermore, as we approach the dissociated limit, 
the assumption that each of the two DFT+$U$ subspaces
interacts relatively weakly with its bath
(in each case, the other atom) becomes increasingly
realistic, represening the best available 
performance of DFT+$U$ using an fixed atomic 
population analysis
(i.e., one that is not dependent on the charge, applied $U$, or
other details of the electronic structure, as Wannier functions 
are~\cite{PhysRevB.82.081102,PhysRevB.85.193101}).

\begin{figure}
\includegraphics[width=1\columnwidth]{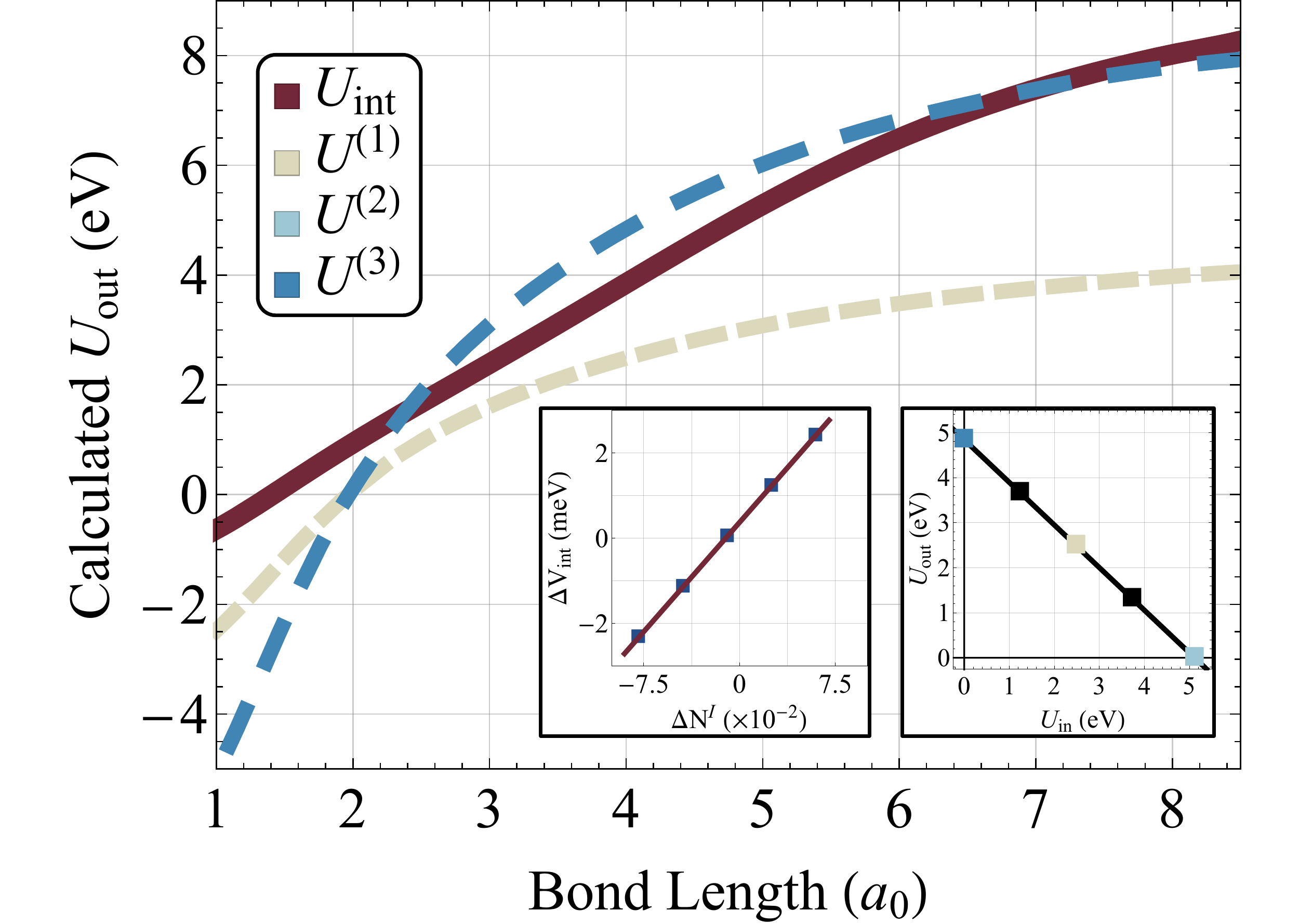}
\caption{(Color online) 
The estimated best $U$ value,
$U_\textrm{int}$ (solid),  for correcting the total-energy
SIE in $H_2^+$, shown with the
$U^{(1)}$ (dashed), 
$U^{(2)}$ (dotted), and 
$U^{(3)}$ (dot-dashed) values.
(Inset left) 
A sample $U_\textrm{out}$ calculation at 
$4$~a$_0$ and $U_\textrm{in}=1.5$~eV. 
(Inset right)
A sample 
$U_\textrm{in}$ vs $U_\textrm{out}$ 
profile used to evaluate 
$U^{(1)}$, $U^{(2)}$, $U^{(3)}$ (highlighted points), 
via Table~\ref{table1}.
$U^{(2)}$ and $U^{(3)}$  
approximately equal   $U_\textrm{int}$ 
in the dissociation limit, 
while $U^{(1)}$ is serendipitously 
more successful at equilibrium and below
due to  subspace overlap and double-counting.}
\label{figure2}
\end{figure}

While conserving the overall charge, 
the external potential $\alpha$
was varied within the range 
$\pm 0.05$~eV and applied to one atom.
DFT+$U$ was applied to both atoms equally,
with $U_\textrm{in}$  sampled 
from $0$~eV up to the value 
that yielded $U_\textrm{out}=0$~eV.
A typical calculation of 
$U_\textrm{out}$ is shown
in the left inset of Fig.~\ref{figure2}.
For each bond-length, 
a $U_\textrm{in}$ versus $U_\textrm{out}$ 
profile was calculated 
according to Eq.~\ref{Eq:linear_response_2},
as illustrated in the right inset of Fig~\ref{figure2}
with due care to error accumulation.
These profiles were found to remain highly linear 
across all bond-lengths 
for this particular system and linear-response methodology, 
and we note that the slope remained 
greater than $-1$,
signifying $d f_\textrm{Hxc}^{\hat{P}} / d U_\textrm{in}
> 0$ and a `resistance' to SIE reduction,
for all but the small bond-lengths $\lesssim 1.3$~a$_0$ strongly
affected by subspace double-counting.
The linear fit to $U_\textrm{out} ( U_\textrm{in} )$
was then used to evaluate 
$U^{(1)}$, $U^{(2)}$, $U^{(3)}$,
according to Table~\ref{table1}, 
and their values are depicted by 
dashed, dotted and dot-dashed lines, respectively, 
in Fig.~\ref{figure2}.
For each bond-length, 
we also estimated, by interpolation, the $U_\textrm{int}$  
(solid line)
required to recover the exact total-energy.
 
The $U^{(2)}$ and $U^{(3)}$ 
schemes, and particularly the former, closely approximate 
the $U_\textrm{int}$ required 
to correct the SIE in the total-energy
 in the dissociated limit,
whereas $U^{(1)}$  
clearly represents  an underestimation by a factor of  $2$, 
as indicated by Table~\ref{table1}.
The numerical situation is reversed within the equilibrium 
bond-length of approximately $2$~a$_0$,
where $U^{(1)}$ appears
to perform better than the alternatives.
We emphasise that the latter  result is  misleading, however, 
since $U^{(1)}$ performs better
at short bond lengths
only due to the cancellation of its factor-of-two magnitude reduction 
with the double-counting effects of spatially
overlapping DFT+$U$ subspaces, as well as the
breakdown, in the strong-bonding regime,
of the subspace-bath separation 
underpinning  DFT+$U$.
This highlights a risk when assessing the relative
merits of correction formulae of this kind solely on the basis 
of numerical results gathered under equilibrium conditions, 
where bonding or overlap effects complicate the analysis.

\begin{figure}[h]
\includegraphics[width=1\columnwidth]{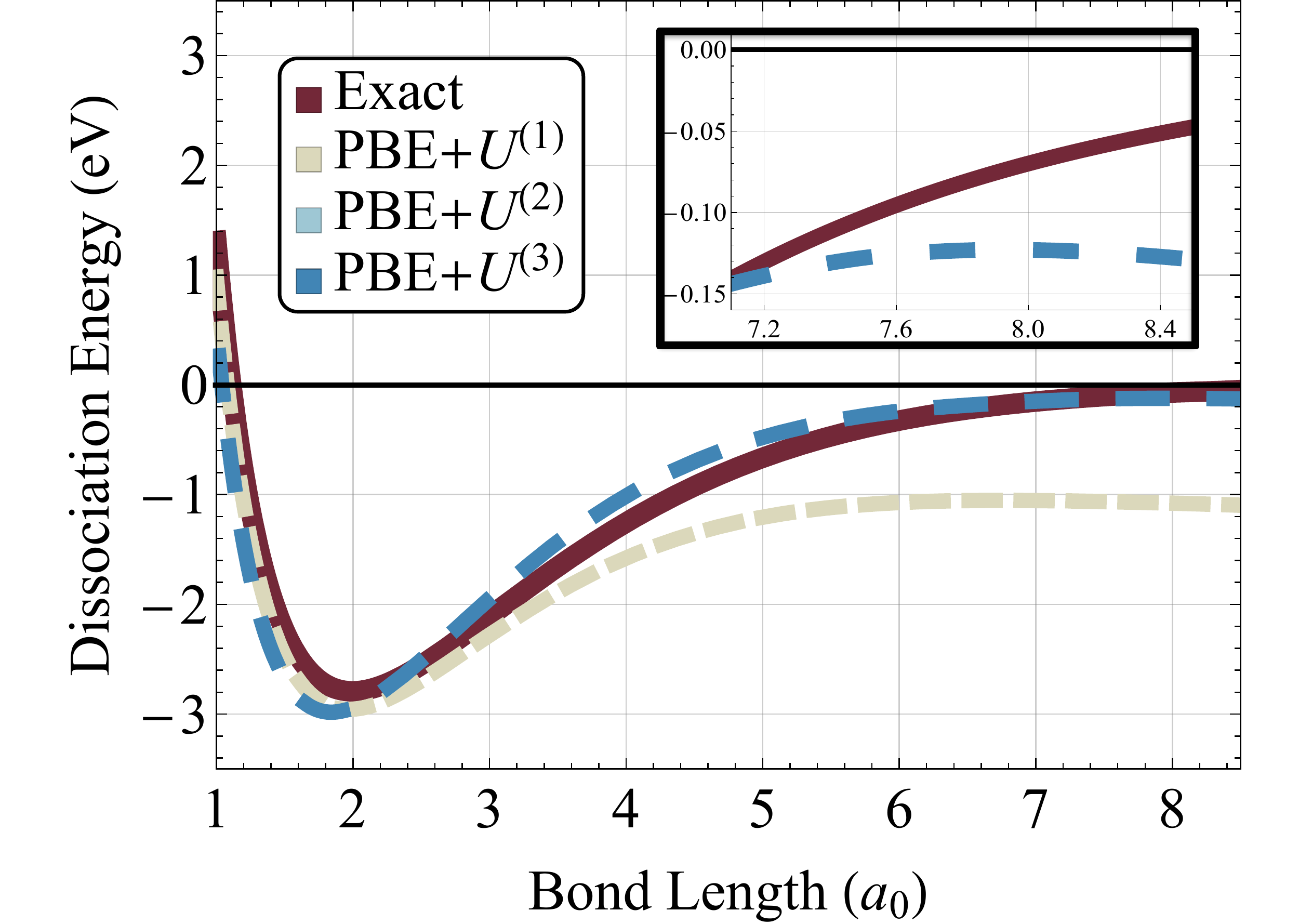}
\caption{(Color online) 
The $H_2^+$ dissociation  curves 
of the exact functional (solid),
PBE+$U^{(1)}$ (dashed), 
PBE+$U^{(2)}$ (dotted), 
and PBE+$U^{(3)}$ (dot-dashed).
In the dissociated limit (inset),  
the $U^{(2)}$ result tends asymptotically to the exact one,
and the  $U^{(3)}$ scheme begins to deviate from it non-negligibly.}
\label{figure3}
\end{figure}

The total-energy based 
dissociation curves of $H_2^+$ were recalculated 
using the bond-length dependent 
$U^{(1)}$ (dashed), 
$U^{(2)}$ (dotted), 
and $U^{(3)}$ (dot-dashed), for comparison with
the exact total-energy (solid) in Fig.~\ref{figure3}.
We note that any attempt to extend our 
bond length interval beyond $8.5$~a$_0$
resulted in numerical instabilities 
due to the near-degeneracy of 
the Kohn-Sham $\sigma$ and $\sigma^*$ 
eigenstates, and present here are  the results only
of well-converging calculations. 
As already suggested by Fig.~\ref{figure2}, 
$U^{(1)}$ fails to correct 
the SIE in the total-energy at bond-lengths further from equilibrium, 
whereas $U^{(2)}$ and $U^{(3)}$  
provide a more universal correction  
of the total-energy, 
becoming  acceptable in the dissociation limit.
The inset of Fig.~\ref{figure3}  
illustrates, however,  that the  
PBE+$U^{(3)}$ scheme, which is numerically equivalent to no 
Hubbard  $U$ self-consistency in this particular system, 
begins to under-perform with respect to  PBE+$U^{(2)}$
in the dissociated limit.
The PBE+$U^{(2)}$ total-energy, 
meanwhile,  seems to converge upon 
the exact total-energy asymptotically. 
Our results confirm that DFT+$U$ 
is  capable of precisely correcting 
the total-energy SIE of a one-electron system 
under ideal population-analysis conditions but only, it seems,
when using 
the simplest self-consistency scheme, $U^{(2)}$.
It is clear, notwithstanding, that DFT+$U$ 
is an efficient and effective corrector for the
SIE  manifested in the total-energy, 
as discussed in detail in 
Refs.~\onlinecite{PhysRevLett.97.103001,:/content/aip/journal/jcp/133/11/10.1063/1.3489110,:/content/aip/journal/jcp/145/5/10.1063/1.4959882}. 

\subsection{Restoration of Koopmans' condition:
DFT+$U_1$+$U_2$}
\label{sec:u1+u2}

Despite the success of DFT+$U$  in SIE-correcting 
the total-energy   using a suitably
calculated $U$ value, 
the fact remains that 
it is incapable of simultaneously correcting the highest 
occupied Kohn-Sham eigenvalue to minus the ionization potential
in compliance with Koopmans' condition,
as indicated in Fig.~\ref{figure1b}. 
This issue has previously been  explored  
in Ref.~\onlinecite{:/content/aip/journal/jcp/145/5/10.1063/1.4959882}, 
and by us in Ref.~\onlinecite{PhysRevB.94.220104} 
where we constructed a generalized, 
two-parameter DFT+$U$ functional, 
comprising separate parameters for the linear ($U_1$) 
and quadratic ($U_2$) terms.
In fact, Eq. 9 of Ref.~\onlinecite{PhysRevB.94.220104}
indicates that if a symmetric system of two one-orbital subspaces
(a very good approximation for $H_2^+$, with approximately
constant subspace occupancies $N$) is Koopmans' compliant
(so that the Koopmans' $U_K = 0$), and it is then corrected 
using DFT+$U$ for the SIE in the total-energy
(it is possible for the interaction strength to be inaccurate,
but for the system still to comply with Koopmans'  condition), then
DFT+$U$ will act to spoil that condition unless  
$U_1 = 2 U_2 ( N - N^2) / ( 1 - 2 N )$.
More pragmatically, we expect the extra degree of freedom 
furnished by $U_1$ to be 
beneficial in cases where  the quadratic
approximation to the subspace-averaged
self-interaction does not remain valid all the way down
to the ionized  state, which is particularly relevant for $H_2^+$
since there that state corresponds to the low-density limit.
For compliance with Koopmans' condition, it seems 
unavoidable that data must be collected from both the
approximate neutral and ionized 
(the total energy of which may be sufficient)
systems, in order
to calculate $U_1$ and $U_2$.

\begin{figure}
\includegraphics[width=1\columnwidth]{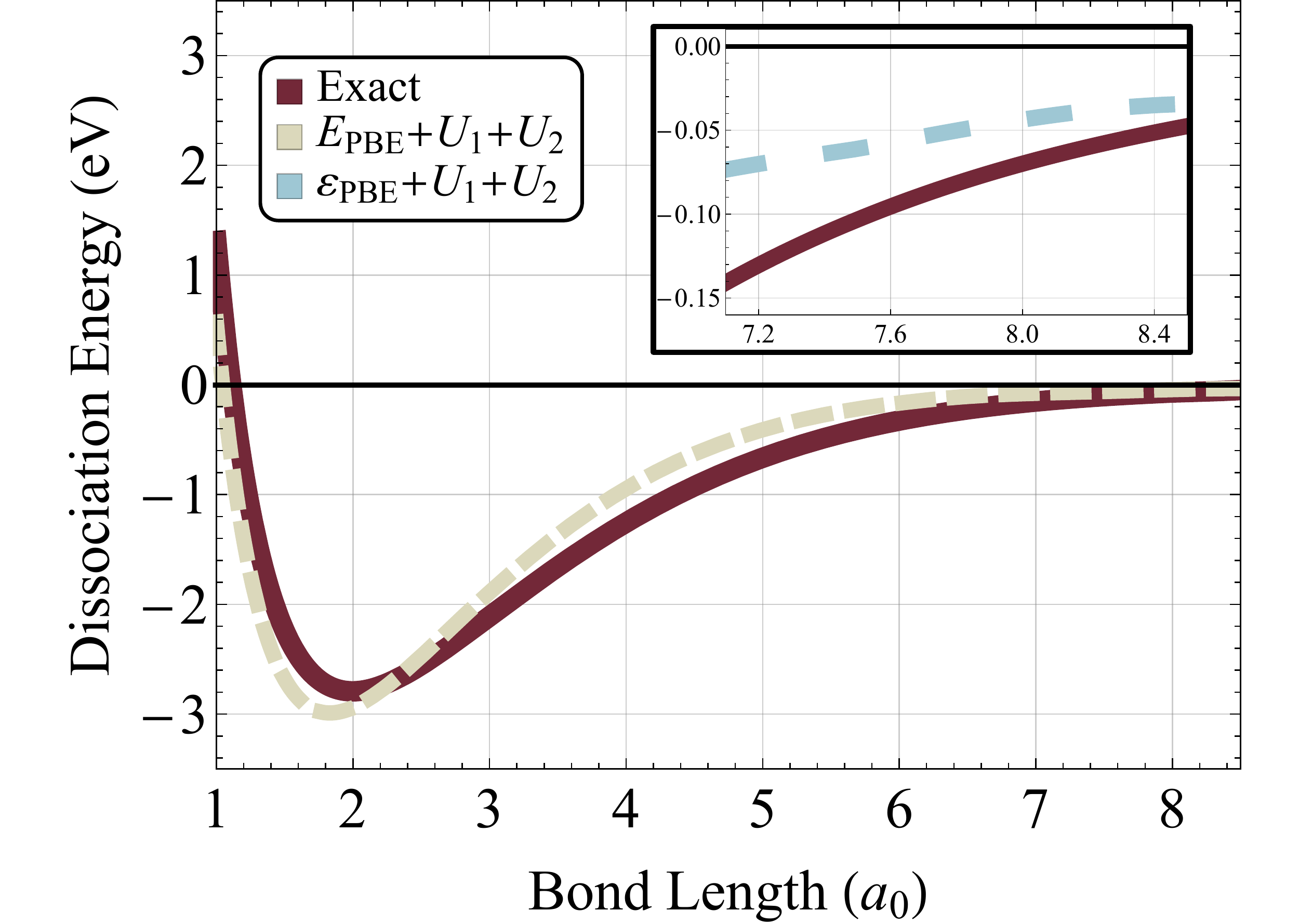}
\caption{(Color online) 
Exact (solid),
total-energy based $E_{\textrm{PBE}+U_1+U_2}$ (dashed), 
and eigenvalue based $\varepsilon_{\textrm{PBE}+U_1+U_2}$ (dotted)
dissociation energy curves for $H_2^+$, calculated
using DFT+$U_1$+$U_2$ as defined in the main text.
This method, combined with self-consistently 
calculated Hubbard $U$ parameters, here 
of $U^{(2)}$ type, enables the simultaneous  
SIE correction of the total-energy and Kohn-Sham eigenvalue
(to precisely the same accuracy for a one-electron system, as shown).}
\label{figure4}
\end{figure}

We carried out  density non-self-consistent 
DFT+$U_1$+$U_2$ calculations on the basis of the
PBE  total energy and occupied Kohn-Sham eigenvalue, 
following the  formulae given in Ref.~\onlinecite{PhysRevB.94.220104}.
To put the method on a first-principles footing, 
we used the self-consistent  
value $U^{(2)}$
to calculate $U_1$ and $U_2$, 
resulting in a density 
correction to the  total-energy  summing to
$\Delta E = U^{(2)} (N - N^2)$.
The corresponding modification 
to the subspace potentials is given by 
$\Delta v_U =  U^{(2)} (N -N^2) - U_K/2 $ where, 
for this system,
$U_K/2 = E_\textrm{ion-ion} - E_\textrm{PBE} + 
\varepsilon_\textrm{PBE}$.
Noting that the correction to the Kohn-Sham eigenvalue
$\varepsilon_\textrm{PBE}$ tends to  $\Delta v_U$
in the dissociated limit where  changes to the occupied Kohn-Sham 
orbital are negligible, there we find that
$\varepsilon_{\textrm{PBE}+U_1+U_2} \equiv 
\varepsilon_\textrm{PBE} + \Delta v_U  
= \varepsilon_\textrm{PBE} + \Delta E - ( E_\textrm{ion-ion} - E_\textrm{PBE} + \varepsilon_\textrm{PBE}) = - ( 
E_\textrm{ion-ion} - 
E_{\textrm{PBE}+U_1+U_2} ) \equiv - \textrm{IP} $, if
IP is the ionization potential, i.e., that 
Koopmans' compliance is restored for a  SIE
correction strength of $U^{(2)}$.
Fig.~\ref{figure4} illustrates 
the result of this simple technique, 
which  simultaneously reconciles the 
total energy $E$ and eigenvalue $\varepsilon$ derived
dissociation curves with the PBE+$U^{(2)}$ dissociation
curve of Fig.~\ref{figure3}, albeit imprecisely   
as this is a non-self-consistent post-processing step.

Our results highlights 
the potential of the DFT+$U_1+U_2$ approach
and its immediate compatibility with self-consistently
calculated Hubbard $U$ parameters.
To our knowledge, the SIE
of approximate DFT 
has not previously been  simultaneously
addressed for the total-energy and the occupied
Kohn-Sham eigenvalue using a
first-principles correction method of DFT+$U$ type,
even for a  one-electron system such as this. 
In the manner in which we have performed it  here, 
non-self-consistent DFT+$U_1+U_2$ 
requires only one total-energy calculation, 
at the ionized state, on top of the usual apparatus
of a linear-response DFT+$U$ calculation,  
in order to simultaneously, albeit approximately, 
correct the total energy
and the highest occupied Kohn-Sham eigenvalue for SIE.
Interesting avenues for the development of this method 
include its extension to multi-electron, heterogeneous,
and non-trivially spin-polarized systems, as well as
 to
perform self-consistency over the density
and to lift the fixed-occupancy approximation, 
as outlined in Ref.~\onlinecite{PhysRevB.94.220104}.
In principle, a further refinement of the
method might entail the self-consistent
linear-response calculation of $U_1$ and $U_2$
separately for the neutral and ionized states.

 %
 \subsection{Binding Curve Parameters}
 
In order to further quantify the results  
of the various Hubbard $U$ self-consistency schemes tested, 
we determined the equilibrium 
bond-length $R_e$, 
dissociation energy $E_D$, 
harmonic frequency $\omega_e$ 
and anharmonicity $\omega_e\chi_e$, 
corresponding to  each, as shown in  Table.~\ref{table2}, 
by fitting a polynomial  
about the energy  minima.
As compared with the 
experimental data of 
Ref.~\onlinecite{HERZBERG1972425}, 
the exact calculations perform well in determining 
the bond-length and harmonicity 
in particular, with errors that reflect the 
inaccuracies due to our fitting scheme, finite 
computational basis set size, core pseudization, and
absent physical effects, as well as experimental factors.
\begin{table}[b]
\begin{tabular*}{\columnwidth}{@{\extracolsep{\fill}}lcccc}
\hline \hline
 &$R_e$  	&$E_D$	 &$\omega_e$&$\omega_e\chi_e$\\
\hline
Experiment~\cite{HERZBERG1972425}
			& 1.988		& 2.6508		& 2321.7		& 66.2 \\
Exact		& 1.997		& 2.7922		& 2323.6	& 59.9 \\
PBE			& 2.138		& 2.9893		& 1912.0	& 37.9 \\
PBE+$U^{(1)}$	& 1.963		& 2.957(3)		& 2346(6)		& 57.7(3)\\
PBE+$U^{(2)}$	& 1.827		& 2.990(3) 	& 2799(5)		& 81.2(4)\\
PBE+$U^{(3)}$	& 1.845		& 2.985(9) 	& 2721(9)		& 76.9(7)\\
PBE+$U_1 + U_2$ & 1.829 	& 2.990(2)		& 2810(3)		& 81.8(6)\\ 
\hline
\hline
\end{tabular*}
\caption{
Equilibrium bond-lengths $R_e$ (bohr), dissociation energy $E_D$ ($eV$), 
harmonic frequencies $\omega_e\ (\textrm{cm}^{-1})$, 
and anharmonicities $\omega_e\chi_e\ (\textrm{cm}^{-1})$ 
for each calculation scheme tested, for comparison with 
experimental values~\cite{HERZBERG1972425}.}
\label{table2}
\end{table}
%

The PBE functional overestimates the 
equilibrium bond-length 
and dissociation energy,  while underestimating 
the harmonic frequency and anharmonicity.
The various DFT+$U$ schemes tested
generally preserve the PBE dissociation energy 
(N.B.,  calculated with respect to the exact-functional one-atom
total-energy, rather than to the PBE local maximum
at $6$-$7$~a$_0$)
but they reduce the bond-length 
and increase the frequency 
and anharmonictiy.
The $U^{(2)}$,  $U^{(3)}$ and $U_1+U_2$ 
schemes over-correct the latter three
and remain as poorly predictive  
of the experimental data as 
 the uncorrected PBE is.
We attribute this to  imperfect
DFT+$U$ population analysis at shorter bond-lengths,
featuring both  double-counting across the two
subspaces and spillage, 
as well as the breakdown of the subspace-bath separation.
The double-counting, in particular, is 
not properly compensated for 
by self-consistently calculated  $U$ parameters, 
since the formula for $U_\textrm{out}$ features no quantification of this
effect.
The $U^{(1)}$ scheme performs well here, 
as reflected also in Fig.~\ref{figure2}, 
approximately recovering the exact 
bond-length, harmonic frequency and anharmonicity.
We emphasise that this is due entirely to  
the $U^{(1)}$ parameter simply being smaller 
be definition, so that the over-correction due to double-counting
is less extreme.
It therefore  coincides with the exact regime serendipitously, 
and not by  design.

\section{The comparability
of DFT+$U$ total energies and the connections between the SCF  
and variational linear-response} 

In this section, we explore the relevance of first-principles calculated
Hubbard $U$ values, particularly at  self-consistency,
to the comparability  of the
DFT+$U$ total energies across different calculations in which
 the Hubbard $U$ is separately calculated due to the variation of 
 external parameters  such as
  stoichiometry and crystallographic geometry.
We also 
clarify the technical similarities and
differences between  SCF
linear-response~\cite{PhysRevB.71.035105} 
and its derived  
variational linear-response method. 

The open question of the rigorous comparability of 
DFT+$U$ total-energies that are
generated by calculations with different
$U$ values, which ordinarily represent external parameters
with the same status as ionic positions, 
is of considerable contemporary relevance.
This is demonstrated by recent progress in calculating  
thermodynamic quantities~\cite{doi:10.1021/cm702327g,CapdevilaCortada201558,PhysRevB.85.155208,PhysRevB.84.045115,PhysRevB.85.115104,PhysRevB.90.115105},
in high-throughput materials informatics~\cite{curtarolo,Curtarolo2012218,PhysRevX.5.011006}, 
 catalysis~\cite{C1CP22128K,doi:10.1021/jp407736f,doi:10.1021/acscatal.6b01907,PhysRevB.88.245204},
and in the study of ion-migration in battery materials~\cite{PhysRevB.70.235121,PhysRevB.83.075112,PhysRevB.93.085135,C2TA00839D,C1EE01782A,Zhou2004181,doi:10.1021/jacs.5b04690,doi:10.1021/acs.chemmater.5b03554,Morgan20075034} 
by means of DFT+$U$  and its related methods.

The variational linear-response definition
$U = d v_\textrm{int} / d N$, so-called as it is
based on the variational response of the  ground-state density, 
demands that the subspace-averaged 
non-interacting response 
$\chi_0 = d N / d v_\textrm{KS} $ is calculated
using the same set of ground-state densities,
parameterized by the external perturbation strength
$\alpha$, as that used for calculating 
the interacting response $\chi = d N / d \alpha$.
We may therefore write, taking the limit of 
small perturbations, that 
\begin{align}
U_\textrm{out}  = \chi_0^{-1} - \chi^{-1}
= \left. \frac{d v_\textrm{KS} \left[ \hat{\rho } \right] }{d N \left[ \hat{\rho } \right] } \right|_{\hat{\rho}_0} 
-
\left. \frac{d v_\textrm{ext}}{d N \left[ \hat{\rho } \right] } \right|_{\hat{\rho}_0} ,
\end{align}
where $\hat{\rho}_0 $ is the unperturbed
 Kohn-Sham density-matrix.
From this, it is clear that $U_\textrm{out} \left[ \hat{\rho}_0 \right] $ 
is a ground-state
density-functional, albeit not one of an explicit algebraic form.
This definition is readily adaptable to orbital-free
DFT, in which there is no Kohn-Sham Hamiltonian to diagonalize  
but only a density (rather than a Kohn-Sham density-matrix)
to optimize, and  where $\hat{v}_\textrm{KS}$ is replaced by the 
total potential. 
If we perform a variational linear-response calculation 
for a given $U_\textrm{in}$, notwithstanding, then the resulting  
$U_\textrm{out} $ may  thought of as
a ground-state density functional parameterised 
by $U_\textrm{in}$.
If we can then uniquely determine  
$U_\textrm{in}$ by applying a self-consistency criterion
such as  $U_\textrm{out} = 0$~eV, 
we will thereby uniquely determine the self-consistent ground-state 
DFT+$U$ density-matrix (up to  unitary transformations)
$\hat{\rho}_0 ( U_\textrm{in} )$, as well as its derived
properties such as the total-energy,  in terms 
of the remaining  parameters, e.g.,  
 ionic positions.

The comparability of total-energies between 
various crystallographic 
or molecular structures with  differing 
self-consistent Hubbard $U$ values,
and the validity of thermodynamic calculations based on 
DFT+$U$, directly follows.
In this way, given the underlying explicit algebraic 
xc functional such as PBE, together with the choice of a set of
 subspaces to target for SIE correction, 
DFT+$U$ is elevated to the status of a 
self-contained  orbital-dependent 
density-functional in its own right,  
incorporating the Hubbard $U$ as a non-algebraic 
but readily computable auxiliary ground-state variable.

The SCF~\cite{PhysRevB.71.035105} 
and variational linear-response 
methods are identical in terms of
their applied external perturbation, 
the use of the Dyson equation
for multi-site models, and issues of DFT+$U$ 
population analysis choice and convergence. 
While they are  equally convenient for use 
with SCF-type DFT solvers,  the variational approach 
 is likely to be more convenient  for 
use with direct-minimization solvers.
They are also perfectly identical insofar as the calculation
of  $\chi $ is concerned.
They differ only in the definition and set of densities used to calculate
the subspace-averaged non-interacting response $\chi_0$.

A calculation of the non-interacting response $\chi_0$
following the first step of the SCF cycle, as in 
state-of-the-art linear-response calculations, 
relies, by construction, 
upon the density (or Kohn-Sham orbitals, 
or density-matrix, as the case may be) not   
 being converged to the variational 
ground-state for each given external parameter $\alpha$. 
A non-optimized density of this type will
typically not correspond  to the  ground-state 
for any value of $\alpha$, although its subspace-total
$N$ will be.
As a result, the  finite-difference 
data points for $\alpha \ne 0$
that build the SCF linear-response $\chi_0$ 
are individually not properties of the ground-states for
their corresponding external potentials,
noting that linear-response does not
imply the sufficiency of first-order 
perturbation theory or first-order screening.

Put another way,  the SCF 
$\chi_0$ (and hence the derived $U$) is not a ground-state 
property of the unperturbed ground-state
density, but instead an excited-state property 
(in the simple sense of non-ground-state, rather than of a  
 resonance) dependent on the eigensystem 
of the non-interacting Kohn-Sham Hamiltonian. 
The comparison of the resulting DFT+$U$ total-energies 
thus remains  well defined in
terms of ground-state densities, 
since the Kohn-Sham eigenspectra are themselves
ground-state properties.
However, by virtue of the SCF Hubbard $U$ not
itself being a purely ground-state density-functional property, 
in general, 
the total energies are also not necessarily so.

The precise effects of the departure from the ground-state 
energy surface in the
calculation of SCF $\chi_0$  have not been 
quantified to date, to the best of our knowledge.
Therefore, while the self-consistency scheme
dubbed $U^{(2)}$ seems to be optimal for use with the
variational linear-response scheme, as we have shown, 
this result does not automatically extend to its SCF progenitor.
Nonetheless, we may expect that the two inequivalent approaches 
for $\chi_0$ should yield similar numerical 
results in practice.

\section{Conclusions}\label{conclusions}

We have developed a simple, variational adaptation of the
widely-used linear-response method for directly calculating 
the Hubbard $U$ of DFT+$U$, in which the $U$ incorporates only
quantities calculated from ground-state densities.
This method puts DFT+$U$ on a  first-principles footing
within the context of direct-minimization
DFT solvers, even the linear-scaling solvers 
of the type now routinely used to simulate systems which
are simultaneously spatially and electronically 
complex~\cite{:/content/aip/journal/jcp/122/8/10.1063/1.1839852,hine2009linear,hine2010linear,Gillan200714,0953-8984-14-11-303,0953-8984-14-11-302,0953-8984-20-6-064208,Genovese2011149,PhysRevB.72.045121,doi:10.1063/1.2841077}

Our formalism simplifies the analysis of parameter self-consistency
schemes considerably and, at least for 
this specific method, there emerges a clear
best choice of self-consistency criterion, $U^{(2)}$, which 
has been explored relatively little in the literature to date. 
We recommend  the use of a more complicated 
criterion, the previously proposed $U^{(3)}$, 
particularly for density-non-self-consistent methods such as 
post-processing DFT+DMFT.
In stringent calculations of the dissociated limit of the
purely SIE-afflicted system PBE $H_2^+$, 
where DFT+$U$ operates under ideal conditions, 
we are able to directly observe that the method
corrects the SIE in the total-energy 
very precisely, as foreseen in Ref.~\onlinecite{PhysRevLett.97.103001}.
It does so entirely from first-principles 
when the $U^{(2)}$ scheme is used.

Our analysis also shows that the comparison of
 thermodynamically relevant DFT+$U$
quantities such as the total-energy 
between dissimilar systems  demanding different  first-principles
$U$ parameters is, at least, well defined.
This comparison evidently becomes one
between purely ground-state properties, moreover,
in the case where the variational
linear-response method is applied together
with parameter self-consistency, but there may well be
other circumstances in which this also holds true.
The  DFT+$U_1+U_2$ method~\cite{PhysRevB.94.220104}, 
put here on a first-principles basis, 
extends the DFT+$U^{(2)}$ full SIE correction of the  $H_2^+$
total-energy to the highest occupied eigenvalue, 
approximately enforcing Koopman's condition.

Finally, we note that to properly account for SIE  
in the total-energy across the bond-length range, 
one would need to fully take into account 
the effects of subspace charge spillage, overlap and double-counting, 
possibly through the use of 
Wannier functions~\cite{PhysRevB.77.085122} 
generated self-consistently with the DFT+$U$
electronic structure~\cite{PhysRevB.82.081102}.
At least as important for correcting SIE in the 
bonding regime, perhaps, is the necessity 
to overcome the breakdown of the 
single-site approximation.
For this, the account of inter-subspace SIE 
offered by the multi-site method 
DFT+$U$+$V$~\cite{campo2010extended}
is a promising avenue for  investigation.

\acknowledgements

This work was enabled by the Royal Irish Academy -- Royal Society 
International Exchange Cost Share Programme (IE131505). 
GT acknowledges support from EPSRC UK 
(EP/I004483/1 and EP/K013610/1). 
GM and DDO'R  acknowledge support from the Science Foundation 
Ireland (SFI) funded centre AMBER (SFI/12/RC/2278).
We thank Edward Linscott, 
Fiona McCarthy, Mark McGrath, Thomas
Wyse Jackson, and Stefano Sanvito for discussions.
All calculations were performed on the Lonsdale cluster 
maintained by the Trinity Centre for High Performance Computing
and funded through grants from Science Foundation Ireland.

%


\begin{thebibliography}{114}%
\makeatletter
\providecommand \@ifxundefined [1]{%
 \@ifx{#1\undefined}
}%
\providecommand \@ifnum [1]{%
 \ifnum #1\expandafter \@firstoftwo
 \else \expandafter \@secondoftwo
 \fi
}%
\providecommand \@ifx [1]{%
 \ifx #1\expandafter \@firstoftwo
 \else \expandafter \@secondoftwo
 \fi
}%
\providecommand \natexlab [1]{#1}%
\providecommand \enquote  [1]{``#1''}%
\providecommand \bibnamefont  [1]{#1}%
\providecommand \bibfnamefont [1]{#1}%
\providecommand \citenamefont [1]{#1}%
\providecommand \href@noop [0]{\@secondoftwo}%
\providecommand \href [0]{\begingroup \@sanitize@url \@href}%
\providecommand \@href[1]{\@@startlink{#1}\@@href}%
\providecommand \@@href[1]{\endgroup#1\@@endlink}%
\providecommand \@sanitize@url [0]{\catcode `\\12\catcode `\$12\catcode
  `\&12\catcode `\#12\catcode `\^12\catcode `\_12\catcode `\%12\relax}%
\providecommand \@@startlink[1]{}%
\providecommand \@@endlink[0]{}%
\providecommand \url  [0]{\begingroup\@sanitize@url \@url }%
\providecommand \@url [1]{\endgroup\@href {#1}{\urlprefix }}%
\providecommand \urlprefix  [0]{URL }%
\providecommand \Eprint [0]{\href }%
\providecommand \doibase [0]{http://dx.doi.org/}%
\providecommand \selectlanguage [0]{\@gobble}%
\providecommand \bibinfo  [0]{\@secondoftwo}%
\providecommand \bibfield  [0]{\@secondoftwo}%
\providecommand \translation [1]{[#1]}%
\providecommand \BibitemOpen [0]{}%
\providecommand \bibitemStop [0]{}%
\providecommand \bibitemNoStop [0]{.\EOS\space}%
\providecommand \EOS [0]{\spacefactor3000\relax}%
\providecommand \BibitemShut  [1]{\csname bibitem#1\endcsname}%
\let\auto@bib@innerbib\@empty
\bibitem [{\citenamefont {Hohenberg}\ and\ \citenamefont
  {Kohn}(1964)}]{PhysRev.136.B864}%
  \BibitemOpen
  \bibfield  {author} {\bibinfo {author} {\bibfnamefont {P.}~\bibnamefont
  {Hohenberg}}\ and\ \bibinfo {author} {\bibfnamefont {W.}~\bibnamefont
  {Kohn}},\ }\bibfield  {title} {\enquote {\bibinfo {title} {Inhomogeneous
  electron gas},}\ }\href {\doibase 10.1103/PhysRev.136.B864} {\bibfield
  {journal} {\bibinfo  {journal} {Phys. Rev.}\ }\textbf {\bibinfo {volume}
  {136}},\ \bibinfo {pages} {B864--B871} (\bibinfo {year} {1964})}\BibitemShut
  {NoStop}%
\bibitem [{\citenamefont {Kohn}\ and\ \citenamefont
  {Sham}(1965)}]{PhysRev.140.A1133}%
  \BibitemOpen
  \bibfield  {author} {\bibinfo {author} {\bibfnamefont {W.}~\bibnamefont
  {Kohn}}\ and\ \bibinfo {author} {\bibfnamefont {L.~J.}\ \bibnamefont
  {Sham}},\ }\bibfield  {title} {\enquote {\bibinfo {title} {Self-consistent
  equations including exchange and correlation effects},}\ }\href {\doibase
  10.1103/PhysRev.140.A1133} {\bibfield  {journal} {\bibinfo  {journal} {Phys.
  Rev.}\ }\textbf {\bibinfo {volume} {140}},\ \bibinfo {pages} {A1133--A1138}
  (\bibinfo {year} {1965})}\BibitemShut {NoStop}%
\bibitem [{\citenamefont {Jones}(2015)}]{RevModPhys.87.897}%
  \BibitemOpen
  \bibfield  {author} {\bibinfo {author} {\bibfnamefont {R.~O.}\ \bibnamefont
  {Jones}},\ }\bibfield  {title} {\enquote {\bibinfo {title}
  {{D}ensity-functional theory: {I}ts origins, rise to prominence, and
  future},}\ }\href {\doibase 10.1103/RevModPhys.87.897} {\bibfield  {journal}
  {\bibinfo  {journal} {Rev. Mod. Phys.}\ }\textbf {\bibinfo {volume} {87}},\
  \bibinfo {pages} {897--923} (\bibinfo {year} {2015})}\BibitemShut {NoStop}%
\bibitem [{\citenamefont {Esch}\ \emph {et~al.}(2005)\citenamefont {Esch},
  \citenamefont {Fabris}, \citenamefont {Zhou}, \citenamefont {Montini},
  \citenamefont {Africh}, \citenamefont {Fornasiero}, \citenamefont {Comelli},\
  and\ \citenamefont {Rosei}}]{Esch752}%
  \BibitemOpen
  \bibfield  {author} {\bibinfo {author} {\bibfnamefont {F.}~\bibnamefont
  {Esch}}, \bibinfo {author} {\bibfnamefont {S.}~\bibnamefont {Fabris}},
  \bibinfo {author} {\bibfnamefont {L.}~\bibnamefont {Zhou}}, \bibinfo {author}
  {\bibfnamefont {T.}~\bibnamefont {Montini}}, \bibinfo {author} {\bibfnamefont
  {C.}~\bibnamefont {Africh}}, \bibinfo {author} {\bibfnamefont
  {P.}~\bibnamefont {Fornasiero}}, \bibinfo {author} {\bibfnamefont
  {G.}~\bibnamefont {Comelli}}, \ and\ \bibinfo {author} {\bibfnamefont
  {R.}~\bibnamefont {Rosei}},\ }\bibfield  {title} {\enquote {\bibinfo {title}
  {{E}lectron localization determines defect formation on ceria substrates},}\
  }\href {\doibase 10.1126/science.1111568} {\bibfield  {journal} {\bibinfo
  {journal} {Science}\ }\textbf {\bibinfo {volume} {309}},\ \bibinfo {pages}
  {752--755} (\bibinfo {year} {2005})}\BibitemShut {NoStop}%
\bibitem [{\citenamefont {Jain}, \citenamefont {Shin},\ and\ \citenamefont
  {Persson}(2016)}]{jain2016computational}%
  \BibitemOpen
  \bibfield  {author} {\bibinfo {author} {\bibfnamefont {A.}~\bibnamefont
  {Jain}}, \bibinfo {author} {\bibfnamefont {Y.}~\bibnamefont {Shin}}, \ and\
  \bibinfo {author} {\bibfnamefont {K.~A.}\ \bibnamefont {Persson}},\
  }\bibfield  {title} {\enquote {\bibinfo {title} {Computational predictions of
  energy materials using density-functional theory},}\ }\href
  {http://dx.doi.org/10.1038/natrevmats.2015.4} {\bibfield  {journal} {\bibinfo
   {journal} {Nat. Rev. Mater.}\ }\textbf {\bibinfo {volume} {1}},\ \bibinfo
  {pages} {15004} (\bibinfo {year} {2016})}\BibitemShut {NoStop}%
\bibitem [{\citenamefont {Perdew}\ and\ \citenamefont
  {Zunger}(1981)}]{PhysRevB.23.5048}%
  \BibitemOpen
  \bibfield  {author} {\bibinfo {author} {\bibfnamefont {J.~P.}\ \bibnamefont
  {Perdew}}\ and\ \bibinfo {author} {\bibfnamefont {A.}~\bibnamefont
  {Zunger}},\ }\bibfield  {title} {\enquote {\bibinfo {title} {Self-interaction
  correction to density-functional approximations for many-electron systems},}\
  }\href {\doibase 10.1103/PhysRevB.23.5048} {\bibfield  {journal} {\bibinfo
  {journal} {Phys. Rev. B}\ }\textbf {\bibinfo {volume} {23}},\ \bibinfo
  {pages} {5048--5079} (\bibinfo {year} {1981})}\BibitemShut {NoStop}%
\bibitem [{\citenamefont {Perdew}, \citenamefont {Burke},\ and\ \citenamefont
  {Ernzerhof}(1996)}]{PhysRevLett.77.3865}%
  \BibitemOpen
  \bibfield  {author} {\bibinfo {author} {\bibfnamefont {J.~P.}\ \bibnamefont
  {Perdew}}, \bibinfo {author} {\bibfnamefont {K.}~\bibnamefont {Burke}}, \
  and\ \bibinfo {author} {\bibfnamefont {M.}~\bibnamefont {Ernzerhof}},\
  }\bibfield  {title} {\enquote {\bibinfo {title} {Generalized gradient
  approximation made simple},}\ }\href {\doibase 10.1103/PhysRevLett.77.3865}
  {\bibfield  {journal} {\bibinfo  {journal} {Phys. Rev. Lett.}\ }\textbf
  {\bibinfo {volume} {77}},\ \bibinfo {pages} {3865--3868} (\bibinfo {year}
  {1996})}\BibitemShut {NoStop}%
\bibitem [{\citenamefont {Becke}(1993)}]{doi:10.1063/1.464913}%
  \BibitemOpen
  \bibfield  {author} {\bibinfo {author} {\bibfnamefont {A.~D.}\ \bibnamefont
  {Becke}},\ }\bibfield  {title} {\enquote {\bibinfo {title}
  {Density-functional thermochemistry. {III}. {T}he role of exact exchange},}\
  }\href {\doibase 10.1063/1.464913} {\bibfield  {journal} {\bibinfo  {journal}
  {J. Chem. Phys.}\ }\textbf {\bibinfo {volume} {98}},\ \bibinfo {pages}
  {5648--5652} (\bibinfo {year} {1993})}\BibitemShut {NoStop}%
\bibitem [{\citenamefont {Yanai}, \citenamefont {Tew},\ and\ \citenamefont
  {Handy}(2004)}]{Yanai200451}%
  \BibitemOpen
  \bibfield  {author} {\bibinfo {author} {\bibfnamefont {T.}~\bibnamefont
  {Yanai}}, \bibinfo {author} {\bibfnamefont {D.~P.}\ \bibnamefont {Tew}}, \
  and\ \bibinfo {author} {\bibfnamefont {N.~C.}\ \bibnamefont {Handy}},\
  }\bibfield  {title} {\enquote {\bibinfo {title} {A new hybrid
  exchange-correlation functional using the {C}oulomb-attenuating method
  ({CAM}-{B3LYP})},}\ }\href {\doibase
  http://dx.doi.org/10.1016/j.cplett.2004.06.011} {\bibfield  {journal}
  {\bibinfo  {journal} {Chem. Phys. Lett.}\ }\textbf {\bibinfo {volume}
  {393}},\ \bibinfo {pages} {51 -- 57} (\bibinfo {year} {2004})}\BibitemShut
  {NoStop}%
\bibitem [{\citenamefont {Wilson}, \citenamefont {Bradley},\ and\ \citenamefont
  {Tozer}(2001)}]{doi:10.1063/1.1412605}%
  \BibitemOpen
  \bibfield  {author} {\bibinfo {author} {\bibfnamefont {P.~J.}\ \bibnamefont
  {Wilson}}, \bibinfo {author} {\bibfnamefont {T.~J.}\ \bibnamefont {Bradley}},
  \ and\ \bibinfo {author} {\bibfnamefont {D.~J.}\ \bibnamefont {Tozer}},\
  }\bibfield  {title} {\enquote {\bibinfo {title} {Hybrid exchange-correlation
  functional determined from thermochemical data and ab initio potentials},}\
  }\href {\doibase 10.1063/1.1412605} {\bibfield  {journal} {\bibinfo
  {journal} {J. Chem. Phys.}\ }\textbf {\bibinfo {volume} {115}},\ \bibinfo
  {pages} {9233--9242} (\bibinfo {year} {2001})}\BibitemShut {NoStop}%
\bibitem [{\citenamefont {Dudarev}\ \emph {et~al.}(1998)\citenamefont
  {Dudarev}, \citenamefont {Botton}, \citenamefont {Savrasov}, \citenamefont
  {Humphreys},\ and\ \citenamefont {Sutton}}]{PhysRevB.57.1505}%
  \BibitemOpen
  \bibfield  {author} {\bibinfo {author} {\bibfnamefont {S.~L.}\ \bibnamefont
  {Dudarev}}, \bibinfo {author} {\bibfnamefont {G.~A.}\ \bibnamefont {Botton}},
  \bibinfo {author} {\bibfnamefont {S.~Y.}\ \bibnamefont {Savrasov}}, \bibinfo
  {author} {\bibfnamefont {C.~J.}\ \bibnamefont {Humphreys}}, \ and\ \bibinfo
  {author} {\bibfnamefont {A.~P.}\ \bibnamefont {Sutton}},\ }\bibfield  {title}
  {\enquote {\bibinfo {title} {Electron-energy-loss spectra and the structural
  stability of nickel oxide: {A}n {LSDA+U} study},}\ }\href {\doibase
  10.1103/PhysRevB.57.1505} {\bibfield  {journal} {\bibinfo  {journal} {Phys.
  Rev. B}\ }\textbf {\bibinfo {volume} {57}},\ \bibinfo {pages} {1505--1509}
  (\bibinfo {year} {1998})}\BibitemShut {NoStop}%
\bibitem [{\citenamefont {Cohen}, \citenamefont {Mori-S{\'a}nchez},\ and\
  \citenamefont {Yang}(2008)}]{Cohen792}%
  \BibitemOpen
  \bibfield  {author} {\bibinfo {author} {\bibfnamefont {A.~J.}\ \bibnamefont
  {Cohen}}, \bibinfo {author} {\bibfnamefont {P.}~\bibnamefont
  {Mori-S{\'a}nchez}}, \ and\ \bibinfo {author} {\bibfnamefont
  {W.}~\bibnamefont {Yang}},\ }\bibfield  {title} {\enquote {\bibinfo {title}
  {Insights into current limitations of density-functional theory},}\ }\href
  {\doibase 10.1126/science.1158722} {\bibfield  {journal} {\bibinfo  {journal}
  {Science}\ }\textbf {\bibinfo {volume} {321}},\ \bibinfo {pages} {792--794}
  (\bibinfo {year} {2008})}\BibitemShut {NoStop}%
\bibitem [{\citenamefont {Perdew}\ \emph {et~al.}(1982)\citenamefont {Perdew},
  \citenamefont {Parr}, \citenamefont {Levy},\ and\ \citenamefont
  {Balduz}}]{PhysRevLett.49.1691}%
  \BibitemOpen
  \bibfield  {author} {\bibinfo {author} {\bibfnamefont {J.~P.}\ \bibnamefont
  {Perdew}}, \bibinfo {author} {\bibfnamefont {R.~G.}\ \bibnamefont {Parr}},
  \bibinfo {author} {\bibfnamefont {M.}~\bibnamefont {Levy}}, \ and\ \bibinfo
  {author} {\bibfnamefont {J.~L.}\ \bibnamefont {Balduz}},\ }\bibfield  {title}
  {\enquote {\bibinfo {title} {Density-functional theory for fractional
  particle number: Derivative discontinuities of the energy},}\ }\href
  {\doibase 10.1103/PhysRevLett.49.1691} {\bibfield  {journal} {\bibinfo
  {journal} {Phys. Rev. Lett.}\ }\textbf {\bibinfo {volume} {49}},\ \bibinfo
  {pages} {1691--1694} (\bibinfo {year} {1982})}\BibitemShut {NoStop}%
\bibitem [{\citenamefont {Droghetti}\ \emph {et~al.}(2016)\citenamefont
  {Droghetti}, \citenamefont {Rungger}, \citenamefont {Das~Pemmaraju},\ and\
  \citenamefont {Sanvito}}]{PhysRevB.93.195208}%
  \BibitemOpen
  \bibfield  {author} {\bibinfo {author} {\bibfnamefont {A.}~\bibnamefont
  {Droghetti}}, \bibinfo {author} {\bibfnamefont {I.}~\bibnamefont {Rungger}},
  \bibinfo {author} {\bibfnamefont {C.}~\bibnamefont {Das~Pemmaraju}}, \ and\
  \bibinfo {author} {\bibfnamefont {S.}~\bibnamefont {Sanvito}},\ }\bibfield
  {title} {\enquote {\bibinfo {title} {Fundamental gap of molecular crystals
  via constrained density-functional theory},}\ }\href {\doibase
  10.1103/PhysRevB.93.195208} {\bibfield  {journal} {\bibinfo  {journal} {Phys.
  Rev. B}\ }\textbf {\bibinfo {volume} {93}},\ \bibinfo {pages} {195208}
  (\bibinfo {year} {2016})}\BibitemShut {NoStop}%
\bibitem [{\citenamefont {Toher}\ \emph {et~al.}(2005)\citenamefont {Toher},
  \citenamefont {Filippetti}, \citenamefont {Sanvito},\ and\ \citenamefont
  {Burke}}]{PhysRevLett.95.146402}%
  \BibitemOpen
  \bibfield  {author} {\bibinfo {author} {\bibfnamefont {C.}~\bibnamefont
  {Toher}}, \bibinfo {author} {\bibfnamefont {A.}~\bibnamefont {Filippetti}},
  \bibinfo {author} {\bibfnamefont {S.}~\bibnamefont {Sanvito}}, \ and\
  \bibinfo {author} {\bibfnamefont {K.}~\bibnamefont {Burke}},\ }\bibfield
  {title} {\enquote {\bibinfo {title} {Self-interaction errors in
  density-functional calculations of electronic transport},}\ }\href {\doibase
  10.1103/PhysRevLett.95.146402} {\bibfield  {journal} {\bibinfo  {journal}
  {Phys. Rev. Lett.}\ }\textbf {\bibinfo {volume} {95}},\ \bibinfo {pages}
  {146402} (\bibinfo {year} {2005})}\BibitemShut {NoStop}%
\bibitem [{\citenamefont {Ke}, \citenamefont {Baranger},\ and\ \citenamefont
  {Yang}(2007)}]{:/content/aip/journal/jcp/126/20/10.1063/1.2743004}%
  \BibitemOpen
  \bibfield  {author} {\bibinfo {author} {\bibfnamefont {S.-H.}\ \bibnamefont
  {Ke}}, \bibinfo {author} {\bibfnamefont {H.~U.}\ \bibnamefont {Baranger}}, \
  and\ \bibinfo {author} {\bibfnamefont {W.}~\bibnamefont {Yang}},\ }\bibfield
  {title} {\enquote {\bibinfo {title} {Role of the exchange-correlation
  potential in ab initio electron transport calculations},}\ }\href
  {http://scitation.aip.org/content/aip/journal/jcp/126/20/10.1063/1.2743004}
  {\bibfield  {journal} {\bibinfo  {journal} {J. Chem. Phys.}\ }\textbf
  {\bibinfo {volume} {126}},\ \bibinfo {eid} {201102} (\bibinfo {year}
  {2007})}\BibitemShut {NoStop}%
\bibitem [{\citenamefont {Zhao}, \citenamefont {Lynch},\ and\ \citenamefont
  {Truhlar}(2004)}]{doi:10.1021/jp049908s}%
  \BibitemOpen
  \bibfield  {author} {\bibinfo {author} {\bibfnamefont {Y.}~\bibnamefont
  {Zhao}}, \bibinfo {author} {\bibfnamefont {B.~J.}\ \bibnamefont {Lynch}}, \
  and\ \bibinfo {author} {\bibfnamefont {D.~G.}\ \bibnamefont {Truhlar}},\
  }\bibfield  {title} {\enquote {\bibinfo {title} {Development and assessment
  of a new hybrid density-functional model for thermochemical kinetics},}\
  }\href {\doibase 10.1021/jp049908s} {\bibfield  {journal} {\bibinfo
  {journal} {J. Phys. Chem. A}\ }\textbf {\bibinfo {volume} {108}},\ \bibinfo
  {pages} {2715--2719} (\bibinfo {year} {2004})}\BibitemShut {NoStop}%
\bibitem [{\citenamefont {Cohen}, \citenamefont {Mori-S\'anchez},\ and\
  \citenamefont {Yang}(2012)}]{doi:10.1021/cr200107z}%
  \BibitemOpen
  \bibfield  {author} {\bibinfo {author} {\bibfnamefont {A.~J.}\ \bibnamefont
  {Cohen}}, \bibinfo {author} {\bibfnamefont {P.}~\bibnamefont
  {Mori-S\'anchez}}, \ and\ \bibinfo {author} {\bibfnamefont {W.}~\bibnamefont
  {Yang}},\ }\bibfield  {title} {\enquote {\bibinfo {title} {Challenges for
  density functional theory},}\ }\href {\doibase 10.1021/cr200107z} {\bibfield
  {journal} {\bibinfo  {journal} {Chem. Rev.}\ }\textbf {\bibinfo {volume}
  {112}},\ \bibinfo {pages} {289--320} (\bibinfo {year} {2012})}\BibitemShut
  {NoStop}%
\bibitem [{\citenamefont {Anisimov}\ and\ \citenamefont
  {Gunnarsson}(1991)}]{PhysRevB.43.7570}%
  \BibitemOpen
  \bibfield  {author} {\bibinfo {author} {\bibfnamefont {V.~I.}\ \bibnamefont
  {Anisimov}}\ and\ \bibinfo {author} {\bibfnamefont {O.}~\bibnamefont
  {Gunnarsson}},\ }\bibfield  {title} {\enquote {\bibinfo {title}
  {Density-functional calculation of effective {C}oulomb interactions in
  metals},}\ }\href {\doibase 10.1103/PhysRevB.43.7570} {\bibfield  {journal}
  {\bibinfo  {journal} {Phys. Rev. B}\ }\textbf {\bibinfo {volume} {43}},\
  \bibinfo {pages} {7570--7574} (\bibinfo {year} {1991})}\BibitemShut {NoStop}%
\bibitem [{\citenamefont {Anisimov}, \citenamefont {Zaanen},\ and\
  \citenamefont {Andersen}(1991)}]{PhysRevB.44.943}%
  \BibitemOpen
  \bibfield  {author} {\bibinfo {author} {\bibfnamefont {V.~I.}\ \bibnamefont
  {Anisimov}}, \bibinfo {author} {\bibfnamefont {J.}~\bibnamefont {Zaanen}}, \
  and\ \bibinfo {author} {\bibfnamefont {O.~K.}\ \bibnamefont {Andersen}},\
  }\bibfield  {title} {\enquote {\bibinfo {title} {Band theory and mott
  insulators: Hubbard \textit{U} instead of stoner \textit{I}},}\ }\href
  {\doibase 10.1103/PhysRevB.44.943} {\bibfield  {journal} {\bibinfo  {journal}
  {Phys. Rev. B}\ }\textbf {\bibinfo {volume} {44}},\ \bibinfo {pages}
  {943--954} (\bibinfo {year} {1991})}\BibitemShut {NoStop}%
\bibitem [{\citenamefont {Anisimov}\ \emph {et~al.}(1993)\citenamefont
  {Anisimov}, \citenamefont {Solovyev}, \citenamefont {Korotin}, \citenamefont
  {Czy\ifmmode~\dot{z}\else \.{z}\fi{}yk},\ and\ \citenamefont
  {Sawatzky}}]{PhysRevB.48.16929}%
  \BibitemOpen
  \bibfield  {author} {\bibinfo {author} {\bibfnamefont {V.~I.}\ \bibnamefont
  {Anisimov}}, \bibinfo {author} {\bibfnamefont {I.~V.}\ \bibnamefont
  {Solovyev}}, \bibinfo {author} {\bibfnamefont {M.~A.}\ \bibnamefont
  {Korotin}}, \bibinfo {author} {\bibfnamefont {M.~T.}\ \bibnamefont
  {Czy\ifmmode~\dot{z}\else \.{z}\fi{}yk}}, \ and\ \bibinfo {author}
  {\bibfnamefont {G.~A.}\ \bibnamefont {Sawatzky}},\ }\bibfield  {title}
  {\enquote {\bibinfo {title} {Density-functional theory and $\mathrm{NiO}$
  photoemission spectra},}\ }\href {\doibase 10.1103/PhysRevB.48.16929}
  {\bibfield  {journal} {\bibinfo  {journal} {Phys. Rev. B}\ }\textbf {\bibinfo
  {volume} {48}},\ \bibinfo {pages} {16929--16934} (\bibinfo {year}
  {1993})}\BibitemShut {NoStop}%
\bibitem [{\citenamefont {Pickett}, \citenamefont {Erwin},\ and\ \citenamefont
  {Ethridge}(1998)}]{PhysRevB.58.1201}%
  \BibitemOpen
  \bibfield  {author} {\bibinfo {author} {\bibfnamefont {W.~E.}\ \bibnamefont
  {Pickett}}, \bibinfo {author} {\bibfnamefont {S.~C.}\ \bibnamefont {Erwin}},
  \ and\ \bibinfo {author} {\bibfnamefont {E.~C.}\ \bibnamefont {Ethridge}},\
  }\bibfield  {title} {\enquote {\bibinfo {title} {Reformulation of the
  $\mathrm{LDA}+\mathrm{U}$ method for a local-orbital basis},}\ }\href
  {\doibase 10.1103/PhysRevB.58.1201} {\bibfield  {journal} {\bibinfo
  {journal} {Phys. Rev. B}\ }\textbf {\bibinfo {volume} {58}},\ \bibinfo
  {pages} {1201--1209} (\bibinfo {year} {1998})}\BibitemShut {NoStop}%
\bibitem [{\citenamefont {Cococcioni}\ and\ \citenamefont
  {de~Gironcoli}(2005)}]{PhysRevB.71.035105}%
  \BibitemOpen
  \bibfield  {author} {\bibinfo {author} {\bibfnamefont {M.}~\bibnamefont
  {Cococcioni}}\ and\ \bibinfo {author} {\bibfnamefont {S.}~\bibnamefont
  {de~Gironcoli}},\ }\bibfield  {title} {\enquote {\bibinfo {title} {Linear
  response approach to the calculation of the effective interaction parameters
  in the $\mathrm{LDA}+\mathrm{U}$ method},}\ }\href {\doibase
  10.1103/PhysRevB.71.035105} {\bibfield  {journal} {\bibinfo  {journal} {Phys.
  Rev. B}\ }\textbf {\bibinfo {volume} {71}},\ \bibinfo {pages} {035105}
  (\bibinfo {year} {2005})}\BibitemShut {NoStop}%
\bibitem [{\citenamefont {O'Regan}\ \emph {et~al.}(2012)\citenamefont
  {O'Regan}, \citenamefont {Hine}, \citenamefont {Payne},\ and\ \citenamefont
  {Mostofi}}]{PhysRevB.85.085107}%
  \BibitemOpen
  \bibfield  {author} {\bibinfo {author} {\bibfnamefont {D.~D.}\ \bibnamefont
  {O'Regan}}, \bibinfo {author} {\bibfnamefont {N.~D.~M.}\ \bibnamefont
  {Hine}}, \bibinfo {author} {\bibfnamefont {M.~C.}\ \bibnamefont {Payne}}, \
  and\ \bibinfo {author} {\bibfnamefont {A.~A.}\ \bibnamefont {Mostofi}},\
  }\bibfield  {title} {\enquote {\bibinfo {title} {Linear-scaling {DFT+U} with
  full local orbital optimization},}\ }\href {\doibase
  10.1103/PhysRevB.85.085107} {\bibfield  {journal} {\bibinfo  {journal} {Phys.
  Rev. B}\ }\textbf {\bibinfo {volume} {85}},\ \bibinfo {pages} {085107}
  (\bibinfo {year} {2012})}\BibitemShut {NoStop}%
\bibitem [{\citenamefont {Kulik}\ \emph {et~al.}(2006)\citenamefont {Kulik},
  \citenamefont {Cococcioni}, \citenamefont {Scherlis},\ and\ \citenamefont
  {Marzari}}]{PhysRevLett.97.103001}%
  \BibitemOpen
  \bibfield  {author} {\bibinfo {author} {\bibfnamefont {H.~J.}\ \bibnamefont
  {Kulik}}, \bibinfo {author} {\bibfnamefont {M.}~\bibnamefont {Cococcioni}},
  \bibinfo {author} {\bibfnamefont {D.~A.}\ \bibnamefont {Scherlis}}, \ and\
  \bibinfo {author} {\bibfnamefont {N.}~\bibnamefont {Marzari}},\ }\bibfield
  {title} {\enquote {\bibinfo {title} {Density-functional theory in
  transition-metal chemistry: {A} self-consistent {H}ubbard {U} approach},}\
  }\href {\doibase 10.1103/PhysRevLett.97.103001} {\bibfield  {journal}
  {\bibinfo  {journal} {Phys. Rev. Lett.}\ }\textbf {\bibinfo {volume} {97}},\
  \bibinfo {pages} {103001} (\bibinfo {year} {2006})}\BibitemShut {NoStop}%
\bibitem [{\citenamefont {Kulik}\ and\ \citenamefont
  {Marzari}(2010)}]{:/content/aip/journal/jcp/133/11/10.1063/1.3489110}%
  \BibitemOpen
  \bibfield  {author} {\bibinfo {author} {\bibfnamefont {H.~J.}\ \bibnamefont
  {Kulik}}\ and\ \bibinfo {author} {\bibfnamefont {N.}~\bibnamefont
  {Marzari}},\ }\bibfield  {title} {\enquote {\bibinfo {title} {Systematic
  study of first-row transition-metal diatomic molecules: {A} self-consistent
  {DFT+U} approach},}\ }\href
  {http://scitation.aip.org/content/aip/journal/jcp/133/11/10.1063/1.3489110}
  {\bibfield  {journal} {\bibinfo  {journal} {J. Chem. Phys.}\ }\textbf
  {\bibinfo {volume} {133}},\ \bibinfo {eid} {114103} (\bibinfo {year}
  {2010})}\BibitemShut {NoStop}%
\bibitem [{\citenamefont {Zhao}, \citenamefont {Ioannidis},\ and\ \citenamefont
  {Kulik}(2016)}]{:/content/aip/journal/jcp/145/5/10.1063/1.4959882}%
  \BibitemOpen
  \bibfield  {author} {\bibinfo {author} {\bibfnamefont {Q.}~\bibnamefont
  {Zhao}}, \bibinfo {author} {\bibfnamefont {E.~I.}\ \bibnamefont {Ioannidis}},
  \ and\ \bibinfo {author} {\bibfnamefont {H.~J.}\ \bibnamefont {Kulik}},\
  }\bibfield  {title} {\enquote {\bibinfo {title} {Global and local curvature
  in density-functional theory},}\ }\href
  {http://scitation.aip.org/content/aip/journal/jcp/145/5/10.1063/1.4959882}
  {\bibfield  {journal} {\bibinfo  {journal} {J. Chem. Phys.}\ }\textbf
  {\bibinfo {volume} {145}},\ \bibinfo {eid} {054109} (\bibinfo {year}
  {2016})}\BibitemShut {NoStop}%
\bibitem [{\citenamefont {Anisimov}, \citenamefont {Aryasetiawan},\ and\
  \citenamefont {Lichtenstein}(1997)}]{0953-8984-9-4-002}%
  \BibitemOpen
  \bibfield  {author} {\bibinfo {author} {\bibfnamefont {V.~I.}\ \bibnamefont
  {Anisimov}}, \bibinfo {author} {\bibfnamefont {F.}~\bibnamefont
  {Aryasetiawan}}, \ and\ \bibinfo {author} {\bibfnamefont {A.~I.}\
  \bibnamefont {Lichtenstein}},\ }\bibfield  {title} {\enquote {\bibinfo
  {title} {First-principles calculations of the electronic structure and
  spectra of strongly correlated systems: the $\mathrm{LDA}+\mathrm{U}$
  method},}\ }\href {http://stacks.iop.org/0953-8984/9/i=4/a=002} {\bibfield
  {journal} {\bibinfo  {journal} {J. Phys. Condens. Matter}\ }\textbf {\bibinfo
  {volume} {9}},\ \bibinfo {pages} {767} (\bibinfo {year} {1997})}\BibitemShut
  {NoStop}%
\bibitem [{\citenamefont {Himmetoglu}\ \emph {et~al.}(2014)\citenamefont
  {Himmetoglu}, \citenamefont {Floris}, \citenamefont {de~Gironcoli},\ and\
  \citenamefont {Cococcioni}}]{QUA:QUA24521}%
  \BibitemOpen
  \bibfield  {author} {\bibinfo {author} {\bibfnamefont {B.}~\bibnamefont
  {Himmetoglu}}, \bibinfo {author} {\bibfnamefont {A.}~\bibnamefont {Floris}},
  \bibinfo {author} {\bibfnamefont {S.}~\bibnamefont {de~Gironcoli}}, \ and\
  \bibinfo {author} {\bibfnamefont {M.}~\bibnamefont {Cococcioni}},\ }\bibfield
   {title} {\enquote {\bibinfo {title} {Hubbard-corrected {DFT} energy
  functionals: {T}he $\mathrm{LDA}+\mathrm{U}$ description of correlated
  systems},}\ }\href {\doibase 10.1002/qua.24521} {\bibfield  {journal}
  {\bibinfo  {journal} {Int. J. Quantum Chem.}\ }\textbf {\bibinfo {volume}
  {114}},\ \bibinfo {pages} {14--49} (\bibinfo {year} {2014})}\BibitemShut
  {NoStop}%
\bibitem [{\citenamefont {Garcia-Lastra}\ \emph {et~al.}(2013)\citenamefont
  {Garcia-Lastra}, \citenamefont {Myrdal}, \citenamefont {Christensen},
  \citenamefont {Thygesen},\ and\ \citenamefont
  {Vegge}}]{doi:10.1021/jp3107809}%
  \BibitemOpen
  \bibfield  {author} {\bibinfo {author} {\bibfnamefont {J.~M.}\ \bibnamefont
  {Garcia-Lastra}}, \bibinfo {author} {\bibfnamefont {J.~S.~G.}\ \bibnamefont
  {Myrdal}}, \bibinfo {author} {\bibfnamefont {R.}~\bibnamefont {Christensen}},
  \bibinfo {author} {\bibfnamefont {K.~S.}\ \bibnamefont {Thygesen}}, \ and\
  \bibinfo {author} {\bibfnamefont {T.}~\bibnamefont {Vegge}},\ }\bibfield
  {title} {\enquote {\bibinfo {title} {$\mathrm{DFT}+\mathrm{U}$ study of
  polaronic conduction in $\mathrm{Li_2O_2}$ and $\mathrm{Li_2CO_3}$:
  Implications for $\mathrm{Li}$-$\mathrm{Air}$ batteries},}\ }\href {\doibase
  10.1021/jp3107809} {\bibfield  {journal} {\bibinfo  {journal} {J. Phys. Chem.
  C}\ }\textbf {\bibinfo {volume} {117}},\ \bibinfo {pages} {5568--5577}
  (\bibinfo {year} {2013})}\BibitemShut {NoStop}%
\bibitem [{\citenamefont {Shishkin}\ and\ \citenamefont
  {Sato}(2016)}]{PhysRevB.93.085135}%
  \BibitemOpen
  \bibfield  {author} {\bibinfo {author} {\bibfnamefont {M.}~\bibnamefont
  {Shishkin}}\ and\ \bibinfo {author} {\bibfnamefont {H.}~\bibnamefont
  {Sato}},\ }\bibfield  {title} {\enquote {\bibinfo {title} {Self-consistent
  parameterization of $\mathrm{DFT}+\mathrm{U}$ framework using linear response
  approach: Application to evaluation of redox potentials of battery
  cathodes},}\ }\href {\doibase 10.1103/PhysRevB.93.085135} {\bibfield
  {journal} {\bibinfo  {journal} {Phys. Rev. B}\ }\textbf {\bibinfo {volume}
  {93}},\ \bibinfo {pages} {085135} (\bibinfo {year} {2016})}\BibitemShut
  {NoStop}%
\bibitem [{\citenamefont {Cole}, \citenamefont {O'Regan},\ and\ \citenamefont
  {Payne}(2012)}]{doi:10.1021/jz3004188}%
  \BibitemOpen
  \bibfield  {author} {\bibinfo {author} {\bibfnamefont {D.~J.}\ \bibnamefont
  {Cole}}, \bibinfo {author} {\bibfnamefont {D.~D.}\ \bibnamefont {O'Regan}}, \
  and\ \bibinfo {author} {\bibfnamefont {M.~C.}\ \bibnamefont {Payne}},\
  }\bibfield  {title} {\enquote {\bibinfo {title} {Ligand discrimination in
  myoglobin from linear-scaling $\mathrm{DFT}+\mathrm{U}$},}\ }\href {\doibase
  10.1021/jz3004188} {\bibfield  {journal} {\bibinfo  {journal} {J. Phys. Chem.
  Lett.}\ }\textbf {\bibinfo {volume} {3}},\ \bibinfo {pages} {1448--1452}
  (\bibinfo {year} {2012})}\BibitemShut {NoStop}%
\bibitem [{\citenamefont {Setvin}\ \emph {et~al.}(2014)\citenamefont {Setvin},
  \citenamefont {Franchini}, \citenamefont {Hao}, \citenamefont {Schmid},
  \citenamefont {Janotti}, \citenamefont {Kaltak}, \citenamefont {Van~de
  Walle}, \citenamefont {Kresse},\ and\ \citenamefont
  {Diebold}}]{PhysRevLett.113.086402}%
  \BibitemOpen
  \bibfield  {author} {\bibinfo {author} {\bibfnamefont {M.}~\bibnamefont
  {Setvin}}, \bibinfo {author} {\bibfnamefont {C.}~\bibnamefont {Franchini}},
  \bibinfo {author} {\bibfnamefont {X.}~\bibnamefont {Hao}}, \bibinfo {author}
  {\bibfnamefont {M.}~\bibnamefont {Schmid}}, \bibinfo {author} {\bibfnamefont
  {A.}~\bibnamefont {Janotti}}, \bibinfo {author} {\bibfnamefont
  {M.}~\bibnamefont {Kaltak}}, \bibinfo {author} {\bibfnamefont {C.~G.}\
  \bibnamefont {Van~de Walle}}, \bibinfo {author} {\bibfnamefont
  {G.}~\bibnamefont {Kresse}}, \ and\ \bibinfo {author} {\bibfnamefont
  {U.}~\bibnamefont {Diebold}},\ }\bibfield  {title} {\enquote {\bibinfo
  {title} {Direct view at excess electrons in $\mathrm{TiO_2}$ rutile and
  anatase},}\ }\href {\doibase 10.1103/PhysRevLett.113.086402} {\bibfield
  {journal} {\bibinfo  {journal} {Phys. Rev. Lett.}\ }\textbf {\bibinfo
  {volume} {113}},\ \bibinfo {pages} {086402} (\bibinfo {year}
  {2014})}\BibitemShut {NoStop}%
\bibitem [{\citenamefont {Himmetoglu}, \citenamefont {Wentzcovitch},\ and\
  \citenamefont {Cococcioni}(2011)}]{PhysRevB.84.115108}%
  \BibitemOpen
  \bibfield  {author} {\bibinfo {author} {\bibfnamefont {B.}~\bibnamefont
  {Himmetoglu}}, \bibinfo {author} {\bibfnamefont {R.~M.}\ \bibnamefont
  {Wentzcovitch}}, \ and\ \bibinfo {author} {\bibfnamefont {M.}~\bibnamefont
  {Cococcioni}},\ }\bibfield  {title} {\enquote {\bibinfo {title}
  {First-principles study of electronic and structural properties of
  $\mathrm{CuO}$},}\ }\href {\doibase 10.1103/PhysRevB.84.115108} {\bibfield
  {journal} {\bibinfo  {journal} {Phys. Rev. B}\ }\textbf {\bibinfo {volume}
  {84}},\ \bibinfo {pages} {115108} (\bibinfo {year} {2011})}\BibitemShut
  {NoStop}%
\bibitem [{\citenamefont {O'Regan}, \citenamefont {Payne},\ and\ \citenamefont
  {Mostofi}(2011)}]{PhysRevB.83.245124}%
  \BibitemOpen
  \bibfield  {author} {\bibinfo {author} {\bibfnamefont {D.~D.}\ \bibnamefont
  {O'Regan}}, \bibinfo {author} {\bibfnamefont {M.~C.}\ \bibnamefont {Payne}},
  \ and\ \bibinfo {author} {\bibfnamefont {A.~A.}\ \bibnamefont {Mostofi}},\
  }\bibfield  {title} {\enquote {\bibinfo {title} {Subspace representations in
  \textit{ab initio} methods for strongly correlated systems},}\ }\href
  {\doibase 10.1103/PhysRevB.83.245124} {\bibfield  {journal} {\bibinfo
  {journal} {Phys. Rev. B}\ }\textbf {\bibinfo {volume} {83}},\ \bibinfo
  {pages} {245124} (\bibinfo {year} {2011})}\BibitemShut {NoStop}%
\bibitem [{\citenamefont {O'Regan}\ \emph {et~al.}(2010)\citenamefont
  {O'Regan}, \citenamefont {Hine}, \citenamefont {Payne},\ and\ \citenamefont
  {Mostofi}}]{PhysRevB.82.081102}%
  \BibitemOpen
  \bibfield  {author} {\bibinfo {author} {\bibfnamefont {D.~D.}\ \bibnamefont
  {O'Regan}}, \bibinfo {author} {\bibfnamefont {N.~D.~M.}\ \bibnamefont
  {Hine}}, \bibinfo {author} {\bibfnamefont {M.~C.}\ \bibnamefont {Payne}}, \
  and\ \bibinfo {author} {\bibfnamefont {A.~A.}\ \bibnamefont {Mostofi}},\
  }\bibfield  {title} {\enquote {\bibinfo {title} {Projector self-consistent
  $\mathrm{DFT}+\mathrm{U}$ using nonorthogonal generalized {W}annier
  functions},}\ }\href {\doibase 10.1103/PhysRevB.82.081102} {\bibfield
  {journal} {\bibinfo  {journal} {Phys. Rev. B}\ }\textbf {\bibinfo {volume}
  {82}},\ \bibinfo {pages} {081102} (\bibinfo {year} {2010})}\BibitemShut
  {NoStop}%
\bibitem [{\citenamefont {Solovyev}\ and\ \citenamefont
  {Dederichs}(1994)}]{PhysRevB.49.6736}%
  \BibitemOpen
  \bibfield  {author} {\bibinfo {author} {\bibfnamefont {I.~V.}\ \bibnamefont
  {Solovyev}}\ and\ \bibinfo {author} {\bibfnamefont {P.~H.}\ \bibnamefont
  {Dederichs}},\ }\bibfield  {title} {\enquote {\bibinfo {title} {\textit{Ab
  initio} calculations of coulomb \textit{U} parameters for transition-metal
  impurities},}\ }\href {\doibase 10.1103/PhysRevB.49.6736} {\bibfield
  {journal} {\bibinfo  {journal} {Phys. Rev. B}\ }\textbf {\bibinfo {volume}
  {49}},\ \bibinfo {pages} {6736--6740} (\bibinfo {year} {1994})}\BibitemShut
  {NoStop}%
\bibitem [{Note1()}]{Note1}%
  \BibitemOpen
  \bibinfo {note} {Calculations were performed using the DFT+$U$
  functionality~\cite {PhysRevB.85.085107} available in the \protect \textsc
  {ONETEP} linear-scaling DFT package~\cite
  {:/content/aip/journal/jcp/122/8/10.1063/1.1839852} with a hard ($0.65$~a$_0$
  cutoff) norm-conserving pseudopotential~\cite {PhysRevB.41.1227}, $10$~a$_0$
  Wannier function cutoff radii, and open boundary conditions~\cite
  {:/content/aip/journal/jcp/110/6/10.1063/1.477923}. DFT+$U$ was applied
  simultaneously to each an atom, using a separate $1s$ orbital subspace
  centred on each, defined using the occupied Kohn-Sham state of the
  pseudopotential for neutral hydrogen. The correct symmetry of $H_2^+$ was
  maintained for all values of $U$ given a symmetric initial guess, i.e., we
  observed no tendency for the charge to localize on a single ion.}\BibitemShut
  {Stop}%
\bibitem [{\citenamefont {Kulik}\ and\ \citenamefont
  {Marzari}(2011{\natexlab{a}})}]{doi:10.1063/1.3660353}%
  \BibitemOpen
  \bibfield  {author} {\bibinfo {author} {\bibfnamefont {H.~J.}\ \bibnamefont
  {Kulik}}\ and\ \bibinfo {author} {\bibfnamefont {N.}~\bibnamefont
  {Marzari}},\ }\bibfield  {title} {\enquote {\bibinfo {title} {{Accurate
  potential energy surfaces with a DFT+U(R) approach}},}\ }\href {\doibase
  10.1063/1.3660353} {\bibfield  {journal} {\bibinfo  {journal} {The Journal of
  Chemical Physics}\ }\textbf {\bibinfo {volume} {135}},\ \bibinfo {pages}
  {194105} (\bibinfo {year} {2011}{\natexlab{a}})}\BibitemShut {NoStop}%
\bibitem [{\citenamefont {Lu}\ and\ \citenamefont
  {Liu}(2014)}]{doi:10.1063/1.4865831}%
  \BibitemOpen
  \bibfield  {author} {\bibinfo {author} {\bibfnamefont {D.}~\bibnamefont
  {Lu}}\ and\ \bibinfo {author} {\bibfnamefont {P.}~\bibnamefont {Liu}},\
  }\bibfield  {title} {\enquote {\bibinfo {title} {Rationalization of the
  hubbard u parameter in ceox from first principles: Unveiling the role of
  local structure in screening},}\ }\href {\doibase 10.1063/1.4865831}
  {\bibfield  {journal} {\bibinfo  {journal} {The Journal of Chemical Physics}\
  }\textbf {\bibinfo {volume} {140}},\ \bibinfo {pages} {084101} (\bibinfo
  {year} {2014})},\ \Eprint
  {http://arxiv.org/abs/http://dx.doi.org/10.1063/1.4865831}
  {http://dx.doi.org/10.1063/1.4865831} \BibitemShut {NoStop}%
\bibitem [{\citenamefont {Aykol}\ and\ \citenamefont
  {Wolverton}(2014)}]{PhysRevB.90.115105}%
  \BibitemOpen
  \bibfield  {author} {\bibinfo {author} {\bibfnamefont {M.}~\bibnamefont
  {Aykol}}\ and\ \bibinfo {author} {\bibfnamefont {C.}~\bibnamefont
  {Wolverton}},\ }\bibfield  {title} {\enquote {\bibinfo {title} {Local
  environment dependent {GGA+U} method for accurate thermochemistry of
  transition metal compounds},}\ }\href {\doibase 10.1103/PhysRevB.90.115105}
  {\bibfield  {journal} {\bibinfo  {journal} {Phys. Rev. B}\ }\textbf {\bibinfo
  {volume} {90}},\ \bibinfo {pages} {115105} (\bibinfo {year}
  {2014})}\BibitemShut {NoStop}%
\bibitem [{\citenamefont {Dabo}\ \emph {et~al.}(2010)\citenamefont {Dabo},
  \citenamefont {Ferretti}, \citenamefont {Poilvert}, \citenamefont {Li},
  \citenamefont {Marzari},\ and\ \citenamefont
  {Cococcioni}}]{PhysRevB.82.115121}%
  \BibitemOpen
  \bibfield  {author} {\bibinfo {author} {\bibfnamefont {I.}~\bibnamefont
  {Dabo}}, \bibinfo {author} {\bibfnamefont {A.}~\bibnamefont {Ferretti}},
  \bibinfo {author} {\bibfnamefont {N.}~\bibnamefont {Poilvert}}, \bibinfo
  {author} {\bibfnamefont {Y.}~\bibnamefont {Li}}, \bibinfo {author}
  {\bibfnamefont {N.}~\bibnamefont {Marzari}}, \ and\ \bibinfo {author}
  {\bibfnamefont {M.}~\bibnamefont {Cococcioni}},\ }\bibfield  {title}
  {\enquote {\bibinfo {title} {Koopmans' condition for density-functional
  theory},}\ }\href {\doibase 10.1103/PhysRevB.82.115121} {\bibfield  {journal}
  {\bibinfo  {journal} {Phys. Rev. B}\ }\textbf {\bibinfo {volume} {82}},\
  \bibinfo {pages} {115121} (\bibinfo {year} {2010})}\BibitemShut {NoStop}%
\bibitem [{\citenamefont {Borghi}\ \emph {et~al.}(2014)\citenamefont {Borghi},
  \citenamefont {Ferretti}, \citenamefont {Nguyen}, \citenamefont {Dabo},\ and\
  \citenamefont {Marzari}}]{PhysRevB.90.075135}%
  \BibitemOpen
  \bibfield  {author} {\bibinfo {author} {\bibfnamefont {G.}~\bibnamefont
  {Borghi}}, \bibinfo {author} {\bibfnamefont {A.}~\bibnamefont {Ferretti}},
  \bibinfo {author} {\bibfnamefont {N.~L.}\ \bibnamefont {Nguyen}}, \bibinfo
  {author} {\bibfnamefont {I.}~\bibnamefont {Dabo}}, \ and\ \bibinfo {author}
  {\bibfnamefont {N.}~\bibnamefont {Marzari}},\ }\bibfield  {title} {\enquote
  {\bibinfo {title} {Koopmans'-compliant functionals and their performance
  against reference molecular data},}\ }\href {\doibase
  10.1103/PhysRevB.90.075135} {\bibfield  {journal} {\bibinfo  {journal} {Phys.
  Rev. B}\ }\textbf {\bibinfo {volume} {90}},\ \bibinfo {pages} {075135}
  (\bibinfo {year} {2014})}\BibitemShut {NoStop}%
\bibitem [{Note2()}]{Note2}%
  \BibitemOpen
  \bibinfo {note} {In all dissociation curves presented, the fully dissociated
  reference energy was fixed to the total-energy of a single exact-functional
  hydrogen atom placed at the midpoint of the dimer, which is equal to the
  occupied Kohn-Sham eigenvalue of the same system, or two half-charged exact
  hydrogen atoms. In this way, only the SIE specific to the PBE dimer is
  analysed, without the SIE present in isolated, half-charged PBE hydrogen
  atoms.}\BibitemShut {Stop}%
\bibitem [{\citenamefont {Moynihan}, \citenamefont {Teobaldi},\ and\
  \citenamefont {O'Regan}(2016)}]{PhysRevB.94.220104}%
  \BibitemOpen
  \bibfield  {author} {\bibinfo {author} {\bibfnamefont {G.}~\bibnamefont
  {Moynihan}}, \bibinfo {author} {\bibfnamefont {G.}~\bibnamefont {Teobaldi}},
  \ and\ \bibinfo {author} {\bibfnamefont {D.~D.}\ \bibnamefont {O'Regan}},\
  }\bibfield  {title} {\enquote {\bibinfo {title} {Inapplicability of exact
  constraints and a minimal two-parameter generalization to the {DFT+U} based
  correction of self-interaction error},}\ }\href {\doibase
  10.1103/PhysRevB.94.220104} {\bibfield  {journal} {\bibinfo  {journal} {Phys.
  Rev. B}\ }\textbf {\bibinfo {volume} {94}},\ \bibinfo {pages} {220104}
  (\bibinfo {year} {2016})}\BibitemShut {NoStop}%
\bibitem [{\citenamefont {Castleton}, \citenamefont {Kullgren},\ and\
  \citenamefont
  {Hermansson}(2007)}]{:/content/aip/journal/jcp/127/24/10.1063/1.2800015}%
  \BibitemOpen
  \bibfield  {author} {\bibinfo {author} {\bibfnamefont {C.~W.~M.}\
  \bibnamefont {Castleton}}, \bibinfo {author} {\bibfnamefont {J.}~\bibnamefont
  {Kullgren}}, \ and\ \bibinfo {author} {\bibfnamefont {K.}~\bibnamefont
  {Hermansson}},\ }\bibfield  {title} {\enquote {\bibinfo {title} {Tuning
  $\mathrm{LDA}+\mathrm{U}$ for electron localization and structure at oxygen
  vacancies in ceria},}\ }\href
  {http://scitation.aip.org/content/aip/journal/jcp/127/24/10.1063/1.2800015}
  {\bibfield  {journal} {\bibinfo  {journal} {J. Chem. Phys.}\ }\textbf
  {\bibinfo {volume} {127}},\ \bibinfo {eid} {244704} (\bibinfo {year}
  {2007})}\BibitemShut {NoStop}%
\bibitem [{\citenamefont {Morgan}\ and\ \citenamefont
  {Watson}(2007)}]{Morgan20075034}%
  \BibitemOpen
  \bibfield  {author} {\bibinfo {author} {\bibfnamefont {B.~J.}\ \bibnamefont
  {Morgan}}\ and\ \bibinfo {author} {\bibfnamefont {G.~W.}\ \bibnamefont
  {Watson}},\ }\bibfield  {title} {\enquote {\bibinfo {title} {A
  $\mathrm{DFT}+\mathrm{U}$ description of oxygen vacancies at the
  $\mathrm{TiO_2}$ rutile (1 1 0) surface},}\ }\href {\doibase
  http://dx.doi.org/10.1016/j.susc.2007.08.025} {\bibfield  {journal} {\bibinfo
   {journal} {Surf. Sci.}\ }\textbf {\bibinfo {volume} {601}},\ \bibinfo
  {pages} {5034 -- 5041} (\bibinfo {year} {2007})}\BibitemShut {NoStop}%
\bibitem [{\citenamefont {Rohrbach}, \citenamefont {Hafner},\ and\
  \citenamefont {Kresse}(2004)}]{PhysRevB.70.125426}%
  \BibitemOpen
  \bibfield  {author} {\bibinfo {author} {\bibfnamefont {A.}~\bibnamefont
  {Rohrbach}}, \bibinfo {author} {\bibfnamefont {J.}~\bibnamefont {Hafner}}, \
  and\ \bibinfo {author} {\bibfnamefont {G.}~\bibnamefont {Kresse}},\
  }\bibfield  {title} {\enquote {\bibinfo {title} {\textit{Ab initio} study of
  the (0001) surfaces of hematite and chromia: {I}nfluence of strong electronic
  correlations},}\ }\href {\doibase 10.1103/PhysRevB.70.125426} {\bibfield
  {journal} {\bibinfo  {journal} {Phys. Rev. B}\ }\textbf {\bibinfo {volume}
  {70}},\ \bibinfo {pages} {125426} (\bibinfo {year} {2004})}\BibitemShut
  {NoStop}%
\bibitem [{\citenamefont {Loschen}\ \emph {et~al.}(2007)\citenamefont
  {Loschen}, \citenamefont {Carrasco}, \citenamefont {Neyman},\ and\
  \citenamefont {Illas}}]{PhysRevB.75.035115}%
  \BibitemOpen
  \bibfield  {author} {\bibinfo {author} {\bibfnamefont {C.}~\bibnamefont
  {Loschen}}, \bibinfo {author} {\bibfnamefont {J.}~\bibnamefont {Carrasco}},
  \bibinfo {author} {\bibfnamefont {K.~M.}\ \bibnamefont {Neyman}}, \ and\
  \bibinfo {author} {\bibfnamefont {F.}~\bibnamefont {Illas}},\ }\bibfield
  {title} {\enquote {\bibinfo {title} {First-principles {LDA+U} and {GGA+U}
  study of cerium oxides: {D}ependence on the effective {U} parameter},}\
  }\href {\doibase 10.1103/PhysRevB.75.035115} {\bibfield  {journal} {\bibinfo
  {journal} {Phys. Rev. B}\ }\textbf {\bibinfo {volume} {75}},\ \bibinfo
  {pages} {035115} (\bibinfo {year} {2007})}\BibitemShut {NoStop}%
\bibitem [{\citenamefont {Wang}, \citenamefont {Maxisch},\ and\ \citenamefont
  {Ceder}(2006)}]{PhysRevB.73.195107}%
  \BibitemOpen
  \bibfield  {author} {\bibinfo {author} {\bibfnamefont {L.}~\bibnamefont
  {Wang}}, \bibinfo {author} {\bibfnamefont {T.}~\bibnamefont {Maxisch}}, \
  and\ \bibinfo {author} {\bibfnamefont {G.}~\bibnamefont {Ceder}},\ }\bibfield
   {title} {\enquote {\bibinfo {title} {Oxidation energies of transition metal
  oxides within the $\mathrm{GGA}+\mathrm{U}$ framework},}\ }\href {\doibase
  10.1103/PhysRevB.73.195107} {\bibfield  {journal} {\bibinfo  {journal} {Phys.
  Rev. B}\ }\textbf {\bibinfo {volume} {73}},\ \bibinfo {pages} {195107}
  (\bibinfo {year} {2006})}\BibitemShut {NoStop}%
\bibitem [{\citenamefont {Ong}\ \emph {et~al.}(2011)\citenamefont {Ong},
  \citenamefont {Chevrier}, \citenamefont {Hautier}, \citenamefont {Jain},
  \citenamefont {Moore}, \citenamefont {Kim}, \citenamefont {Ma},\ and\
  \citenamefont {Ceder}}]{C1EE01782A}%
  \BibitemOpen
  \bibfield  {author} {\bibinfo {author} {\bibfnamefont {S.~P.}\ \bibnamefont
  {Ong}}, \bibinfo {author} {\bibfnamefont {V.~L.}\ \bibnamefont {Chevrier}},
  \bibinfo {author} {\bibfnamefont {G.}~\bibnamefont {Hautier}}, \bibinfo
  {author} {\bibfnamefont {A.}~\bibnamefont {Jain}}, \bibinfo {author}
  {\bibfnamefont {C.}~\bibnamefont {Moore}}, \bibinfo {author} {\bibfnamefont
  {S.}~\bibnamefont {Kim}}, \bibinfo {author} {\bibfnamefont {X.}~\bibnamefont
  {Ma}}, \ and\ \bibinfo {author} {\bibfnamefont {G.}~\bibnamefont {Ceder}},\
  }\bibfield  {title} {\enquote {\bibinfo {title} {Voltage{,} stability and
  diffusion barrier differences between sodium-ion and lithium-ion
  intercalation materials},}\ }\href {\doibase 10.1039/C1EE01782A} {\bibfield
  {journal} {\bibinfo  {journal} {Energy Environ. Sci.}\ }\textbf {\bibinfo
  {volume} {4}},\ \bibinfo {pages} {3680--3688} (\bibinfo {year}
  {2011})}\BibitemShut {NoStop}%
\bibitem [{\citenamefont {Zhou}\ \emph
  {et~al.}(2004{\natexlab{a}})\citenamefont {Zhou}, \citenamefont {Cococcioni},
  \citenamefont {Marianetti}, \citenamefont {Morgan},\ and\ \citenamefont
  {Ceder}}]{PhysRevB.70.235121}%
  \BibitemOpen
  \bibfield  {author} {\bibinfo {author} {\bibfnamefont {F.}~\bibnamefont
  {Zhou}}, \bibinfo {author} {\bibfnamefont {M.}~\bibnamefont {Cococcioni}},
  \bibinfo {author} {\bibfnamefont {C.~A.}\ \bibnamefont {Marianetti}},
  \bibinfo {author} {\bibfnamefont {D.}~\bibnamefont {Morgan}}, \ and\ \bibinfo
  {author} {\bibfnamefont {G.}~\bibnamefont {Ceder}},\ }\bibfield  {title}
  {\enquote {\bibinfo {title} {First-principles prediction of redox potentials
  in transition-metal compounds with {LDA+U}},}\ }\href {\doibase
  10.1103/PhysRevB.70.235121} {\bibfield  {journal} {\bibinfo  {journal} {Phys.
  Rev. B}\ }\textbf {\bibinfo {volume} {70}},\ \bibinfo {pages} {235121}
  (\bibinfo {year} {2004}{\natexlab{a}})}\BibitemShut {NoStop}%
\bibitem [{\citenamefont {Zhou}\ \emph
  {et~al.}(2004{\natexlab{b}})\citenamefont {Zhou}, \citenamefont {Cococcioni},
  \citenamefont {Kang},\ and\ \citenamefont {Ceder}}]{Zhou20041144}%
  \BibitemOpen
  \bibfield  {author} {\bibinfo {author} {\bibfnamefont {F.}~\bibnamefont
  {Zhou}}, \bibinfo {author} {\bibfnamefont {M.}~\bibnamefont {Cococcioni}},
  \bibinfo {author} {\bibfnamefont {K.}~\bibnamefont {Kang}}, \ and\ \bibinfo
  {author} {\bibfnamefont {G.}~\bibnamefont {Ceder}},\ }\bibfield  {title}
  {\enquote {\bibinfo {title} {The {L}i intercalation potential of
  $\mathrm{LiMPO_4}$ and $\mathrm{LiMSiO_4}$ olivines with {M} = {Fe, Mn, Co,
  Ni}},}\ }\href {\doibase http://dx.doi.org/10.1016/j.elecom.2004.09.007}
  {\bibfield  {journal} {\bibinfo  {journal} {Electrochem. Commun.}\ }\textbf
  {\bibinfo {volume} {6}},\ \bibinfo {pages} {1144 -- 1148} (\bibinfo {year}
  {2004}{\natexlab{b}})}\BibitemShut {NoStop}%
\bibitem [{\citenamefont {Zhou}\ \emph
  {et~al.}(2004{\natexlab{c}})\citenamefont {Zhou}, \citenamefont {Kang},
  \citenamefont {Maxisch}, \citenamefont {Ceder},\ and\ \citenamefont
  {Morgan}}]{Zhou2004181}%
  \BibitemOpen
  \bibfield  {author} {\bibinfo {author} {\bibfnamefont {F.}~\bibnamefont
  {Zhou}}, \bibinfo {author} {\bibfnamefont {K.}~\bibnamefont {Kang}}, \bibinfo
  {author} {\bibfnamefont {T.}~\bibnamefont {Maxisch}}, \bibinfo {author}
  {\bibfnamefont {G.}~\bibnamefont {Ceder}}, \ and\ \bibinfo {author}
  {\bibfnamefont {D.}~\bibnamefont {Morgan}},\ }\bibfield  {title} {\enquote
  {\bibinfo {title} {The electronic structure and band gap of
  $\mathrm{LiFePO_4}$ and $\mathrm{LiMnPO_4}$},}\ }\href {\doibase
  http://dx.doi.org/10.1016/j.ssc.2004.07.055} {\bibfield  {journal} {\bibinfo
  {journal} {Solid State Commun.}\ }\textbf {\bibinfo {volume} {132}},\
  \bibinfo {pages} {181 -- 186} (\bibinfo {year}
  {2004}{\natexlab{c}})}\BibitemShut {NoStop}%
\bibitem [{\citenamefont {Ganduglia-Pirovano}, \citenamefont {Da~Silva},\ and\
  \citenamefont {Sauer}(2009)}]{PhysRevLett.102.026101}%
  \BibitemOpen
  \bibfield  {author} {\bibinfo {author} {\bibfnamefont {M.~V.}\ \bibnamefont
  {Ganduglia-Pirovano}}, \bibinfo {author} {\bibfnamefont {J.~L.~F.}\
  \bibnamefont {Da~Silva}}, \ and\ \bibinfo {author} {\bibfnamefont
  {J.}~\bibnamefont {Sauer}},\ }\bibfield  {title} {\enquote {\bibinfo {title}
  {Density-functional calculations of the structure of near-surface oxygen
  vacancies and electron localization on $\mathrm{CeO_2}(111)$},}\ }\href
  {\doibase 10.1103/PhysRevLett.102.026101} {\bibfield  {journal} {\bibinfo
  {journal} {Phys. Rev. Lett.}\ }\textbf {\bibinfo {volume} {102}},\ \bibinfo
  {pages} {026101} (\bibinfo {year} {2009})}\BibitemShut {NoStop}%
\bibitem [{\citenamefont {Da~Silva}\ \emph {et~al.}(2007)\citenamefont
  {Da~Silva}, \citenamefont {Ganduglia-Pirovano}, \citenamefont {Sauer},
  \citenamefont {Bayer},\ and\ \citenamefont {Kresse}}]{PhysRevB.75.045121}%
  \BibitemOpen
  \bibfield  {author} {\bibinfo {author} {\bibfnamefont {J.~L.~F.}\
  \bibnamefont {Da~Silva}}, \bibinfo {author} {\bibfnamefont {M.~V.}\
  \bibnamefont {Ganduglia-Pirovano}}, \bibinfo {author} {\bibfnamefont
  {J.}~\bibnamefont {Sauer}}, \bibinfo {author} {\bibfnamefont
  {V.}~\bibnamefont {Bayer}}, \ and\ \bibinfo {author} {\bibfnamefont
  {G.}~\bibnamefont {Kresse}},\ }\bibfield  {title} {\enquote {\bibinfo {title}
  {Hybrid functionals applied to rare-earth oxides: The example of ceria},}\
  }\href {\doibase 10.1103/PhysRevB.75.045121} {\bibfield  {journal} {\bibinfo
  {journal} {Phys. Rev. B}\ }\textbf {\bibinfo {volume} {75}},\ \bibinfo
  {pages} {045121} (\bibinfo {year} {2007})}\BibitemShut {NoStop}%
\bibitem [{\citenamefont {He}\ and\ \citenamefont
  {Millis}(2015)}]{PhysRevB.91.195138}%
  \BibitemOpen
  \bibfield  {author} {\bibinfo {author} {\bibfnamefont {Z.}~\bibnamefont
  {He}}\ and\ \bibinfo {author} {\bibfnamefont {A.~J.}\ \bibnamefont
  {Millis}},\ }\bibfield  {title} {\enquote {\bibinfo {title} {Strain control
  of electronic phase in rare-earth nickelates},}\ }\href {\doibase
  10.1103/PhysRevB.91.195138} {\bibfield  {journal} {\bibinfo  {journal} {Phys.
  Rev. B}\ }\textbf {\bibinfo {volume} {91}},\ \bibinfo {pages} {195138}
  (\bibinfo {year} {2015})}\BibitemShut {NoStop}%
\bibitem [{\citenamefont {Bjaalie}\ \emph {et~al.}(2015)\citenamefont
  {Bjaalie}, \citenamefont {Verma}, \citenamefont {Himmetoglu}, \citenamefont
  {Janotti}, \citenamefont {Raghavan}, \citenamefont {Protasenko},
  \citenamefont {Steenbergen}, \citenamefont {Jena}, \citenamefont {Stemmer},\
  and\ \citenamefont {Van~de Walle}}]{PhysRevB.92.085111}%
  \BibitemOpen
  \bibfield  {author} {\bibinfo {author} {\bibfnamefont {L.}~\bibnamefont
  {Bjaalie}}, \bibinfo {author} {\bibfnamefont {A.}~\bibnamefont {Verma}},
  \bibinfo {author} {\bibfnamefont {B.}~\bibnamefont {Himmetoglu}}, \bibinfo
  {author} {\bibfnamefont {A.}~\bibnamefont {Janotti}}, \bibinfo {author}
  {\bibfnamefont {S.}~\bibnamefont {Raghavan}}, \bibinfo {author}
  {\bibfnamefont {V.}~\bibnamefont {Protasenko}}, \bibinfo {author}
  {\bibfnamefont {E.~H.}\ \bibnamefont {Steenbergen}}, \bibinfo {author}
  {\bibfnamefont {D.}~\bibnamefont {Jena}}, \bibinfo {author} {\bibfnamefont
  {S.}~\bibnamefont {Stemmer}}, \ and\ \bibinfo {author} {\bibfnamefont
  {C.~G.}\ \bibnamefont {Van~de Walle}},\ }\bibfield  {title} {\enquote
  {\bibinfo {title} {Determination of the {M}ott-{H}ubbard gap in
  $\mathrm{GdTiO_3}$},}\ }\href {\doibase 10.1103/PhysRevB.92.085111}
  {\bibfield  {journal} {\bibinfo  {journal} {Phys. Rev. B}\ }\textbf {\bibinfo
  {volume} {92}},\ \bibinfo {pages} {085111} (\bibinfo {year}
  {2015})}\BibitemShut {NoStop}%
\bibitem [{\citenamefont {Isaacs}\ and\ \citenamefont
  {Marianetti}(2016)}]{PhysRevB.94.035120}%
  \BibitemOpen
  \bibfield  {author} {\bibinfo {author} {\bibfnamefont {E.~B.}\ \bibnamefont
  {Isaacs}}\ and\ \bibinfo {author} {\bibfnamefont {C.~A.}\ \bibnamefont
  {Marianetti}},\ }\bibfield  {title} {\enquote {\bibinfo {title} {Electronic
  correlations in monolayer $\mathrm{VS_2}$},}\ }\href {\doibase
  10.1103/PhysRevB.94.035120} {\bibfield  {journal} {\bibinfo  {journal} {Phys.
  Rev. B}\ }\textbf {\bibinfo {volume} {94}},\ \bibinfo {pages} {035120}
  (\bibinfo {year} {2016})}\BibitemShut {NoStop}%
\bibitem [{\citenamefont {Curtarolo}\ \emph
  {et~al.}(2013{\natexlab{a}})\citenamefont {Curtarolo}, \citenamefont {Hart},
  \citenamefont {Nardelli}, \citenamefont {Mingo}, \citenamefont {Sanvito},\
  and\ \citenamefont {Levy}}]{nmat3568}%
  \BibitemOpen
  \bibfield  {author} {\bibinfo {author} {\bibfnamefont {S.}~\bibnamefont
  {Curtarolo}}, \bibinfo {author} {\bibfnamefont {G.~L.~W.}\ \bibnamefont
  {Hart}}, \bibinfo {author} {\bibfnamefont {M.~B.}\ \bibnamefont {Nardelli}},
  \bibinfo {author} {\bibfnamefont {N.}~\bibnamefont {Mingo}}, \bibinfo
  {author} {\bibfnamefont {S.}~\bibnamefont {Sanvito}}, \ and\ \bibinfo
  {author} {\bibfnamefont {O.}~\bibnamefont {Levy}},\ }\bibfield  {title}
  {\enquote {\bibinfo {title} {The high-throughput highway to computational
  materials design},}\ }\href {http://dx.doi.org/10.1038/nmat3568} {\bibfield
  {journal} {\bibinfo  {journal} {Nat. Mater.}\ }\textbf {\bibinfo {volume}
  {12}},\ \bibinfo {pages} {191--201} (\bibinfo {year}
  {2013}{\natexlab{a}})}\BibitemShut {NoStop}%
\bibitem [{\citenamefont {Scherlis}\ \emph {et~al.}(2007)\citenamefont
  {Scherlis}, \citenamefont {Cococcioni}, \citenamefont {Sit},\ and\
  \citenamefont {Marzari}}]{doi:10.1021/jp070549l}%
  \BibitemOpen
  \bibfield  {author} {\bibinfo {author} {\bibfnamefont {D.~A.}\ \bibnamefont
  {Scherlis}}, \bibinfo {author} {\bibfnamefont {M.}~\bibnamefont
  {Cococcioni}}, \bibinfo {author} {\bibfnamefont {P.}~\bibnamefont {Sit}}, \
  and\ \bibinfo {author} {\bibfnamefont {N.}~\bibnamefont {Marzari}},\
  }\bibfield  {title} {\enquote {\bibinfo {title} {Simulation of heme using
  $\mathrm{DFT}+\mathrm{U}$: {A} step toward accurate spin-state energetics},}\
  }\href {\doibase 10.1021/jp070549l} {\bibfield  {journal} {\bibinfo
  {journal} {J. Phys. Chem. B}\ }\textbf {\bibinfo {volume} {111}},\ \bibinfo
  {pages} {7384--7391} (\bibinfo {year} {2007})}\BibitemShut {NoStop}%
\bibitem [{\citenamefont {Capdevila-Cortada}, \citenamefont {\L{}odziana},\
  and\ \citenamefont {L\'opez}(2016)}]{doi:10.1021/acscatal.6b01907}%
  \BibitemOpen
  \bibfield  {author} {\bibinfo {author} {\bibfnamefont {M.}~\bibnamefont
  {Capdevila-Cortada}}, \bibinfo {author} {\bibfnamefont {Z.}~\bibnamefont
  {\L{}odziana}}, \ and\ \bibinfo {author} {\bibfnamefont {N.}~\bibnamefont
  {L\'opez}},\ }\bibfield  {title} {\enquote {\bibinfo {title} {Performance of
  {DFT+U} approaches in the study of catalytic materials},}\ }\href {\doibase
  10.1021/acscatal.6b01907} {\bibfield  {journal} {\bibinfo  {journal} {ACS
  Catalysis}\ }\textbf {\bibinfo {volume} {6}},\ \bibinfo {pages} {8370--8379}
  (\bibinfo {year} {2016})}\BibitemShut {NoStop}%
\bibitem [{\citenamefont {Nawa}\ \emph {et~al.}(2016)\citenamefont {Nawa},
  \citenamefont {Kitaoka}, \citenamefont {Nakamura}, \citenamefont {Imamura},
  \citenamefont {Akiyama}, \citenamefont {Ito},\ and\ \citenamefont
  {Weinert}}]{PhysRevB.94.035136}%
  \BibitemOpen
  \bibfield  {author} {\bibinfo {author} {\bibfnamefont {K.}~\bibnamefont
  {Nawa}}, \bibinfo {author} {\bibfnamefont {Y.}~\bibnamefont {Kitaoka}},
  \bibinfo {author} {\bibfnamefont {K.}~\bibnamefont {Nakamura}}, \bibinfo
  {author} {\bibfnamefont {H.}~\bibnamefont {Imamura}}, \bibinfo {author}
  {\bibfnamefont {T.}~\bibnamefont {Akiyama}}, \bibinfo {author} {\bibfnamefont
  {T.}~\bibnamefont {Ito}}, \ and\ \bibinfo {author} {\bibfnamefont
  {M.}~\bibnamefont {Weinert}},\ }\bibfield  {title} {\enquote {\bibinfo
  {title} {Search for the ground-state electronic configurations of correlated
  organometallic metallocenes from constraint density-functional theory},}\
  }\href {\doibase 10.1103/PhysRevB.94.035136} {\bibfield  {journal} {\bibinfo
  {journal} {Phys. Rev. B}\ }\textbf {\bibinfo {volume} {94}},\ \bibinfo
  {pages} {035136} (\bibinfo {year} {2016})}\BibitemShut {NoStop}%
\bibitem [{\citenamefont {Aryasetiawan}\ \emph {et~al.}(2006)\citenamefont
  {Aryasetiawan}, \citenamefont {Karlsson}, \citenamefont {Jepsen},\ and\
  \citenamefont {Sch\"onberger}}]{PhysRevB.74.125106}%
  \BibitemOpen
  \bibfield  {author} {\bibinfo {author} {\bibfnamefont {F.}~\bibnamefont
  {Aryasetiawan}}, \bibinfo {author} {\bibfnamefont {K.}~\bibnamefont
  {Karlsson}}, \bibinfo {author} {\bibfnamefont {O.}~\bibnamefont {Jepsen}}, \
  and\ \bibinfo {author} {\bibfnamefont {U.}~\bibnamefont {Sch\"onberger}},\
  }\bibfield  {title} {\enquote {\bibinfo {title} {Calculations of {H}ubbard
  {U} from first-principles},}\ }\href {\doibase 10.1103/PhysRevB.74.125106}
  {\bibfield  {journal} {\bibinfo  {journal} {Phys. Rev. B}\ }\textbf {\bibinfo
  {volume} {74}},\ \bibinfo {pages} {125106} (\bibinfo {year}
  {2006})}\BibitemShut {NoStop}%
\bibitem [{\citenamefont {Karlsson}, \citenamefont {Aryasetiawan},\ and\
  \citenamefont {Jepsen}(2010)}]{PhysRevB.81.245113}%
  \BibitemOpen
  \bibfield  {author} {\bibinfo {author} {\bibfnamefont {K.}~\bibnamefont
  {Karlsson}}, \bibinfo {author} {\bibfnamefont {F.}~\bibnamefont
  {Aryasetiawan}}, \ and\ \bibinfo {author} {\bibfnamefont {O.}~\bibnamefont
  {Jepsen}},\ }\bibfield  {title} {\enquote {\bibinfo {title} {Method for
  calculating the electronic structure of correlated materials from a truly
  first-principles $\mathrm{LDA}+\mathrm{U}$ scheme},}\ }\href {\doibase
  10.1103/PhysRevB.81.245113} {\bibfield  {journal} {\bibinfo  {journal} {Phys.
  Rev. B}\ }\textbf {\bibinfo {volume} {81}},\ \bibinfo {pages} {245113}
  (\bibinfo {year} {2010})}\BibitemShut {NoStop}%
\bibitem [{\citenamefont {Cococcioni}(2012)}]{cococcioni4}%
  \BibitemOpen
  \bibfield  {author} {\bibinfo {author} {\bibfnamefont {M.}~\bibnamefont
  {Cococcioni}},\ }\bibfield  {title} {\enquote {\bibinfo {title} {The {LDA+U}
  approach: {A} simple {H}ubbard correction for correlated ground states},}\
  }\href {http://www.cond-mat.de/events/correl12/manuscripts/cococcioni.pdf}
  {\bibfield  {journal} {\bibinfo  {journal} {Lecture Notes}\ } (\bibinfo
  {year} {2012})}\BibitemShut {NoStop}%
\bibitem [{\citenamefont {Jr}\ and\ \citenamefont
  {Cococcioni}(2010)}]{campo2010extended}%
  \BibitemOpen
  \bibfield  {author} {\bibinfo {author} {\bibfnamefont {V.~L.~C.}\
  \bibnamefont {Jr}}\ and\ \bibinfo {author} {\bibfnamefont {M.}~\bibnamefont
  {Cococcioni}},\ }\bibfield  {title} {\enquote {\bibinfo {title} {{Extended
  DFT + U + V method with on-site and inter-site electronic interactions}},}\
  }\href {http://stacks.iop.org/0953-8984/22/i=5/a=055602} {\bibfield
  {journal} {\bibinfo  {journal} {Journal of Physics: Condensed Matter}\
  }\textbf {\bibinfo {volume} {22}},\ \bibinfo {pages} {055602} (\bibinfo
  {year} {2010})}\BibitemShut {NoStop}%
\bibitem [{Note3()}]{Note3}%
  \BibitemOpen
  \bibinfo {note} {In this work, since we find it necessary to use only
  single-site DFT+$U$ with no +$V$ term, we treat the two $1s$ atomic subspaces
  as decoupled, each comprising the majority of the screening bath for the
  other. Hence we use a pair of scalar Dyson equations (identical by symmetry,
  i.e., only one is treated numerically) to calculate the Hubbard $U$, rather
  than selecting the diagonal of the $2 \times 2$ site-indexed Hubbard
  $U$.}\BibitemShut {Stop}%
\bibitem [{\citenamefont {Skylaris}\ \emph {et~al.}(2005)\citenamefont
  {Skylaris}, \citenamefont {Haynes}, \citenamefont {Mostofi},\ and\
  \citenamefont {Payne}}]{:/content/aip/journal/jcp/122/8/10.1063/1.1839852}%
  \BibitemOpen
  \bibfield  {author} {\bibinfo {author} {\bibfnamefont {C.-K.}\ \bibnamefont
  {Skylaris}}, \bibinfo {author} {\bibfnamefont {P.~D.}\ \bibnamefont
  {Haynes}}, \bibinfo {author} {\bibfnamefont {A.~A.}\ \bibnamefont {Mostofi}},
  \ and\ \bibinfo {author} {\bibfnamefont {M.~C.}\ \bibnamefont {Payne}},\
  }\bibfield  {title} {\enquote {\bibinfo {title} {Introducing {ONETEP}:
  Linear-scaling density-functional simulations on parallel computers},}\
  }\href
  {http://scitation.aip.org/content/aip/journal/jcp/122/8/10.1063/1.1839852}
  {\bibfield  {journal} {\bibinfo  {journal} {J. Chem. Phys.}\ }\textbf
  {\bibinfo {volume} {122}},\ \bibinfo {eid} {084119} (\bibinfo {year}
  {2005})}\BibitemShut {NoStop}%
\bibitem [{\citenamefont {Hine}\ \emph {et~al.}(2009)\citenamefont {Hine},
  \citenamefont {Haynes}, \citenamefont {Mostofi}, \citenamefont {Skylaris},\
  and\ \citenamefont {Payne}}]{hine2009linear}%
  \BibitemOpen
  \bibfield  {author} {\bibinfo {author} {\bibfnamefont {N.~D.}\ \bibnamefont
  {Hine}}, \bibinfo {author} {\bibfnamefont {P.~D.}\ \bibnamefont {Haynes}},
  \bibinfo {author} {\bibfnamefont {A.~A.}\ \bibnamefont {Mostofi}}, \bibinfo
  {author} {\bibfnamefont {C.-K.}\ \bibnamefont {Skylaris}}, \ and\ \bibinfo
  {author} {\bibfnamefont {M.~C.}\ \bibnamefont {Payne}},\ }\bibfield  {title}
  {\enquote {\bibinfo {title} {Linear-scaling density-functional theory with
  tens of thousands of atoms: {E}xpanding the scope and scale of calculations
  with {ONETEP}},}\ }\href
  {http://www.sciencedirect.com/science/article/pii/S0010465508004414}
  {\bibfield  {journal} {\bibinfo  {journal} {Comput. Phys. Commun.}\ }\textbf
  {\bibinfo {volume} {180}},\ \bibinfo {pages} {1041--1053} (\bibinfo {year}
  {2009})}\BibitemShut {NoStop}%
\bibitem [{\citenamefont {Hine}\ \emph {et~al.}(2010)\citenamefont {Hine},
  \citenamefont {Haynes}, \citenamefont {Mostofi},\ and\ \citenamefont
  {Payne}}]{hine2010linear}%
  \BibitemOpen
  \bibfield  {author} {\bibinfo {author} {\bibfnamefont {N.}~\bibnamefont
  {Hine}}, \bibinfo {author} {\bibfnamefont {P.}~\bibnamefont {Haynes}},
  \bibinfo {author} {\bibfnamefont {A.}~\bibnamefont {Mostofi}}, \ and\
  \bibinfo {author} {\bibfnamefont {M.}~\bibnamefont {Payne}},\ }\bibfield
  {title} {\enquote {\bibinfo {title} {Linear-scaling density-functional
  simulations of charged point defects in $\mathrm{Al2O3}$ using hierarchical
  sparse matrix algebra},}\ }\href
  {http://scitation.aip.org/content/aip/journal/jcp/133/11/10.1063/1.3492379}
  {\bibfield  {journal} {\bibinfo  {journal} {J. Chem. Phys.}\ }\textbf
  {\bibinfo {volume} {133}},\ \bibinfo {pages} {114111} (\bibinfo {year}
  {2010})}\BibitemShut {NoStop}%
\bibitem [{\citenamefont {Gillan}\ \emph {et~al.}(2007)\citenamefont {Gillan},
  \citenamefont {Bowler}, \citenamefont {Torralba},\ and\ \citenamefont
  {Miyazaki}}]{Gillan200714}%
  \BibitemOpen
  \bibfield  {author} {\bibinfo {author} {\bibfnamefont {M.}~\bibnamefont
  {Gillan}}, \bibinfo {author} {\bibfnamefont {D.}~\bibnamefont {Bowler}},
  \bibinfo {author} {\bibfnamefont {A.}~\bibnamefont {Torralba}}, \ and\
  \bibinfo {author} {\bibfnamefont {T.}~\bibnamefont {Miyazaki}},\ }\bibfield
  {title} {\enquote {\bibinfo {title} {Order-n first-principles calculations
  with the conquest code},}\ }\href {\doibase
  http://dx.doi.org/10.1016/j.cpc.2007.02.075} {\bibfield  {journal} {\bibinfo
  {journal} {Computer Physics Communications}\ }\textbf {\bibinfo {volume}
  {177}},\ \bibinfo {pages} {14 -- 18} (\bibinfo {year} {2007})}\BibitemShut
  {NoStop}%
\bibitem [{\citenamefont {Bowler}, \citenamefont {Miyazaki},\ and\
  \citenamefont {Gillan}(2002)}]{0953-8984-14-11-303}%
  \BibitemOpen
  \bibfield  {author} {\bibinfo {author} {\bibfnamefont {D.~R.}\ \bibnamefont
  {Bowler}}, \bibinfo {author} {\bibfnamefont {T.}~\bibnamefont {Miyazaki}}, \
  and\ \bibinfo {author} {\bibfnamefont {M.~J.}\ \bibnamefont {Gillan}},\
  }\bibfield  {title} {\enquote {\bibinfo {title} {Recent progress in linear
  scaling ab initio electronic structure techniques},}\ }\href
  {http://stacks.iop.org/0953-8984/14/i=11/a=303} {\bibfield  {journal}
  {\bibinfo  {journal} {Journal of Physics: Condensed Matter}\ }\textbf
  {\bibinfo {volume} {14}},\ \bibinfo {pages} {2781} (\bibinfo {year}
  {2002})}\BibitemShut {NoStop}%
\bibitem [{\citenamefont {Soler}\ \emph {et~al.}(2002)\citenamefont {Soler},
  \citenamefont {Artacho}, \citenamefont {Gale}, \citenamefont {Garc\'ia},
  \citenamefont {Junquera}, \citenamefont {Ordej\'on},\ and\ \citenamefont
  {S\'anchez-Portal}}]{0953-8984-14-11-302}%
  \BibitemOpen
  \bibfield  {author} {\bibinfo {author} {\bibfnamefont {J.~M.}\ \bibnamefont
  {Soler}}, \bibinfo {author} {\bibfnamefont {E.}~\bibnamefont {Artacho}},
  \bibinfo {author} {\bibfnamefont {J.~D.}\ \bibnamefont {Gale}}, \bibinfo
  {author} {\bibfnamefont {A.}~\bibnamefont {Garc\'ia}}, \bibinfo {author}
  {\bibfnamefont {J.}~\bibnamefont {Junquera}}, \bibinfo {author}
  {\bibfnamefont {P.}~\bibnamefont {Ordej\'on}}, \ and\ \bibinfo {author}
  {\bibfnamefont {D.}~\bibnamefont {S\'anchez-Portal}},\ }\bibfield  {title}
  {\enquote {\bibinfo {title} {The {SIESTA} method for ab initio order-{N}
  materials simulation},}\ }\href
  {http://stacks.iop.org/0953-8984/14/i=11/a=302} {\bibfield  {journal}
  {\bibinfo  {journal} {J. Phys.: Condens. Matter}\ }\textbf {\bibinfo {volume}
  {14}},\ \bibinfo {pages} {2745} (\bibinfo {year} {2002})}\BibitemShut
  {NoStop}%
\bibitem [{\citenamefont {Artacho}\ \emph {et~al.}(2008)\citenamefont
  {Artacho}, \citenamefont {Anglada}, \citenamefont {Di\'eguez}, \citenamefont
  {Gale}, \citenamefont {Garc\'ia}, \citenamefont {Junquera}, \citenamefont
  {Martin}, \citenamefont {Ordej\'on}, \citenamefont {Pruneda}, \citenamefont
  {S\'anchez-Portal},\ and\ \citenamefont {Soler}}]{0953-8984-20-6-064208}%
  \BibitemOpen
  \bibfield  {author} {\bibinfo {author} {\bibfnamefont {E.}~\bibnamefont
  {Artacho}}, \bibinfo {author} {\bibfnamefont {E.}~\bibnamefont {Anglada}},
  \bibinfo {author} {\bibfnamefont {O.}~\bibnamefont {Di\'eguez}}, \bibinfo
  {author} {\bibfnamefont {J.~D.}\ \bibnamefont {Gale}}, \bibinfo {author}
  {\bibfnamefont {A.}~\bibnamefont {Garc\'ia}}, \bibinfo {author}
  {\bibfnamefont {J.}~\bibnamefont {Junquera}}, \bibinfo {author}
  {\bibfnamefont {R.~M.}\ \bibnamefont {Martin}}, \bibinfo {author}
  {\bibfnamefont {P.}~\bibnamefont {Ordej\'on}}, \bibinfo {author}
  {\bibfnamefont {J.~M.}\ \bibnamefont {Pruneda}}, \bibinfo {author}
  {\bibfnamefont {D.}~\bibnamefont {S\'anchez-Portal}}, \ and\ \bibinfo
  {author} {\bibfnamefont {J.~M.}\ \bibnamefont {Soler}},\ }\bibfield  {title}
  {\enquote {\bibinfo {title} {The {SIESTA} method; developments and
  applicability},}\ }\href {http://stacks.iop.org/0953-8984/20/i=6/a=064208}
  {\bibfield  {journal} {\bibinfo  {journal} {J. Phys.: Condens. Matter}\
  }\textbf {\bibinfo {volume} {20}},\ \bibinfo {pages} {064208} (\bibinfo
  {year} {2008})}\BibitemShut {NoStop}%
\bibitem [{\citenamefont {Genovese}\ \emph {et~al.}(2011)\citenamefont
  {Genovese}, \citenamefont {Videau}, \citenamefont {Ospici}, \citenamefont
  {Deutsch}, \citenamefont {Goedecker},\ and\ \citenamefont
  {M\'ehaut}}]{Genovese2011149}%
  \BibitemOpen
  \bibfield  {author} {\bibinfo {author} {\bibfnamefont {L.}~\bibnamefont
  {Genovese}}, \bibinfo {author} {\bibfnamefont {B.}~\bibnamefont {Videau}},
  \bibinfo {author} {\bibfnamefont {M.}~\bibnamefont {Ospici}}, \bibinfo
  {author} {\bibfnamefont {T.}~\bibnamefont {Deutsch}}, \bibinfo {author}
  {\bibfnamefont {S.}~\bibnamefont {Goedecker}}, \ and\ \bibinfo {author}
  {\bibfnamefont {J.-F.}\ \bibnamefont {M\'ehaut}},\ }\bibfield  {title}
  {\enquote {\bibinfo {title} {Daubechies wavelets for high performance
  electronic structure calculations: The {B}ig{DFT} project},}\ }\href
  {\doibase http://dx.doi.org/10.1016/j.crme.2010.12.003} {\bibfield  {journal}
  {\bibinfo  {journal} {CR Mecanique}\ }\textbf {\bibinfo {volume} {339}},\
  \bibinfo {pages} {149 -- 164} (\bibinfo {year} {2011})}\BibitemShut {NoStop}%
\bibitem [{\citenamefont {Ozaki}\ and\ \citenamefont
  {Kino}(2005)}]{PhysRevB.72.045121}%
  \BibitemOpen
  \bibfield  {author} {\bibinfo {author} {\bibfnamefont {T.}~\bibnamefont
  {Ozaki}}\ and\ \bibinfo {author} {\bibfnamefont {H.}~\bibnamefont {Kino}},\
  }\bibfield  {title} {\enquote {\bibinfo {title} {Efficient projector
  expansion for the \textit{ab initio} lcao method},}\ }\href {\doibase
  10.1103/PhysRevB.72.045121} {\bibfield  {journal} {\bibinfo  {journal} {Phys.
  Rev. B}\ }\textbf {\bibinfo {volume} {72}},\ \bibinfo {pages} {045121}
  (\bibinfo {year} {2005})}\BibitemShut {NoStop}%
\bibitem [{\citenamefont {Weber}\ \emph {et~al.}(2008)\citenamefont {Weber},
  \citenamefont {VandeVondele}, \citenamefont {Hutter},\ and\ \citenamefont
  {Niklasson}}]{doi:10.1063/1.2841077}%
  \BibitemOpen
  \bibfield  {author} {\bibinfo {author} {\bibfnamefont {V.}~\bibnamefont
  {Weber}}, \bibinfo {author} {\bibfnamefont {J.}~\bibnamefont {VandeVondele}},
  \bibinfo {author} {\bibfnamefont {J.}~\bibnamefont {Hutter}}, \ and\ \bibinfo
  {author} {\bibfnamefont {A.~M.~N.}\ \bibnamefont {Niklasson}},\ }\bibfield
  {title} {\enquote {\bibinfo {title} {Direct energy functional minimization
  under orthogonality constraints},}\ }\href {\doibase 10.1063/1.2841077}
  {\bibfield  {journal} {\bibinfo  {journal} {The Journal of Chemical Physics}\
  }\textbf {\bibinfo {volume} {128}},\ \bibinfo {pages} {084113} (\bibinfo
  {year} {2008})}\BibitemShut {NoStop}%
\bibitem [{\citenamefont {Vuckovic}\ \emph {et~al.}(2015)\citenamefont
  {Vuckovic}, \citenamefont {Wagner}, \citenamefont {Mirtschink},\ and\
  \citenamefont {Gori-Giorgi}}]{vuckovic2015hydrogen}%
  \BibitemOpen
  \bibfield  {author} {\bibinfo {author} {\bibfnamefont {S.}~\bibnamefont
  {Vuckovic}}, \bibinfo {author} {\bibfnamefont {L.~O.}\ \bibnamefont
  {Wagner}}, \bibinfo {author} {\bibfnamefont {A.}~\bibnamefont {Mirtschink}},
  \ and\ \bibinfo {author} {\bibfnamefont {P.}~\bibnamefont {Gori-Giorgi}},\
  }\bibfield  {title} {\enquote {\bibinfo {title} {Hydrogen molecule
  dissociation curve with functionals based on the strictly correlated
  regime},}\ }\href {http://pubs.acs.org/doi/abs/10.1021/acs.jctc.5b00387}
  {\bibfield  {journal} {\bibinfo  {journal} {J. Chem. Theory Comput.}\
  }\textbf {\bibinfo {volume} {11}},\ \bibinfo {pages} {3153--3162} (\bibinfo
  {year} {2015})}\BibitemShut {NoStop}%
\bibitem [{\citenamefont {Wu}, \citenamefont {Cheng},\ and\ \citenamefont
  {Van~Voorhis}(2007)}]{:/content/aip/journal/jcp/127/16/10.1063/1.2800022}%
  \BibitemOpen
  \bibfield  {author} {\bibinfo {author} {\bibfnamefont {Q.}~\bibnamefont
  {Wu}}, \bibinfo {author} {\bibfnamefont {C.-L.}\ \bibnamefont {Cheng}}, \
  and\ \bibinfo {author} {\bibfnamefont {T.}~\bibnamefont {Van~Voorhis}},\
  }\bibfield  {title} {\enquote {\bibinfo {title} {Configuration interaction
  based on constrained density functional theory: A multireference method},}\
  }\href
  {http://scitation.aip.org/content/aip/journal/jcp/127/16/10.1063/1.2800022}
  {\bibfield  {journal} {\bibinfo  {journal} {J. Chem. Phys.}\ }\textbf
  {\bibinfo {volume} {127}},\ \bibinfo {eid} {164119} (\bibinfo {year}
  {2007})}\BibitemShut {NoStop}%
\bibitem [{\citenamefont {Miyake}\ and\ \citenamefont
  {Aryasetiawan}(2008)}]{PhysRevB.77.085122}%
  \BibitemOpen
  \bibfield  {author} {\bibinfo {author} {\bibfnamefont {T.}~\bibnamefont
  {Miyake}}\ and\ \bibinfo {author} {\bibfnamefont {F.}~\bibnamefont
  {Aryasetiawan}},\ }\bibfield  {title} {\enquote {\bibinfo {title} {Screened
  {C}oulomb interaction in the maximally localized {W}annier basis},}\ }\href
  {\doibase 10.1103/PhysRevB.77.085122} {\bibfield  {journal} {\bibinfo
  {journal} {Phys. Rev. B}\ }\textbf {\bibinfo {volume} {77}},\ \bibinfo
  {pages} {085122} (\bibinfo {year} {2008})}\BibitemShut {NoStop}%
\bibitem [{\citenamefont {Miyake}, \citenamefont {Aryasetiawan},\ and\
  \citenamefont {Imada}(2009)}]{PhysRevB.80.155134}%
  \BibitemOpen
  \bibfield  {author} {\bibinfo {author} {\bibfnamefont {T.}~\bibnamefont
  {Miyake}}, \bibinfo {author} {\bibfnamefont {F.}~\bibnamefont
  {Aryasetiawan}}, \ and\ \bibinfo {author} {\bibfnamefont {M.}~\bibnamefont
  {Imada}},\ }\bibfield  {title} {\enquote {\bibinfo {title} {\textit{Ab
  initio} procedure for constructing effective models of correlated materials
  with entangled band structure},}\ }\href {\doibase
  10.1103/PhysRevB.80.155134} {\bibfield  {journal} {\bibinfo  {journal} {Phys.
  Rev. B}\ }\textbf {\bibinfo {volume} {80}},\ \bibinfo {pages} {155134}
  (\bibinfo {year} {2009})}\BibitemShut {NoStop}%
\bibitem [{\citenamefont {O'Regan}\ and\ \citenamefont
  {Teobaldi}(2016)}]{PhysRevB.94.035159}%
  \BibitemOpen
  \bibfield  {author} {\bibinfo {author} {\bibfnamefont {D.~D.}\ \bibnamefont
  {O'Regan}}\ and\ \bibinfo {author} {\bibfnamefont {G.}~\bibnamefont
  {Teobaldi}},\ }\bibfield  {title} {\enquote {\bibinfo {title} {Optimization
  of constrained density functional theory},}\ }\href {\doibase
  10.1103/PhysRevB.94.035159} {\bibfield  {journal} {\bibinfo  {journal} {Phys.
  Rev. B}\ }\textbf {\bibinfo {volume} {94}},\ \bibinfo {pages} {035159}
  (\bibinfo {year} {2016})}\BibitemShut {NoStop}%
\bibitem [{\citenamefont {Ong}, \citenamefont {Chevrier},\ and\ \citenamefont
  {Ceder}(2011)}]{PhysRevB.83.075112}%
  \BibitemOpen
  \bibfield  {author} {\bibinfo {author} {\bibfnamefont {S.~P.}\ \bibnamefont
  {Ong}}, \bibinfo {author} {\bibfnamefont {V.~L.}\ \bibnamefont {Chevrier}}, \
  and\ \bibinfo {author} {\bibfnamefont {G.}~\bibnamefont {Ceder}},\ }\bibfield
   {title} {\enquote {\bibinfo {title} {Comparison of small polaron migration
  and phase separation in olivine $\mathrm{LiMnPO_4}$ and $\mathrm{LiFePO_4}$
  using hybrid density-functional theory},}\ }\href {\doibase
  10.1103/PhysRevB.83.075112} {\bibfield  {journal} {\bibinfo  {journal} {Phys.
  Rev. B}\ }\textbf {\bibinfo {volume} {83}},\ \bibinfo {pages} {075112}
  (\bibinfo {year} {2011})}\BibitemShut {NoStop}%
\bibitem [{\citenamefont {Kulik}\ and\ \citenamefont
  {Marzari}(2008)}]{kulik2008self}%
  \BibitemOpen
  \bibfield  {author} {\bibinfo {author} {\bibfnamefont {H.~J.}\ \bibnamefont
  {Kulik}}\ and\ \bibinfo {author} {\bibfnamefont {N.}~\bibnamefont
  {Marzari}},\ }\bibfield  {title} {\enquote {\bibinfo {title} {A
  self-consistent {H}ubbard \textit{U} density-functional theory approach to
  the addition-elimination reactions of hydrocarbons on bare
  $\mathrm{FeO+}$},}\ }\href
  {http://scitation.aip.org/content/aip/journal/jcp/129/13/10.1063/1.2987444}
  {\bibfield  {journal} {\bibinfo  {journal} {J. Chem. Phys.}\ }\textbf
  {\bibinfo {volume} {129}},\ \bibinfo {pages} {134314} (\bibinfo {year}
  {2008})}\BibitemShut {NoStop}%
\bibitem [{\citenamefont {Kulik}\ and\ \citenamefont
  {Marzari}(2011{\natexlab{b}})}]{kulik2011transition}%
  \BibitemOpen
  \bibfield  {author} {\bibinfo {author} {\bibfnamefont {H.~J.}\ \bibnamefont
  {Kulik}}\ and\ \bibinfo {author} {\bibfnamefont {N.}~\bibnamefont
  {Marzari}},\ }\bibfield  {title} {\enquote {\bibinfo {title}
  {Transition-metal dioxides: {A} case for the intersite term in
  {H}ubbard-model functionals},}\ }\href
  {http://scitation.aip.org/content/aip/journal/jcp/134/9/10.1063/1.3559452}
  {\bibfield  {journal} {\bibinfo  {journal} {J. Chem. Phys.}\ }\textbf
  {\bibinfo {volume} {134}},\ \bibinfo {pages} {094103} (\bibinfo {year}
  {2011}{\natexlab{b}})}\BibitemShut {NoStop}%
\bibitem [{\citenamefont {Hsu}\ \emph {et~al.}(2011)\citenamefont {Hsu},
  \citenamefont {Blaha}, \citenamefont {Cococcioni},\ and\ \citenamefont
  {Wentzcovitch}}]{PhysRevLett.106.118501}%
  \BibitemOpen
  \bibfield  {author} {\bibinfo {author} {\bibfnamefont {H.}~\bibnamefont
  {Hsu}}, \bibinfo {author} {\bibfnamefont {P.}~\bibnamefont {Blaha}}, \bibinfo
  {author} {\bibfnamefont {M.}~\bibnamefont {Cococcioni}}, \ and\ \bibinfo
  {author} {\bibfnamefont {R.~M.}\ \bibnamefont {Wentzcovitch}},\ }\bibfield
  {title} {\enquote {\bibinfo {title} {Spin-state crossover and hyperfine
  interactions of ferric iron in $\mathrm{MgSiO_3}$ perovskite},}\ }\href
  {\doibase 10.1103/PhysRevLett.106.118501} {\bibfield  {journal} {\bibinfo
  {journal} {Phys. Rev. Lett.}\ }\textbf {\bibinfo {volume} {106}},\ \bibinfo
  {pages} {118501} (\bibinfo {year} {2011})}\BibitemShut {NoStop}%
\bibitem [{\citenamefont {Youmbi}\ and\ \citenamefont
  {Calvayrac}(2014)}]{Youmbi20141}%
  \BibitemOpen
  \bibfield  {author} {\bibinfo {author} {\bibfnamefont {B.~S.}\ \bibnamefont
  {Youmbi}}\ and\ \bibinfo {author} {\bibfnamefont {F.}~\bibnamefont
  {Calvayrac}},\ }\bibfield  {title} {\enquote {\bibinfo {title} {Structure of
  $\mathrm{CoO}$(001) surface from $\mathrm{DFT}+\mathrm{U}$ calculations},}\
  }\href {\doibase http://dx.doi.org/10.1016/j.susc.2013.10.012} {\bibfield
  {journal} {\bibinfo  {journal} {Surf. Sci.}\ }\textbf {\bibinfo {volume}
  {621}},\ \bibinfo {pages} {1 -- 6} (\bibinfo {year} {2014})}\BibitemShut
  {NoStop}%
\bibitem [{\citenamefont {Aharbil}\ \emph {et~al.}(2016)\citenamefont
  {Aharbil}, \citenamefont {Labrim}, \citenamefont {Benmokhtar}, \citenamefont
  {Haddouch}, \citenamefont {Bahmad},\ and\ \citenamefont
  {Laanab}}]{2053-1591-3-8-086104}%
  \BibitemOpen
  \bibfield  {author} {\bibinfo {author} {\bibfnamefont {Y.}~\bibnamefont
  {Aharbil}}, \bibinfo {author} {\bibfnamefont {H.}~\bibnamefont {Labrim}},
  \bibinfo {author} {\bibfnamefont {S.}~\bibnamefont {Benmokhtar}}, \bibinfo
  {author} {\bibfnamefont {M.~A.}\ \bibnamefont {Haddouch}}, \bibinfo {author}
  {\bibfnamefont {L.}~\bibnamefont {Bahmad}}, \ and\ \bibinfo {author}
  {\bibfnamefont {L.}~\bibnamefont {Laanab}},\ }\bibfield  {title} {\enquote
  {\bibinfo {title} {Self-consistent ($\mathrm{DFT}+\mathrm{U}$) study of
  electronic, structural and magnetic properties in $\mathrm{A_2NiMoO_6}$
  ({A = Ba, Sr}) compounds},}\ }\href
  {http://stacks.iop.org/2053-1591/3/i=8/a=086104} {\bibfield  {journal}
  {\bibinfo  {journal} {Mater. Res. Express}\ }\textbf {\bibinfo {volume}
  {3}},\ \bibinfo {pages} {086104} (\bibinfo {year} {2016})}\BibitemShut
  {NoStop}%
\bibitem [{\citenamefont {Gani}\ and\ \citenamefont
  {Kulik}(2016)}]{doi:10.1021/acs.jctc.6b00937}%
  \BibitemOpen
  \bibfield  {author} {\bibinfo {author} {\bibfnamefont {T.~Z.~H.}\
  \bibnamefont {Gani}}\ and\ \bibinfo {author} {\bibfnamefont {H.~J.}\
  \bibnamefont {Kulik}},\ }\bibfield  {title} {\enquote {\bibinfo {title}
  {Where does the density localize? {C}onvergent behavior for global hybrids,
  range separation, and {DFT+U}},}\ }\href {\doibase 10.1021/acs.jctc.6b00937}
  {\bibfield  {journal} {\bibinfo  {journal} {J. Chem. Theory Comput.}\
  }\textbf {\bibinfo {volume} {12}},\ \bibinfo {pages} {5931--5945} (\bibinfo
  {year} {2016})}\BibitemShut {NoStop}%
\bibitem [{\citenamefont {Hsu}\ \emph {et~al.}(2010)\citenamefont {Hsu},
  \citenamefont {Umemoto}, \citenamefont {Blaha},\ and\ \citenamefont
  {Wentzcovitch}}]{Hsu201019}%
  \BibitemOpen
  \bibfield  {author} {\bibinfo {author} {\bibfnamefont {H.}~\bibnamefont
  {Hsu}}, \bibinfo {author} {\bibfnamefont {K.}~\bibnamefont {Umemoto}},
  \bibinfo {author} {\bibfnamefont {P.}~\bibnamefont {Blaha}}, \ and\ \bibinfo
  {author} {\bibfnamefont {R.~M.}\ \bibnamefont {Wentzcovitch}},\ }\bibfield
  {title} {\enquote {\bibinfo {title} {Spin states and hyperfine interactions
  of iron in ({Mg,Fe})$\mathrm{SiO_3}$ perovskite under pressure},}\ }\href
  {\doibase http://dx.doi.org/10.1016/j.epsl.2010.02.031} {\bibfield  {journal}
  {\bibinfo  {journal} {Earth and Planetary Science Letters}\ }\textbf
  {\bibinfo {volume} {294}},\ \bibinfo {pages} {19 -- 26} (\bibinfo {year}
  {2010})}\BibitemShut {NoStop}%
\bibitem [{\citenamefont {Mattioli}\ \emph {et~al.}(2008)\citenamefont
  {Mattioli}, \citenamefont {Filippone}, \citenamefont {Alippi},\ and\
  \citenamefont {Amore~Bonapasta}}]{PhysRevB.78.241201}%
  \BibitemOpen
  \bibfield  {author} {\bibinfo {author} {\bibfnamefont {G.}~\bibnamefont
  {Mattioli}}, \bibinfo {author} {\bibfnamefont {F.}~\bibnamefont {Filippone}},
  \bibinfo {author} {\bibfnamefont {P.}~\bibnamefont {Alippi}}, \ and\ \bibinfo
  {author} {\bibfnamefont {A.}~\bibnamefont {Amore~Bonapasta}},\ }\bibfield
  {title} {\enquote {\bibinfo {title} {\textit{Ab initio} study of the
  electronic states induced by oxygen vacancies in rutile and anatase
  $\mathrm{TiO_2}$},}\ }\href {\doibase 10.1103/PhysRevB.78.241201} {\bibfield
  {journal} {\bibinfo  {journal} {Phys. Rev. B}\ }\textbf {\bibinfo {volume}
  {78}},\ \bibinfo {pages} {241201} (\bibinfo {year} {2008})}\BibitemShut
  {NoStop}%
\bibitem [{\citenamefont {Kowalski}\ \emph {et~al.}(2010)\citenamefont
  {Kowalski}, \citenamefont {Camellone}, \citenamefont {Nair}, \citenamefont
  {Meyer},\ and\ \citenamefont {Marx}}]{PhysRevLett.105.146405}%
  \BibitemOpen
  \bibfield  {author} {\bibinfo {author} {\bibfnamefont {P.~M.}\ \bibnamefont
  {Kowalski}}, \bibinfo {author} {\bibfnamefont {M.~F.}\ \bibnamefont
  {Camellone}}, \bibinfo {author} {\bibfnamefont {N.~N.}\ \bibnamefont {Nair}},
  \bibinfo {author} {\bibfnamefont {B.}~\bibnamefont {Meyer}}, \ and\ \bibinfo
  {author} {\bibfnamefont {D.}~\bibnamefont {Marx}},\ }\bibfield  {title}
  {\enquote {\bibinfo {title} {Charge localization dynamics induced by oxygen
  vacancies on the $\mathrm{TiO_2}(110)$ surface},}\ }\href {\doibase
  10.1103/PhysRevLett.105.146405} {\bibfield  {journal} {\bibinfo  {journal}
  {Phys. Rev. Lett.}\ }\textbf {\bibinfo {volume} {105}},\ \bibinfo {pages}
  {146405} (\bibinfo {year} {2010})}\BibitemShut {NoStop}%
\bibitem [{\citenamefont {Mattioli}\ \emph {et~al.}(2010)\citenamefont
  {Mattioli}, \citenamefont {Alippi}, \citenamefont {Filippone}, \citenamefont
  {Caminiti},\ and\ \citenamefont {Amore~Bonapasta}}]{doi:10.1021/jp1041316}%
  \BibitemOpen
  \bibfield  {author} {\bibinfo {author} {\bibfnamefont {G.}~\bibnamefont
  {Mattioli}}, \bibinfo {author} {\bibfnamefont {P.}~\bibnamefont {Alippi}},
  \bibinfo {author} {\bibfnamefont {F.}~\bibnamefont {Filippone}}, \bibinfo
  {author} {\bibfnamefont {R.}~\bibnamefont {Caminiti}}, \ and\ \bibinfo
  {author} {\bibfnamefont {A.}~\bibnamefont {Amore~Bonapasta}},\ }\bibfield
  {title} {\enquote {\bibinfo {title} {Deep versus shallow behavior of
  intrinsic defects in rutile and anatase $\mathrm{TiO_2}$ polymorphs},}\
  }\href {\doibase 10.1021/jp1041316} {\bibfield  {journal} {\bibinfo
  {journal} {J. Phys. Chem. C}\ }\textbf {\bibinfo {volume} {114}},\ \bibinfo
  {pages} {21694--21704} (\bibinfo {year} {2010})}\BibitemShut {NoStop}%
\bibitem [{\citenamefont {Gougoussis}\ \emph {et~al.}(2009)\citenamefont
  {Gougoussis}, \citenamefont {Calandra}, \citenamefont {Seitsonen},\ and\
  \citenamefont {Mauri}}]{PhysRevB.80.075102}%
  \BibitemOpen
  \bibfield  {author} {\bibinfo {author} {\bibfnamefont {C.}~\bibnamefont
  {Gougoussis}}, \bibinfo {author} {\bibfnamefont {M.}~\bibnamefont
  {Calandra}}, \bibinfo {author} {\bibfnamefont {A.~P.}\ \bibnamefont
  {Seitsonen}}, \ and\ \bibinfo {author} {\bibfnamefont {F.}~\bibnamefont
  {Mauri}},\ }\bibfield  {title} {\enquote {\bibinfo {title} {First-principles
  calculations of x-ray absorption in a scheme based on ultrasoft
  pseudopotentials: From $\ensuremath{\alpha}$-quartz to high-${T}_{c}$
  compounds},}\ }\href {\doibase 10.1103/PhysRevB.80.075102} {\bibfield
  {journal} {\bibinfo  {journal} {Phys. Rev. B}\ }\textbf {\bibinfo {volume}
  {80}},\ \bibinfo {pages} {075102} (\bibinfo {year} {2009})}\BibitemShut
  {NoStop}%
\bibitem [{\citenamefont {Mann}\ \emph {et~al.}(2016)\citenamefont {Mann},
  \citenamefont {Lee}, \citenamefont {Cococcioni}, \citenamefont {Smit},\ and\
  \citenamefont {Neaton}}]{doi:10.1063/1.4947240}%
  \BibitemOpen
  \bibfield  {author} {\bibinfo {author} {\bibfnamefont {G.~W.}\ \bibnamefont
  {Mann}}, \bibinfo {author} {\bibfnamefont {K.}~\bibnamefont {Lee}}, \bibinfo
  {author} {\bibfnamefont {M.}~\bibnamefont {Cococcioni}}, \bibinfo {author}
  {\bibfnamefont {B.}~\bibnamefont {Smit}}, \ and\ \bibinfo {author}
  {\bibfnamefont {J.~B.}\ \bibnamefont {Neaton}},\ }\bibfield  {title}
  {\enquote {\bibinfo {title} {First-principles {H}ubbard {U} approach for
  small molecule binding in metal-organic frameworks},}\ }\href {\doibase
  10.1063/1.4947240} {\bibfield  {journal} {\bibinfo  {journal} {J. Chem.
  Phys.}\ }\textbf {\bibinfo {volume} {144}},\ \bibinfo {pages} {174104}
  (\bibinfo {year} {2016})}\BibitemShut {NoStop}%
\bibitem [{\citenamefont {O'Regan}, \citenamefont {Payne},\ and\ \citenamefont
  {Mostofi}(2012)}]{PhysRevB.85.193101}%
  \BibitemOpen
  \bibfield  {author} {\bibinfo {author} {\bibfnamefont {D.~D.}\ \bibnamefont
  {O'Regan}}, \bibinfo {author} {\bibfnamefont {M.~C.}\ \bibnamefont {Payne}},
  \ and\ \bibinfo {author} {\bibfnamefont {A.~A.}\ \bibnamefont {Mostofi}},\
  }\bibfield  {title} {\enquote {\bibinfo {title} {Generalized wannier
  functions: A comparison of molecular electric dipole polarizabilities},}\
  }\href@noop {} {\bibfield  {journal} {\bibinfo  {journal} {Phys. Rev. B}\
  }\textbf {\bibinfo {volume} {85}},\ \bibinfo {pages} {193101} (\bibinfo
  {year} {2012})}\BibitemShut {NoStop}%
\bibitem [{\citenamefont {Herzberg}\ and\ \citenamefont
  {Jungen}(1972)}]{HERZBERG1972425}%
  \BibitemOpen
  \bibfield  {author} {\bibinfo {author} {\bibfnamefont {G.}~\bibnamefont
  {Herzberg}}\ and\ \bibinfo {author} {\bibfnamefont {C.}~\bibnamefont
  {Jungen}},\ }\bibfield  {title} {\enquote {\bibinfo {title} {Rydberg series
  and ionization potential of the $\mathrm{H_2}$ molecule},}\ }\href {\doibase
  http://dx.doi.org/10.1016/0022-2852(72)90064-1} {\bibfield  {journal}
  {\bibinfo  {journal} {J. Mol. Spectrosc.}\ }\textbf {\bibinfo {volume}
  {41}},\ \bibinfo {pages} {425 -- 486} (\bibinfo {year} {1972})}\BibitemShut
  {NoStop}%
\bibitem [{\citenamefont {Ping~Ong}\ \emph {et~al.}(2008)\citenamefont
  {Ping~Ong}, \citenamefont {Wang}, \citenamefont {Kang},\ and\ \citenamefont
  {Ceder}}]{doi:10.1021/cm702327g}%
  \BibitemOpen
  \bibfield  {author} {\bibinfo {author} {\bibfnamefont {S.}~\bibnamefont
  {Ping~Ong}}, \bibinfo {author} {\bibfnamefont {L.}~\bibnamefont {Wang}},
  \bibinfo {author} {\bibfnamefont {B.}~\bibnamefont {Kang}}, \ and\ \bibinfo
  {author} {\bibfnamefont {G.}~\bibnamefont {Ceder}},\ }\bibfield  {title}
  {\enquote {\bibinfo {title} {$\mathrm{Li−Fe−P−O_2}$ phase diagram from
  first principles calculations},}\ }\href {\doibase 10.1021/cm702327g}
  {\bibfield  {journal} {\bibinfo  {journal} {Chem. Mater.}\ }\textbf {\bibinfo
  {volume} {20}},\ \bibinfo {pages} {1798--1807} (\bibinfo {year}
  {2008})}\BibitemShut {NoStop}%
\bibitem [{\citenamefont {Capdevila-Cortada}, \citenamefont
  {Garc\'ia-Melchor},\ and\ \citenamefont
  {L\'opez}(2015)}]{CapdevilaCortada201558}%
  \BibitemOpen
  \bibfield  {author} {\bibinfo {author} {\bibfnamefont {M.}~\bibnamefont
  {Capdevila-Cortada}}, \bibinfo {author} {\bibfnamefont {M.}~\bibnamefont
  {Garc\'ia-Melchor}}, \ and\ \bibinfo {author} {\bibfnamefont
  {N.}~\bibnamefont {L\'opez}},\ }\bibfield  {title} {\enquote {\bibinfo
  {title} {Unraveling the structure sensitivity in methanol conversion on
  $\mathrm{CeO_2}$: {A} {DFT+U} study},}\ }\href {\doibase
  http://dx.doi.org/10.1016/j.jcat.2015.04.016} {\bibfield  {journal} {\bibinfo
   {journal} {J. Catal.}\ }\textbf {\bibinfo {volume} {327}},\ \bibinfo {pages}
  {58 -- 64} (\bibinfo {year} {2015})}\BibitemShut {NoStop}%
\bibitem [{\citenamefont {Hautier}\ \emph {et~al.}(2012)\citenamefont
  {Hautier}, \citenamefont {Ong}, \citenamefont {Jain}, \citenamefont {Moore},\
  and\ \citenamefont {Ceder}}]{PhysRevB.85.155208}%
  \BibitemOpen
  \bibfield  {author} {\bibinfo {author} {\bibfnamefont {G.}~\bibnamefont
  {Hautier}}, \bibinfo {author} {\bibfnamefont {S.~P.}\ \bibnamefont {Ong}},
  \bibinfo {author} {\bibfnamefont {A.}~\bibnamefont {Jain}}, \bibinfo {author}
  {\bibfnamefont {C.~J.}\ \bibnamefont {Moore}}, \ and\ \bibinfo {author}
  {\bibfnamefont {G.}~\bibnamefont {Ceder}},\ }\bibfield  {title} {\enquote
  {\bibinfo {title} {Accuracy of density-functional theory in predicting
  formation energies of ternary oxides from binary oxides and its implication
  on phase stability},}\ }\href {\doibase 10.1103/PhysRevB.85.155208}
  {\bibfield  {journal} {\bibinfo  {journal} {Phys. Rev. B}\ }\textbf {\bibinfo
  {volume} {85}},\ \bibinfo {pages} {155208} (\bibinfo {year}
  {2012})}\BibitemShut {NoStop}%
\bibitem [{\citenamefont {Jain}\ \emph {et~al.}(2011)\citenamefont {Jain},
  \citenamefont {Hautier}, \citenamefont {Ong}, \citenamefont {Moore},
  \citenamefont {Fischer}, \citenamefont {Persson},\ and\ \citenamefont
  {Ceder}}]{PhysRevB.84.045115}%
  \BibitemOpen
  \bibfield  {author} {\bibinfo {author} {\bibfnamefont {A.}~\bibnamefont
  {Jain}}, \bibinfo {author} {\bibfnamefont {G.}~\bibnamefont {Hautier}},
  \bibinfo {author} {\bibfnamefont {S.~P.}\ \bibnamefont {Ong}}, \bibinfo
  {author} {\bibfnamefont {C.~J.}\ \bibnamefont {Moore}}, \bibinfo {author}
  {\bibfnamefont {C.~C.}\ \bibnamefont {Fischer}}, \bibinfo {author}
  {\bibfnamefont {K.~A.}\ \bibnamefont {Persson}}, \ and\ \bibinfo {author}
  {\bibfnamefont {G.}~\bibnamefont {Ceder}},\ }\bibfield  {title} {\enquote
  {\bibinfo {title} {Formation enthalpies by mixing {GGA} and {GGA+U}
  calculations},}\ }\href {\doibase 10.1103/PhysRevB.84.045115} {\bibfield
  {journal} {\bibinfo  {journal} {Phys. Rev. B}\ }\textbf {\bibinfo {volume}
  {84}},\ \bibinfo {pages} {045115} (\bibinfo {year} {2011})}\BibitemShut
  {NoStop}%
\bibitem [{\citenamefont {Stevanovi\ifmmode~\acute{c}\else \'{c}\fi{}}\ \emph
  {et~al.}(2012)\citenamefont {Stevanovi\ifmmode~\acute{c}\else \'{c}\fi{}},
  \citenamefont {Lany}, \citenamefont {Zhang},\ and\ \citenamefont
  {Zunger}}]{PhysRevB.85.115104}%
  \BibitemOpen
  \bibfield  {author} {\bibinfo {author} {\bibfnamefont {V.}~\bibnamefont
  {Stevanovi\ifmmode~\acute{c}\else \'{c}\fi{}}}, \bibinfo {author}
  {\bibfnamefont {S.}~\bibnamefont {Lany}}, \bibinfo {author} {\bibfnamefont
  {X.}~\bibnamefont {Zhang}}, \ and\ \bibinfo {author} {\bibfnamefont
  {A.}~\bibnamefont {Zunger}},\ }\bibfield  {title} {\enquote {\bibinfo {title}
  {Correcting density functional theory for accurate predictions of compound
  enthalpies of formation: Fitted elemental-phase reference energies},}\ }\href
  {\doibase 10.1103/PhysRevB.85.115104} {\bibfield  {journal} {\bibinfo
  {journal} {Phys. Rev. B}\ }\textbf {\bibinfo {volume} {85}},\ \bibinfo
  {pages} {115104} (\bibinfo {year} {2012})}\BibitemShut {NoStop}%
\bibitem [{\citenamefont {Curtarolo}\ \emph
  {et~al.}(2013{\natexlab{b}})\citenamefont {Curtarolo}, \citenamefont {Hart},
  \citenamefont {Nardelli}, \citenamefont {Mingo}, \citenamefont {Sanvito},\
  and\ \citenamefont {Levy}}]{curtarolo}%
  \BibitemOpen
  \bibfield  {author} {\bibinfo {author} {\bibfnamefont {S.}~\bibnamefont
  {Curtarolo}}, \bibinfo {author} {\bibfnamefont {G.~L.~W.}\ \bibnamefont
  {Hart}}, \bibinfo {author} {\bibfnamefont {M.~B.}\ \bibnamefont {Nardelli}},
  \bibinfo {author} {\bibfnamefont {N.}~\bibnamefont {Mingo}}, \bibinfo
  {author} {\bibfnamefont {S.}~\bibnamefont {Sanvito}}, \ and\ \bibinfo
  {author} {\bibfnamefont {O.}~\bibnamefont {Levy}},\ }\bibfield  {title}
  {\enquote {\bibinfo {title} {The high-throughput highway to computational
  materials design},}\ }\href {http://dx.doi.org/10.1038/nmat3568} {\bibfield
  {journal} {\bibinfo  {journal} {Nat. Mater.}\ }\textbf {\bibinfo {volume}
  {12}},\ \bibinfo {pages} {191--201} (\bibinfo {year}
  {2013}{\natexlab{b}})}\BibitemShut {NoStop}%
\bibitem [{\citenamefont {Curtarolo}\ \emph {et~al.}(2012)\citenamefont
  {Curtarolo}, \citenamefont {Setyawan}, \citenamefont {Hart}, \citenamefont
  {Jahnatek}, \citenamefont {Chepulskii}, \citenamefont {Taylor}, \citenamefont
  {Wang}, \citenamefont {Xue}, \citenamefont {Yang}, \citenamefont {Levy},
  \citenamefont {Mehl}, \citenamefont {Stokes}, \citenamefont {Demchenko},\
  and\ \citenamefont {Morgan}}]{Curtarolo2012218}%
  \BibitemOpen
  \bibfield  {author} {\bibinfo {author} {\bibfnamefont {S.}~\bibnamefont
  {Curtarolo}}, \bibinfo {author} {\bibfnamefont {W.}~\bibnamefont {Setyawan}},
  \bibinfo {author} {\bibfnamefont {G.~L.}\ \bibnamefont {Hart}}, \bibinfo
  {author} {\bibfnamefont {M.}~\bibnamefont {Jahnatek}}, \bibinfo {author}
  {\bibfnamefont {R.~V.}\ \bibnamefont {Chepulskii}}, \bibinfo {author}
  {\bibfnamefont {R.~H.}\ \bibnamefont {Taylor}}, \bibinfo {author}
  {\bibfnamefont {S.}~\bibnamefont {Wang}}, \bibinfo {author} {\bibfnamefont
  {J.}~\bibnamefont {Xue}}, \bibinfo {author} {\bibfnamefont {K.}~\bibnamefont
  {Yang}}, \bibinfo {author} {\bibfnamefont {O.}~\bibnamefont {Levy}}, \bibinfo
  {author} {\bibfnamefont {M.~J.}\ \bibnamefont {Mehl}}, \bibinfo {author}
  {\bibfnamefont {H.~T.}\ \bibnamefont {Stokes}}, \bibinfo {author}
  {\bibfnamefont {D.~O.}\ \bibnamefont {Demchenko}}, \ and\ \bibinfo {author}
  {\bibfnamefont {D.}~\bibnamefont {Morgan}},\ }\bibfield  {title} {\enquote
  {\bibinfo {title} {Aflow: An automatic framework for high-throughput
  materials discovery},}\ }\href {\doibase
  http://dx.doi.org/10.1016/j.commatsci.2012.02.005} {\bibfield  {journal}
  {\bibinfo  {journal} {Comput. Mater. Sci.}\ }\textbf {\bibinfo {volume}
  {58}},\ \bibinfo {pages} {218 -- 226} (\bibinfo {year} {2012})}\BibitemShut
  {NoStop}%
\bibitem [{\citenamefont {Agapito}, \citenamefont {Curtarolo},\ and\
  \citenamefont {Buongiorno~Nardelli}(2015)}]{PhysRevX.5.011006}%
  \BibitemOpen
  \bibfield  {author} {\bibinfo {author} {\bibfnamefont {L.~A.}\ \bibnamefont
  {Agapito}}, \bibinfo {author} {\bibfnamefont {S.}~\bibnamefont {Curtarolo}},
  \ and\ \bibinfo {author} {\bibfnamefont {M.}~\bibnamefont
  {Buongiorno~Nardelli}},\ }\bibfield  {title} {\enquote {\bibinfo {title}
  {Reformulation of {DFT+U} as a pseudohybrid {H}ubbard density functional for
  accelerated materials discovery},}\ }\href {\doibase
  10.1103/PhysRevX.5.011006} {\bibfield  {journal} {\bibinfo  {journal} {Phys.
  Rev. X}\ }\textbf {\bibinfo {volume} {5}},\ \bibinfo {pages} {011006}
  (\bibinfo {year} {2015})}\BibitemShut {NoStop}%
\bibitem [{\citenamefont {Toroker}\ \emph {et~al.}(2011)\citenamefont
  {Toroker}, \citenamefont {Kanan}, \citenamefont {Alidoust}, \citenamefont
  {Isseroff}, \citenamefont {Liao},\ and\ \citenamefont {Carter}}]{C1CP22128K}%
  \BibitemOpen
  \bibfield  {author} {\bibinfo {author} {\bibfnamefont {M.~C.}\ \bibnamefont
  {Toroker}}, \bibinfo {author} {\bibfnamefont {D.~K.}\ \bibnamefont {Kanan}},
  \bibinfo {author} {\bibfnamefont {N.}~\bibnamefont {Alidoust}}, \bibinfo
  {author} {\bibfnamefont {L.~Y.}\ \bibnamefont {Isseroff}}, \bibinfo {author}
  {\bibfnamefont {P.}~\bibnamefont {Liao}}, \ and\ \bibinfo {author}
  {\bibfnamefont {E.~A.}\ \bibnamefont {Carter}},\ }\bibfield  {title}
  {\enquote {\bibinfo {title} {First principles scheme to evaluate band edge
  positions in potential transition metal oxide photocatalysts and
  photoelectrodes},}\ }\href {\doibase 10.1039/C1CP22128K} {\bibfield
  {journal} {\bibinfo  {journal} {Phys. Chem. Chem. Phys.}\ }\textbf {\bibinfo
  {volume} {13}},\ \bibinfo {pages} {16644--16654} (\bibinfo {year}
  {2011})}\BibitemShut {NoStop}%
\bibitem [{\citenamefont {O'Rourke}\ and\ \citenamefont
  {Bowler}(2014)}]{doi:10.1021/jp407736f}%
  \BibitemOpen
  \bibfield  {author} {\bibinfo {author} {\bibfnamefont {C.}~\bibnamefont
  {O'Rourke}}\ and\ \bibinfo {author} {\bibfnamefont {D.~R.}\ \bibnamefont
  {Bowler}},\ }\bibfield  {title} {\enquote {\bibinfo {title} {Intrinsic oxygen
  vacancy and extrinsic aluminum dopant interplay: A route to the restoration
  of defective $\mathrm{TiO_2}$},}\ }\href {\doibase 10.1021/jp407736f}
  {\bibfield  {journal} {\bibinfo  {journal} {J. Phys. Chem. C}\ }\textbf
  {\bibinfo {volume} {118}},\ \bibinfo {pages} {7261--7271} (\bibinfo {year}
  {2014})}\BibitemShut {NoStop}%
\bibitem [{\citenamefont {Yan}\ and\ \citenamefont
  {N\o{}rskov}(2013)}]{PhysRevB.88.245204}%
  \BibitemOpen
  \bibfield  {author} {\bibinfo {author} {\bibfnamefont {J.}~\bibnamefont
  {Yan}}\ and\ \bibinfo {author} {\bibfnamefont {J.~K.}\ \bibnamefont
  {N\o{}rskov}},\ }\bibfield  {title} {\enquote {\bibinfo {title} {Calculated
  formation and reaction energies of $3d$ transition metal oxides using a
  hierarchy of exchange-correlation functionals},}\ }\href {\doibase
  10.1103/PhysRevB.88.245204} {\bibfield  {journal} {\bibinfo  {journal} {Phys.
  Rev. B}\ }\textbf {\bibinfo {volume} {88}},\ \bibinfo {pages} {245204}
  (\bibinfo {year} {2013})}\BibitemShut {NoStop}%
\bibitem [{\citenamefont {Baldoni}\ \emph {et~al.}(2013)\citenamefont
  {Baldoni}, \citenamefont {Craco}, \citenamefont {Seifert},\ and\
  \citenamefont {Leoni}}]{C2TA00839D}%
  \BibitemOpen
  \bibfield  {author} {\bibinfo {author} {\bibfnamefont {M.}~\bibnamefont
  {Baldoni}}, \bibinfo {author} {\bibfnamefont {L.}~\bibnamefont {Craco}},
  \bibinfo {author} {\bibfnamefont {G.}~\bibnamefont {Seifert}}, \ and\
  \bibinfo {author} {\bibfnamefont {S.}~\bibnamefont {Leoni}},\ }\bibfield
  {title} {\enquote {\bibinfo {title} {A two-electron mechanism of lithium
  insertion into layered $\alpha$-$\mathrm{MoO_3}$: a {DFT} and {DFT+U}
  study},}\ }\href {\doibase 10.1039/C2TA00839D} {\bibfield  {journal}
  {\bibinfo  {journal} {J. Mater. Chem. A}\ }\textbf {\bibinfo {volume} {1}},\
  \bibinfo {pages} {1778--1784} (\bibinfo {year} {2013})}\BibitemShut {NoStop}%
\bibitem [{\citenamefont {Huang}\ \emph {et~al.}(2015)\citenamefont {Huang},
  \citenamefont {Wilson}, \citenamefont {Wang}, \citenamefont {Fang},
  \citenamefont {Buffington}, \citenamefont {Stein},\ and\ \citenamefont
  {Truhlar}}]{doi:10.1021/jacs.5b04690}%
  \BibitemOpen
  \bibfield  {author} {\bibinfo {author} {\bibfnamefont {S.}~\bibnamefont
  {Huang}}, \bibinfo {author} {\bibfnamefont {B.~E.}\ \bibnamefont {Wilson}},
  \bibinfo {author} {\bibfnamefont {B.}~\bibnamefont {Wang}}, \bibinfo {author}
  {\bibfnamefont {Y.}~\bibnamefont {Fang}}, \bibinfo {author} {\bibfnamefont
  {K.}~\bibnamefont {Buffington}}, \bibinfo {author} {\bibfnamefont
  {A.}~\bibnamefont {Stein}}, \ and\ \bibinfo {author} {\bibfnamefont {D.~G.}\
  \bibnamefont {Truhlar}},\ }\bibfield  {title} {\enquote {\bibinfo {title}
  {Y-doped $\mathrm{Li_8ZrO_6}$: A {L}i-ion battery cathode material with high
  capacity},}\ }\href {\doibase 10.1021/jacs.5b04690} {\bibfield  {journal}
  {\bibinfo  {journal} {J. Am. Chem. Soc.}\ }\textbf {\bibinfo {volume}
  {137}},\ \bibinfo {pages} {10992--11003} (\bibinfo {year}
  {2015})}\BibitemShut {NoStop}%
\bibitem [{\citenamefont {Huang}\ \emph {et~al.}(2016)\citenamefont {Huang},
  \citenamefont {Wilson}, \citenamefont {Smyrl}, \citenamefont {Truhlar},\ and\
  \citenamefont {Stein}}]{doi:10.1021/acs.chemmater.5b03554}%
  \BibitemOpen
  \bibfield  {author} {\bibinfo {author} {\bibfnamefont {S.}~\bibnamefont
  {Huang}}, \bibinfo {author} {\bibfnamefont {B.~E.}\ \bibnamefont {Wilson}},
  \bibinfo {author} {\bibfnamefont {W.~H.}\ \bibnamefont {Smyrl}}, \bibinfo
  {author} {\bibfnamefont {D.~G.}\ \bibnamefont {Truhlar}}, \ and\ \bibinfo
  {author} {\bibfnamefont {A.}~\bibnamefont {Stein}},\ }\bibfield  {title}
  {\enquote {\bibinfo {title} {Transition-metal-doped $\mathrm{M-Li_8ZrO_6}$
  ({M = Mn, Fe, Co, Ni, Cu, Ce}) as high-specific-capacity {L}i-ion battery
  cathode materials: {S}ynthesis, electrochemistry, and quantum mechanical
  characterization},}\ }\href {\doibase 10.1021/acs.chemmater.5b03554}
  {\bibfield  {journal} {\bibinfo  {journal} {Chem. Mater.}\ }\textbf {\bibinfo
  {volume} {28}},\ \bibinfo {pages} {746--755} (\bibinfo {year}
  {2016})}\BibitemShut {NoStop}%
\bibitem [{\citenamefont {Rappe}\ \emph {et~al.}(1990)\citenamefont {Rappe},
  \citenamefont {Rabe}, \citenamefont {Kaxiras},\ and\ \citenamefont
  {Joannopoulos}}]{PhysRevB.41.1227}%
  \BibitemOpen
  \bibfield  {author} {\bibinfo {author} {\bibfnamefont {A.~M.}\ \bibnamefont
  {Rappe}}, \bibinfo {author} {\bibfnamefont {K.~M.}\ \bibnamefont {Rabe}},
  \bibinfo {author} {\bibfnamefont {E.}~\bibnamefont {Kaxiras}}, \ and\
  \bibinfo {author} {\bibfnamefont {J.~D.}\ \bibnamefont {Joannopoulos}},\
  }\bibfield  {title} {\enquote {\bibinfo {title} {Optimized
  pseudopotentials},}\ }\href {\doibase 10.1103/PhysRevB.41.1227} {\bibfield
  {journal} {\bibinfo  {journal} {Phys. Rev. B}\ }\textbf {\bibinfo {volume}
  {41}},\ \bibinfo {pages} {1227--1230} (\bibinfo {year} {1990})}\BibitemShut
  {NoStop}%
\bibitem [{\citenamefont {Martyna}\ and\ \citenamefont
  {Tuckerman}(1999)}]{:/content/aip/journal/jcp/110/6/10.1063/1.477923}%
  \BibitemOpen
  \bibfield  {author} {\bibinfo {author} {\bibfnamefont {G.~J.}\ \bibnamefont
  {Martyna}}\ and\ \bibinfo {author} {\bibfnamefont {M.~E.}\ \bibnamefont
  {Tuckerman}},\ }\bibfield  {title} {\enquote {\bibinfo {title} {A reciprocal
  space based method for treating long range interactions in ab initio and
  force-field-based calculations in clusters},}\ }\href {\doibase
  dx.doi.org/10.1063/1.477923} {\bibfield  {journal} {\bibinfo  {journal} {J.
  Chem. Phys.}\ }\textbf {\bibinfo {volume} {110}},\ \bibinfo {pages}
  {2810--2821} (\bibinfo {year} {1999})}\BibitemShut {NoStop}%
\end{thebibliography}
\end{document}